\documentclass[aps,amsmath,showpacs,prd,nofootinbib]{revtex4}
\usepackage{graphicx}

\def\l{\left}
\def\r{\right}
\def\pa{\partial}
\def\be{\begin{equation}}
\def\ee{\end{equation}}
\def\bea{\begin{eqnarray}}
\def\eea{\end{eqnarray}}

\begin{document}

\title{Oscillons and Quasi-breathers in the $\phi^4$ Klein-Gordon model}

\author{Gyula Fodor$^1$, P\'eter Forg\'acs$^{1,2}$, Philippe Grandcl\'ement$^{2,3}$,
 Istv\'an R\'acz$^1$}
\affiliation{$^1$ MTA RMKI, H-1525 Budapest 114, P.O.Box 49, Hungary,
{$^2$LMPT, CNRS-UMR 6083, Universit\'e de Tours, Parc de Grandmont,
37200 Tours, FRANCE,}\\
$^3$LUTH,
CNRS-UMR 8102, Observatoire de Paris-Meudon, place Jules Janssen, 92195 Meudon Cedex, FRANCE}

 \begin{abstract}

Strong numerical evidence is presented for the existence of a continuous family of
time-periodic solutions with ``weak'' spatial localization
of the spherically symmetric non-linear Klein-Gordon equation
in 3+1 dimensions.
These solutions are ``weakly'' localized
in space in that they have slowly decaying oscillatory tails and
can be interpreted as localized standing waves (quasi-breathers).
By a detailed analysis of long-lived metastable states (oscillons)
formed during the time evolution it is demonstrated that the
oscillon states can be quantitatively described by the
weakly localized quasi-breathers.
It is found that the quasi-breathers and their oscillon counterparts exist
for a whole continuum of frequencies.
\end{abstract}

 \maketitle

\section{Introduction}\label{s:intro}

Nonlinear wave equations (NLWE) lie at the heart of many fields in physics including
hydrodynamics, classical integrable systems, Field Theories, etc.
Let us give a prototype class of NLWE for a real scalar field, $\phi$ in $1+n$ dimensional
space-time:
\be\label{NLWE}
-\frac{\pa^2\phi}{\pa t^2} + \Delta^{(n)} \phi = F(\phi)\,,\quad
\hbox{where}\quad {\Delta}^{(n)}=\sum_{i=1}^{n}\frac{\pa^2}{\pa x_i^2}\,,
\ee
where the real function, $F(\phi)$, defining the theory is given
for example as $F(\phi)=\phi(\phi^2-1)$ for the ``canonical'' $\phi^4$ model, and
$F(\phi)=\sin(\phi)$ for the Sine-Gordon (SG) model.

A particularly interesting class of solutions of NLWE is the class of nonsingular ones
exhibiting spatial localization. Such spatially localized solutions have finite energy and can
correspond to static particle-like objects or to various traveling waves.
In Field Theory localized static solutions have been quite intensively studied in a great
number of models in
various space-time dimensions, see e.g.\ the recent book of Sutcliffe and Manton
\cite{Sutcliffe-Manton}.

Spatially localized solutions with a non-trivial time dependence
(i.e. not simply in uniform motion) of NLWE are much harder to find.
In fact the simple qualitative argument stating that
{\sl ``anything that can radiate does radiate''}, indicates that
the time evolution of well localized Cauchy data leads {\sl generically} to either a
{\sl stable static} solution plus radiation fields, or the originally localized fields
disperse completely
due to the continuous loss of energy through radiation. Since Derrick-type scaling
arguments exclude the existence of localized static solutions in scalar field theories
given by Eq.\ (\ref{NLWE}) in more than two spatial dimensions, for $n>2$ one does not
expect to find localized
states at all at the end of a time evolution.
Simplifying somewhat the above, one could say that in the absence of static localized
solutions,
localized initial data cannot stay forever localized. In fact for localized initial data
there is a time scale which is the crossing time, $\tau_c$, (the time it takes for a wave
propagating with characteristic speed to cross the localized region) and a
priori one would expect the rapid dispersion of initially localized Cauchy data
within a few units of $\tau_c$.
One of the rare examples of a time-periodic solution, staying localized
forever is the famous ``breather'' in the $1+1$ dimensional Sine-Gordon (SG)
model.

In the $1+1$ dimensional ``canonical'' (with a double well potential) $\phi^4$ theory
the pioneering work, based on perturbation theory by Dashen, Hasslacher and Neveu
\cite{DHN} indicated the possible existence of breather-like solutions.
A completely independent numerical study by Kudryavtsev
\cite{Kudryav} has also indicated that suitable initial data
evolve into breather-type states.
These results stimulated a number of investigations about the possible existence of non-radiative
solutions in the $1+1$ dimensional $\phi^4$ model.
After a long history Segur and Kruskal \cite{Segur} and Vuillermot \cite{Vui} have
finally established that in spite of the above mentioned perturbative and numerical indications
time-periodic spatially localized finite energy solutions (breathers)
do not exist.
Even if genuine localized breathers in the $1+1$ dimensional $\phi^4$ model are absent,
in view of the perturbative and numerical evidence for the existence of long living
states ``close'' to genuine breathers, it is a natural question how to describe them.
Boyd has made a detailed study of time periodic solutions which are only {\sl weakly localized}
in space (i.e. the field $\phi$ possesses a slowly decreasing oscillatory tail) but as long
as the amplitude of the oscillatory ``wings'' are small they still
have a well defined core. Boyd has dubbed such solutions  ``nanopterons'' (small wings),
we refer to his book for a detailed review \cite{Boyd}.

Quite interestingly Bogoluvskii and Makhankov have found numerical evidence
for the existence of spatially localized breather-type states
in the spherically symmetric sector of $\phi^4$ theory in $3+1$ dimensions \cite{BogMak}.
These breather-like objects observed during the time evolution of some initial data
are called nowadays ``oscillons''.
Most of the observed oscillon states
are unstable having only a {\sl finite lifetime}. They lose their energy by radiating it
(slowly) to infinity.
More recent investigations started by Gleiser \cite{Gleiser, CopelGM95}
have revealed that oscillons do form in a fairly large class of scalar theories in various
spatial dimensions via the collapse of field configurations (initial data)
that interpolate between two vacuum states of a double well potential.
Such a spherically symmetric configuration corresponds to a bubble, where the interpolating
region is the bubble wall that separates the two vacuum states
at some characteristic radius. These works have led to a renewed interest in the subject.

It has been found that oscillons have extremely long lifetimes
which is already quite remarkable and makes them of quite some
interest. These long living oscillon states seem to occur in a
rather generic way in various field theories involving scalar
fields in even higher dimensional space-times \cite{Gleiser04} and
according to \cite{Farhi05} oscillons are also present in the
nonabelian SU(2) bosonic sector of the standard model of
electroweak interactions at least for certain values of the
pertinent couplings. Such oscillons might have important effects
on the inflationary scenario \cite{scenario} as they could form in
large numbers retaining a considerable amount of energy. In a recent study \cite{Graham06}
of a $1+1$ dimensional scalar theory on an expanding background exhibited
very long oscillon lifetimes, while in Ref.\ \cite{PietteZakr98},
\cite{Hindmarsh06} extremely long living
oscillons have been exhibited in a $1+2$ dimensional sine-Gordon model.

The sophisticated numerical simulations by Honda and Choptuik of
the Cauchy problem for spherically symmetric configurations in the
$\phi^4$ theory in $3+1$ dimensions \cite{Honda}, \cite{HondaPhD},
have revealed some interesting new features
of the oscillons. In particular in Ref.\ \cite{Honda} it has been found
that by a suitable fine-tuning of the initial data the lifetime of the oscillons
could be increased seemingly indefinitely, and it has been conjectured that actually
an infinitely long lived, i.e.\ non-radiative, spatially localized solution exists.
Furthermore the existence of such a solution would even provide the explanation of
the ``raison d'\^etre'' and of the observed genericness of long lived oscillons.
%This would clearly be quite remarkable.
The eventual existence
of a non-radiative breather in this simple and ``generic'' $\phi^4$ model
in $3+1$ dimensions would be clearly of quite some importance.
It should be noted, that
in quite a few spatially discrete models, localized time-periodic ``discrete breathers'' have
been shown to exist and they are being intensively studied \cite{FlachWillis98}.

One of the motivation of our work has been to clarify if non-radiative breathers
indeed exist and to find them directly by studying time-periodic solutions of the NLWE.
Our numerical results led us to conclude that no localized (with finite energy) time-periodic
solutions exist in the $\phi^4$ model.
On the other hand this study has led us to understand the oscillon phenomenon better and we present a simple
but quantitatively correct scenario explaining some important properties of the oscillons
(such as their existence and their long lifetimes).
Our scenario is based on the existence of a special class of time periodic solutions which are
weakly localized in space. Such solutions (which have infinite energy)
will be referred to as quasi-breathers (QB).

In the present paper we make a detailed study of oscillons in the (already much studied)
$\phi^4$ model in $3+1$ dimensions.
Using a previously developed and well tested time evolution code
where space is compactified, thereby avoiding the problem of artificial boundaries
\cite{FodorG04}, we compute some long-time evolutions of Gaussian-type initial data.
We observe that long living ($6000-7000$ in natural units) oscillon states are formed
from generic initial data. These oscillons radiate slowly their energy and
for short (as compared to their total lifetime) timescales they can be characterized
by a typical frequency, $\omega$. This frequency increases slowly during the lifetime
of the oscillon and when $\omega$ reaches a critical value $\omega_{\rm c}\approx1.365$
there is a rapid decay.
By fine-tuning the initial data one can achieve that the oscillon state instead of
rapidly decaying evolves into a near time-periodic state, whose frequency is nearly constant
in time and $1.365<\omega<1.412$.
The existence of such near-periodic states (referred to as resonant oscillons)
has been already reported in Ref.\ \cite{Honda},
and we also observe an increase of the lifetime of these states
without any apparent limit by fine-tuning the parameters of the
initial data to more and more significant digits.
There are, however, also some discrepancies.
We find that clearly distinct near-periodic states for various values of the pulsation frequency
$1.365<\omega<1.412$ exist.
According to our results there is little doubt that for
any value of $\omega$ in this range a corresponding near-periodic oscillon state exists.
Our data clearly shows that the near-periodic states also radiate, although very weakly.
The radiation becomes weaker and weaker as $\omega\to\sqrt{2}$.

On the other hand we have implemented a multi domain spectral method in order to find directly
stationary, time periodic solutions of the NLWE and compare them to the
long living oscillon states obtained from the time evolution.
This makes it possible to attack the problem of finding directly the putative time
periodic breather(s). (One could easily generalize our method for the quasi-periodic case.)
We find that there is a large family  of time
periodic solutions which are only {\sl weakly localized} in space, in that they
have a well defined core, and an oscillatory tail decreasing as $\propto1/r$.
We single out a special family among them by minimizing the amplitude of their oscillatory tail
which definition comes close to minimizing the energy density of the oscillatory tail.
These solutions are the closest
to a breather and for that reason we call them quasi-breathers.
They seem to exist for any frequency, $0<\omega<\sqrt{2}$, although in this paper
we exhibit QBs only with $1.30<\omega<\sqrt{2}$ .
The amplitude of the oscillatory tail of the QBs becomes arbitrarily small as
the frequency approaches the continuum threshold defined by the mass of the field,
$\omega\to\sqrt{2}$.
Our numerical evidence speaks clearly against the possible existence of a truly localized,
breather-like solution periodic in time, for the frequency range  $1.30\leq\omega\leq1.412$
contrary to the claims of Ref.\ \cite{Honda}.
In view of these conflicting numerical findings it is now highly desirable to try to find an
analytical proof or disproof of the existence of a localized non-radiative
solution to settle this issue. We do not expect the situation being qualitatively different
from the $1+1$ dimensional case, and although the proofs of Refs.\ \cite{Segur, Vui}
are not applicable for the $3+1$ dimensional case, we see no reason that their negative
conclusion would be altered.

More importantly, we believe to have made a step towards understanding the mechanism behind the
existence of such long-living oscillonic states without
the need to invoke genuine breather solutions, which even if they would exist
would be clearly non-generic, while the QBs seem to be generic.
The total energy of the quasi-breathers is divergent, due to the lack of
sufficient (exponential) spatial localization, hence they are not of direct physical relevance.
Nevertheless a careful numerical study shows that the {\sl oscillons} produced during the time
evolution of some suitable Cauchy data are quantitatively very well described by the
{\sl quasi-breathers}.
By comparing the Fourier decomposition of an oscillon state at some instant, $t$
characterized by a frequency, $\omega(t)$, obtained during the time evolution
with that of the corresponding QB, we have obtained convincing evidence
that the localized part of the oscillon corresponds to the core
of the QB of frequency $\omega(t)$. What is more, the oscillatory tail of the QB describes
very well the standing wave part of the oscillon.
Our oscillon scenario is based on this
analysis and leads us to propose that any oscillon contains the core and a significant part
of the oscillatory tail of the corresponding QB. The time evolution of an oscillon
can be approximatively described as an adiabatic evolution through a sequence of QBs
with a slowly changing frequency  $\omega(t)$.
This oscillon scenario is based on the existence and of the genericness of
the QBs.

This paper is organized as follows. In Sec.\ \ref{s:evol} we study the time evolution of
localized, Gaussian-type initial data in $\phi^4$ theory and investigate some important
aspects of the oscillon solutions.
In Sec.\ \ref{s:mode} we present an infinite system of coupled ordinary differential
equations (ODE's) obtained by the Fourier-mode decomposition of the NLWE
Eq.\ (\ref{e:evol}) and discuss some of its properties. Section \ref{s:numerics} is
devoted to the description of the spectral methods used to solve this system.
In particular, we carefully deal with the asymptotic behavior of the Fourier-modes.
Various convergence tests are exhibited. The quasi-breathers are discussed in Sec.
\ref{s:results}, the
results on our oscillon scenario are discussed in Sec.\ \ref{s:compar}
and conclusions are drawn
in Sec. \ref{s:conclu}.

%%%%%%%%%%%%%%%%%%%%%%%%%%%%%%%%%%%%%%%%%%%%%%%%%%%%%%%%%%%%%%%%%%%%%%%%%%

\section{Time evolution}\label{s:evol}

%%%%%%%%%%%%%%%%%%%%%%%%%%%%%%%%%%%%%%%%%%%%%%%%%%%%%%%%%%%%%%%%%%%%%%%%%%

\subsection{The nonlinear wave equation of the $\phi^4$ theory}

We consider the following $\phi^4$ theory in $1+3$ dimensions whose action can be written as
\be\label{action}
S=\int dt\, d^3\! x \left[(\partial_t\phi)^2-(\partial_i\phi)^2-(\phi^2-1)^2 \right] \,,
\ee
where $\phi$ is a real scalar field, $\partial_t=\partial/\partial t$,  $\partial_i=\partial/\partial x^i$ and $i=1,2,3$.
In this paper we shall restrict ourselves to spherically symmetric field configurations,
when the corresponding NLWE is given by
 \be\label{e:evol}
 -\phi_{,tt} + \Delta \phi = \phi \l(\phi^2-1\r)\,,\quad
{\rm where}\quad\Delta=\pa^2_r+\frac{2}{r}\pa_r\,.
 \ee

The energy corresponding to the action (\ref{action}) can be written as
\be
E = 4 \pi \int\limits_0^{\infty} {\mathrm d}r\,r^2\xi\,,
\ee
where $\xi$ denotes the energy density
\be\label{e:density}
\xi = \frac{1}{2} \l(\partial_t \phi\r)^2
+ \frac{1}{2} \l(\partial_r \phi\r)^2
+\frac{1}{4}\l(\phi^2-1\r)^2.
\ee
It is easy to see that the finiteness of the total energy is guaranteed
by $\phi\to\pm 1+{\cal O}(r^{-3/2})$ as $r\to\infty$.

\subsection{Numerical techniques}

We briefly outline here the main ideas for the implementation of our evolution code
to solve numerically the Cauchy problem for Eq.\ (\ref{e:evol}).
Assuming  $\phi\to1$ as $r\to\infty$ we introduce the new field, $\hat\phi$ as
\begin{equation}
\phi(t,r)=\frac{\hat\phi(t,r)}{r}-1\,.\label{e:ph}
\end{equation}
Then the NLWE Eq.\ (\ref{e:evol}) takes the form:
\begin{eqnarray}
r^2\left({\partial ^{2}_r}{\hat\phi}- {\partial ^{2}_t}{\hat\phi}\right)
&=& \hat\phi\left({\hat\phi}-r\right)\left({\hat\phi}-2{r}\right)\,.
\label{e:fe2}
\end{eqnarray}
The next step is to compactify in the spacelike directions
by a suitable coordinate transformation of $r$. This way we
can guarantee that our computational grid, associated with a finite-difference scheme,
covers the  entire physical spacetime, at least in principle.
Specifically, a new radial coordinate, $R$, is introduced in the following way:
\begin{equation}
r=\frac{2R}{\kappa(1-R^2)}\,, \label{e:iTR2}
\end{equation}
where $\kappa$ is an arbitrary positive constant. In the new radial
coordinate, $R$, the entire Minkowski spacetime is covered by the coordinate domain
$0 \leq R < 1$ while spacelike infinity is represented by the `hypersurface' $R=1$.
The $R=const$ `lines' represent world-lines of `static observers', i.e. integral
curves of the vector field $(\partial/ \partial t)^a$.
In the compactified representation the field equation, (\ref{e:fe2}), reads as
\begin{equation}
R^2\left(
\frac{4\Omega^4}{\kappa^2(1+R^2)^2}{\partial ^{2}_R {\hat\phi}} -
{\partial ^{2}_t {\hat\phi}} -\frac{4\Omega^3 R(R^2+3)} {\kappa(1+R^2)^3}
{\partial_R {\hat\phi} }\right)=
{\hat\phi}\left(\Omega{\hat\phi}-R\right)\left(\Omega{\hat\phi}-2R\right)
\label{e:fe3}
\end{equation}
where $\Omega$ is given by
\begin{equation}
\Omega=\frac{\kappa}{2}(1-R^2).\label{Om}
\end{equation}

We remark that the spacelike compactification used here is a simplified
variant of the conformal transformation used in \cite{FodorG04}.
There instead of the $t=const.$ hypersurfaces the initial data are
specified on hyperboloids.
Furthermore Minkowski spacetime is compactified mapping null infinity
to finite coordinate values.
Since in the present case the scalar field, $\phi$, is massive,
i.e. never reaches null infinity the hyperboloidal compactification is not
essential.

In order to obtain a system of first order equations we introduce the
independent variables
$\hat\phi_t=\partial_t\hat\phi$ and $\hat\phi_R=\partial_R\hat\phi$.
Then equation (\ref{e:fe3}) can be rewritten as
\begin{equation}
{\partial_t {\hat\phi_t}}=
\frac{4\Omega^4}{\kappa^2(1+R^2)^2}{\partial_R {\hat\phi_R}}
-\frac{4\Omega^3 R(R^2+3)} {\kappa(1+R^2)^3}{\hat\phi_R}
+{\hat\phi}\left(\frac{\Omega}{R}{\hat\phi}-1\right)
\left(\frac{\Omega}{R}{\hat\phi}-2\right) \ ,
\label{e:fe4}
\end{equation}
which together with the integrability
condition $\partial_t\hat\phi_R=\partial_R\hat\phi_t$ and the defining equation
$\partial_t\hat\phi=\hat\phi_t$ form a strongly hyperbolic system of first order
differential equations for the three
variables $\hat\phi$, $\hat\phi_t$ and $\hat\phi_R$
(see e.g. \cite{CouraH62}).
The initial value problem for such a first order system is known to be
well-posed \cite{GustaKO95}.
Note that the relation $\partial_R\hat\phi=\hat\phi_R$ is preserved by the evolution
equations, and therefore it corresponds to a constraint equation.

In order to solve the initial value problem for equation (\ref{e:fe4}) we
discretize the independent variables $t$ and $R$.
A simple uniform grid with steps $\Delta t$ and $\Delta R$ is introduced.
Spatial derivatives are calculated by symmetric fourth order stencils.
Time integration is done
using the `method of lines' in a fourth order Runge-Kutta scheme,
following the recipes proposed by Gustafsson {\it et al}
\cite{GustaKO95}.
A dissipative term proportional to the sixth derivative of the field is added in
order to stabilize the evolution. Since this dissipative term is also chosen to be
proportional to $(\Delta R)^5$, it does not reduce the order of the applied
numerical method, in other words, its influence is decreased by the increase of the
used resolution.
The numerical methods and the actual numerical code we use for calculating time
evolution in this paper are also based on those developed in \cite{FodorG04}.

A few non-physical grid points are introduced for both negative radii $R<0$ and for
the region ``beyond infinity'' $R>1$.
Instead of calculating the time evolution of the $R<0$ points,
the symmetry property of $\phi$ about the origin $R=0$ is used to set the function
values at each time step.
Similarly, $\phi$, being a massive field, decays exponentially towards infinity,
consequently all the field values $\hat\phi$, $\hat\phi_t$ and $\hat\phi_R$
are set to zero for $R\geq 1$ during the entire evolution.
This takes care of the spacelike infinity $\Omega=0$ in equation (\ref{e:fe4}).
Therefore it is possible to use symmetric stencils exclusively when
calculating spatial derivatives.
The grid point at the origin $R=0$ needs special treatment since the last term as
it stands on the right hand side of equation (\ref{e:fe4}) is apparently singular.
However, this term, when evaluated in terms of the original (non-singular) field
value $\phi$ have zero limit value at $R=0$.

Although compactifying in spatial direction restricts the coordinate $R$ to a finite
domain, grid points in our numerical representation get separated by larger and larger
physical distances as we approach $R=1$.
This far region, where shells of outgoing radiation cannot be represented properly,
moves out to higher and higher physical radii as $\Delta R$ decreases.
Numerical simulations with increasing number of grid points demonstrate that
wave packets of outgoing massive fields
get absorbed in the transitional region without getting reflected back
into the inner domain.
In this way our numerical simulations still give a good description of the field behavior
precisely in the central region for very long time periods.
The simple but physically non-uniform grid together with the help of the dissipation
term appears to absorb outgoing radiation in a similar way to the explicit adiabatic
dumping term method applied by Gleiser and Sornborger \cite{GleiserM00}.
Furthermore, because of the very low inward light velocity in the asymptotic region
$R\approx 1$ the field behavior in the central region is correctly given
for a long time period even after the appearance of numerical errors at $R\approx 1$.

Simulating time evolution of oscillons up to their typical maximal lifetime of
$t=7000$ using spatial resolution of $2^{13}$ points took usually a week on
personal computers.
However, because of the need of several runs when fine-tuning parameters in the
initial data, we mostly used typical resolutions of $2^{12}$ spatial points.
The parameter $\kappa$ in the coordinate transformation (\ref{e:iTR2}) was set to
$\kappa=0.05$ in our simulations in order to concentrate approximately the same number
of grid points to the central oscillon region and to the far away region where the
massive fields form high frequency expanding shells.
A Courant factor of $\frac{\Delta t}{\Delta R}=1$ turned out to be appropriate to
obtain stable simulations with our choice of $\kappa$.

The convergence tests confirmed that our code does provide a fourth order
representation of the selected evolution equations.
Moreover, we monitored the energy
conservation and the preservation of the constraint equation
$\partial_R\hat\phi=\hat\phi_R$.
Most importantly, we compared the field values which can be deduced by
making use of the Green's function and by the adaptation of our
particular numerical code to the case of massive linear Klein-Gordon
fields \cite{FodorR03}.
The coincidence between the values in the central region
provided by these two
independent methods for long time evolutions ($t\approx10^4$ measured
in mass units) made it apparent that the  phenomena described below
should be considered as true physical properties of the investigated
non-linear field configurations.

\subsection{Oscillons}

\begin{figure}[!ht]
\includegraphics[width=12cm]{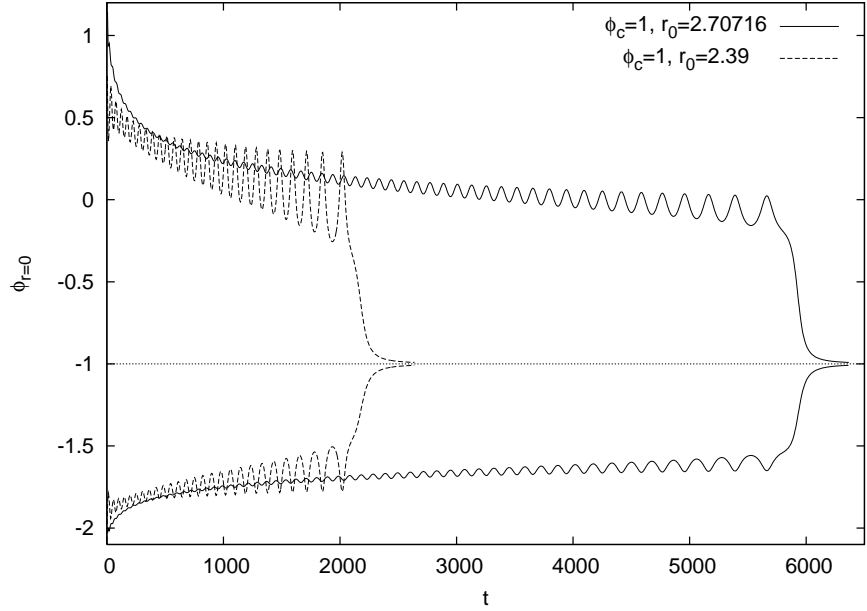}
\caption{\label{f:nontune1}
The upper and lower envelope of the oscillations of the field
$\phi(r=0)$ is shown for the evolution of Gaussian initial data of the type
(\ref{e:ff}) with two different sets of initial parameters.
It is important to note that the field $\phi$ oscillates with a non-constant
period $4.6<T<5$.
For generic oscillons the period $T$ shows a decreasing tendency towards
$T\approx 4.6$.
}
\end{figure}

Following Refs.\ \cite{CopelGM95,Honda} we start with the following Gaussian-type
initial data:
\begin{equation} \phi|_{t=0} = \phi_\infty+
(\phi_c-\phi_\infty)\cdot\exp(-r^2/r_0^2)\,,\quad  \pa_t\phi{\mid}_{t=0} = 0\,,
\label{e:ff}
\end{equation}
with $\phi_c$ and $\phi_\infty$ being the field values at the center
$r=0$ and at infinity $r=\infty$ while $r_0$ is the characteristic
size of the bubble at which the field values interpolates between
$\phi_c$ and $\phi_\infty$. By fixing $\phi_\infty=-1$ as in \cite{CopelGM95,Honda}
but varying $r_0$ and $\phi_c$, Eq.\ (\ref{e:ff}) provides a two-parameter family
of smooth and suitably localized initial data.
For a large open subset of the possible initial parameters $\phi_c$ and $r_0$, after a
short transitional period the field evolves into a long living localized nearly periodic
state, named \emph{oscillon} by Copeland, Gleiser and M\"uller.
Although these configurations live much longer than the dynamical time
scale expected from the linearized version of the problem (i.e. light crossing time),
their lifetime is clearly not infinite.
The energy of these oscillating states is slowly but definitely decreasing in time,
and after a certain time period they quickly disintegrate.
For the time dependence of the energy in a compact region see Fig.\ 3 of
\cite{CopelGM95}.
We illustrate on Fig.\ \ref{f:nontune1} two such oscillon states with rather different
lifetimes.

The parameter dependence of the lifetime is illustrated on
Figs.\ 6,\ 7 and 8 of \cite{CopelGM95}.
We note that since the final decaying period is relatively short, furthermore its time
dependence is almost the same for each particular choice of initial data, the
lifetime plots are quite insensitive
of the precise definition of how one measures the lifetime of a given configuration.
In our calculations the lifetime $\tau$ was defined by observing when the
value of the oscillating field at the center $r=0$ falls (and remains) below a certain
prescribed value (e.g. $\phi=-0.95$).

\begin{figure}[!ht]
\includegraphics[width=12cm]{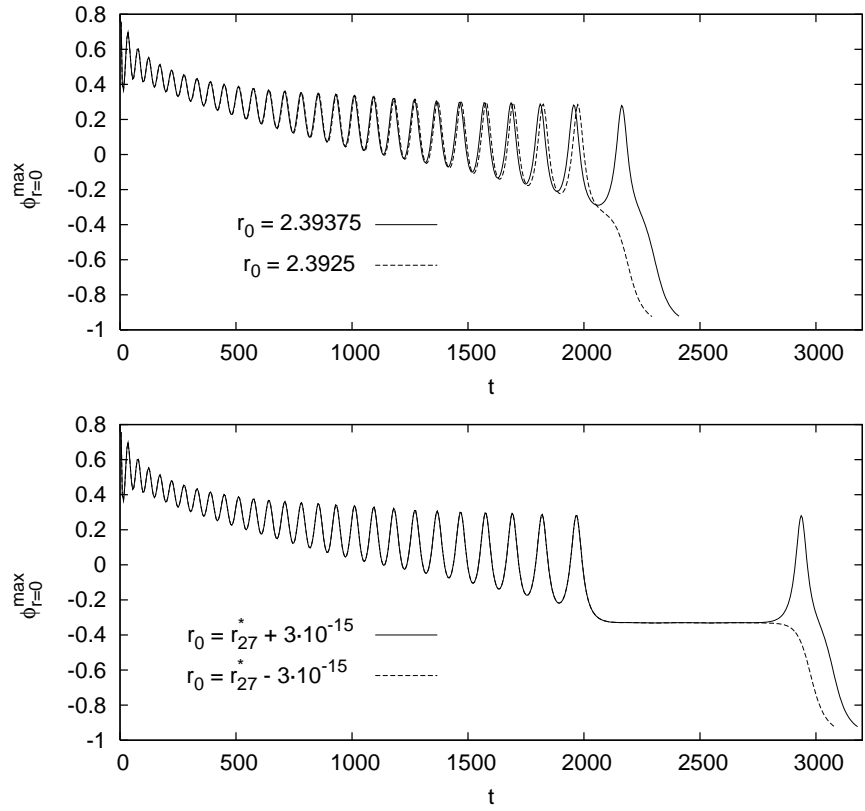}
\caption{\label{f:csx4}
Upper envelope of the field value for different oscillon states close to the $27$th
peak of the lifetime curve belonging to $\phi_c=1$.
These states are characterized by initial value $r_0$ which is near
$r_{27}^{*}\approx 2.39297$.
The \emph{near-periodic state} is approximately at $2100<t<2800$ for the
fine-tuned oscillons in the lower plot.
}
\end{figure}

Already Copeland, Gleiser and M\"uller have noticed that a delicate
fine structure appears in the lifetime plot.
This peculiar dependence on the precise value of the initial parameter $r_0$ has
motivated the detailed investigation of Honda and Choptuik
(see Figs.\ 4 and 5 of Ref.\ \cite{Honda}).
The calculations have shown that fixing the value $\phi_c$ the lifetime increases
without any apparent upper
limit when the parameter $r_0$ approaches some element of a large set of discrete
resonance values $r_i^{*}$.
For example, in the case $\phi_c=1$ Honda and Choptuik have found $125$ peaks
on the lifetime plot between $2<r_0<5$.
These fine-tuned oscillons at a later stage during their evolution develop into a
state which is very close to a periodic (non-radiating) one.
On Fig. \ref{f:csx4} we show the field value at the center for typical
oscillons close to a chosen peak.
The seemingly stable almost periodic stages of the fine-tuned oscillon configurations
will be referred as \emph{near-periodic states}.
The closer the initial parameters are to the critical value the longer the lifetime of the
near-periodic state becomes.
Although the period $T$ of the underlying high frequency oscillations remains constant
to a very good approximation during a chosen near-periodic state, different
near-periodic states, corresponding to various peaks on the lifetime curve, oscillate
with clearly distinct periods in the range $4.446<T<4.556$.
Generic oscillon states and even the not near-periodic initial part of fine-tuned
oscillons pulsate with longer period, in the range $4.6<T<5$.

Generic oscillons with initial data between two neighboring resonance values
$r^{*}_i<r_0<r^{*}_{i+1}$ may have already quite a long lifetime without developing
into a near-periodic state.
Plotting the field value at the center $r=0$ one can see a low frequency modulation
of the amplitude (called shape mode by Honda and Choptuik) of the high frequency
basic oscillating mode (see upper plot of Fig. \ref{f:csx4}).
Oscillons between the next two resonance values $r^{*}_{i+1}<r_0<r^{*}_{i+2}$
are distinguished from those in the previous interval by the fact that they
possess exactly one more or one less peak associated with these low frequency
oscillations on the envelope of the field value $\phi(t,0)$ before they disperse.
Longer living \emph{supercritical} states arise when one closely approaches a critical
value from one of the possible two directions.
Then the last low frequency modulation peak gets shifted out to a later and later
time as one goes towards the resonance, making room for a
near-periodic state between the last two modulations.
Close to the resonance, but on the other side of it, the same long living
near-periodic state appears, now called \emph{subcritical}, with the only difference that
at the end the field disperses without forming a last low frequency modulation peak.

\subsection{Results}

\begin{figure}[!ht]
\includegraphics[width=12cm]{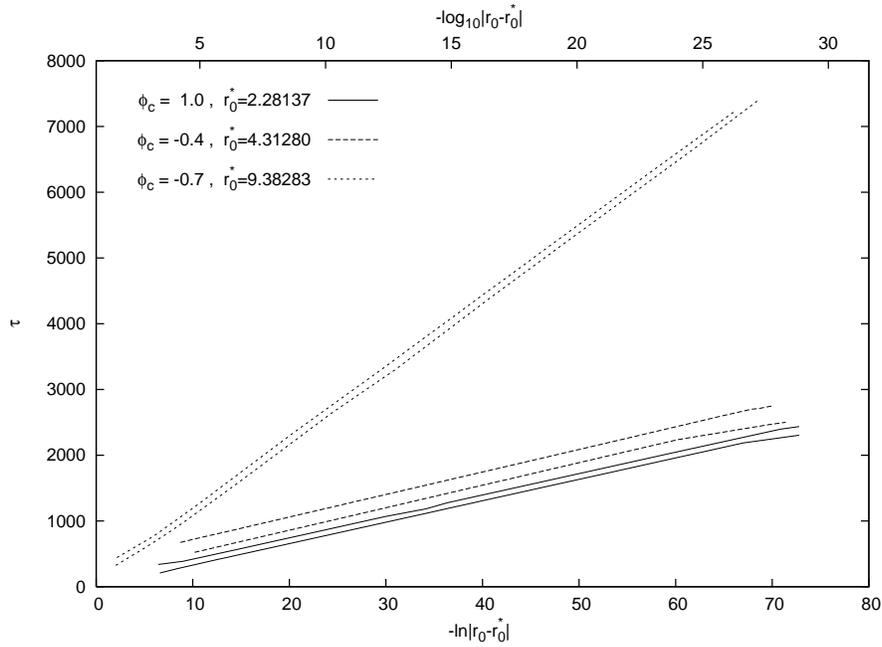}
\caption{\label{f:lt}
The oscillon lifetime, $\tau$, versus $-\ln|r_0-r_0^*|$  is shown for states close to
three different resonances.
The lower lines plotted with a given line type represent subcritical, $r_0<r_0^*$,
states while the upper lines with the same line type correspond to supercritical ,
$r_0>r_0^*$, solutions.
}
\end{figure}

It was shown in Ref.\ \cite{Honda} that close to a resonance the oscillon lifetime
$\tau$ obeys a scaling law
\begin{equation}
\tau \sim \gamma \ln|r_0-r_0^*|+\delta \label{f:crit}\ ,
\end{equation}
where the scaling exponent $\gamma$ has specific values for each resonance, while the
constant $\delta$ is also different for subcritical and supercritical states.
Although the lifetime appears to increase without any limit by fine-tuning the
initial parameter, in practice it is very difficult to achieve very long lifetimes
because one cannot represent numbers very close to the resonance value
due to the limitation implied by the applied machine precision.
Achieving longer lifetime for the near-periodic state is possible by using high
precision arithmetics, although then there is a considerable increase of the computation
time which limits the applicability of this approach.
Using ``long double'' variables on an SGI machine, or applying  the software
``doubledouble''\cite{doubledouble} on a personal computer, we could calculate with
twice as many significant digits than standard double precision computer variables
can represent (i.e. $32$ instead of $16$).
This way we could improve the fine-tuning and double the observed
lifetime of near-periodic states.
On Fig.\,\ref{f:lt} we plot the scaling law of the lifetime for oscillons near three
different resonances.
Instead of choosing three resonances with $\phi_c=1$, we calculated
states close to the \emph{first peak} on the lifetime curves corresponding to three
different values of $\phi_c$.
The reason for this was that the normal, slowly but evidently decaying, oscillon state
is the shortest near the first peak (i.e. no modulation on the contour curve),
thereby we could concentrate computational resources on the near-periodic state.

In order to clarify what we mean by fine-tuning the parameter $r_0$ to $16$ (or $32$)
digits and what is the actual error of the quantities, we present a table on the
precise location of the first peak for $\phi_c=1$  when performing fine-tuning with
five different numerical resolutions.
For each spatial resolution we could achieve approximately the same lifetime of
approximately $\tau=1100$ when the parameter $r_0$ approximated a resolution
dependent value to $16$ digits.
The convergence of the data indicates that the actual position of the peak is at
$r_0^{*}=2.281370594$ with an error of $10^{-9}$.

\begin{table}[htb]
\begin{tabular}{c||c|c|c|c|}
$i$ & $n_i$ & $r_0^{*(i)}$ & $\delta_i$ & $c_i$ \\
\hline
\hline
$8$  & $2^{8}$  & $2.281990488596033$ & $6.2\ 10^{-4}$ & \\
\hline
$9$  & $2^{9}$  & $2.281392051715203$ & $2.1\ 10^{-5}$ &  \\
\hline
$10$ & $2^{10}$ & $2.281371382459355$ & $7.9\ 10^{-7}$ & $4.86$ \\
\hline
$11$ & $2^{11}$ & $2.281370625452998$ & $3.1\ 10^{-8}$ & $4.77$ \\
\hline
$12$ & $2^{12}$ & $2.281370594875569$ &  & $4.63$ \\
\end{tabular}
\caption{
The position $r_0^{*(i)}$ of the first peak for $\phi_c=1$ using various resolutions.
The number of spatial grid points used for a specific fine-tuning is $n_i=2^i$.
The error is estimated as $\delta_i=|r_0^{*(i)}-r_0^{*(12)}|$.
The convergence factor is defined as
$c_i=\log_2|(r_0^{*(i-2)}-r_0^{*(i-1)})/(r_0^{*(i-1)}-r_0^{*(i)})|$.
}
\end{table}

\begin{figure}[!ht]
\includegraphics[width=12cm]{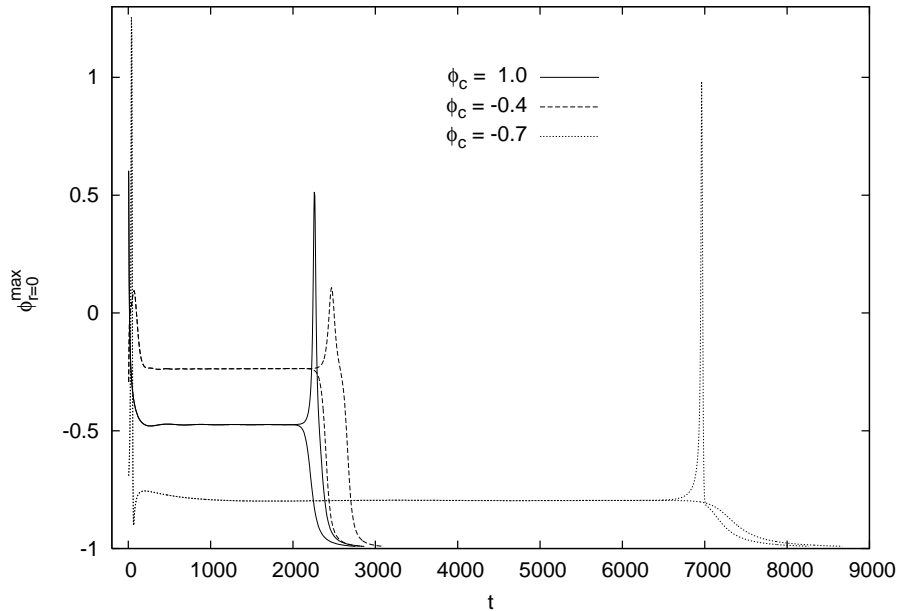}
\caption{\label{f:max}
Upper envelopes of the oscillations of the field value at $r=0$ for three
pairs of sub and supercritical oscillon
states close to the first resonances corresponding to three given $\phi_c$
initial parameters.
The fine-tuning was performed using $32$ digit arithmetics.
}
\end{figure}

\begin{figure}[!ht]
\includegraphics[width=12cm]{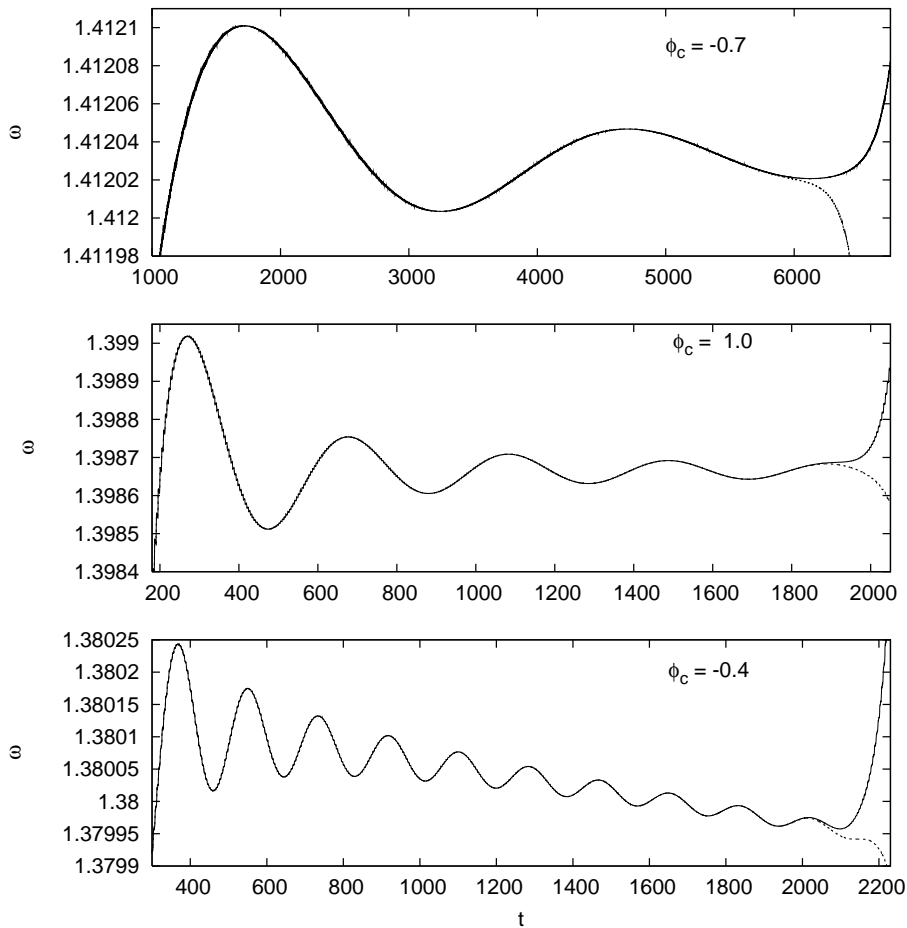}
\caption{\label{f:fr}
The pulsation frequency, $\omega$ in function of the time for the three
pairs of oscillons shown on the previous figure during their
corresponding near-periodic states.
}
\end{figure}

Our numerical simulations clearly show that the different resonance peaks on the lifetime
plot correspond to different near-periodic states.
In fact, a one-parameter family of distinct near-periodic states appears to exist.
This statement is in marked contrast with the claim of Honda and Choptuik in \cite{Honda},
where in Sec. III-A they claim that the oscillation period is almost the
same for all oscillons
and is roughly $T \simeq 4.6$, which corresponds to a pulsation frequency of
$\omega \simeq 1.366$.
From our numerical analysis we can clearly see that the period of the oscillation
depends on the resonance considered.
On Fig.\ \ref{f:max} the upper envelope of the oscillating field
value at the origin is shown as function of time for oscillon states close to the same
three resonances as on Fig.\ \ref{f:lt}.
On Fig.\ \ref{f:fr} the time dependence of the frequency at the origin is shown for
the same three states during the time period where the oscillation is
almost periodic (i.e. during the near-periodic states).
The frequency $\omega$ of the oscillations along with its time dependence
at some radius $r=\bar r$ has been determined from our numerical results
by minimizing the following integral for the oscillation
period $T=2\pi/\omega$
\begin{equation}
\int_{t-t_0}^{t+t_0} \left[\phi(t,\bar
r)-\phi(t+T,\bar r)\right]^2dt
\end{equation}
using some suitably chosen integration interval determined by $t_0$.
This procedure, applying polynomial interpolation, yields significantly
more precise frequency values than the direct use of the Fast Fourier
Transform method, especially when the time step by which our evolution
code outputs data is not extremely short.
Another advantage of our procedure is it being much less sensitive to
the particular choice of the sampling interval (in this case $2t_0$)
than the FFT algorithm.
The relatively small value of $t_0$ (for example $t_0=10$) makes it
possible to monitor relatively sharp changes in the time
dependence of the frequency.

All three near-periodic states show a low frequency change of $\omega$
with a decaying amplitude,
with the lowest frequency modulation belonging to the state with the highest
$\omega$.
The amplitudes on Fig.\ \ref{f:max} would show a similarly decaying slight
modulation if we would plot them individually in time intervals where the
oscillations are almost time-periodic.
These observations suggest that near-periodic states of frequency $\omega$
also contain a superposition of states with frequencies
$\omega\pm\Delta_\omega$ where $\Delta_\omega/\omega\ll1$.
In the next sections we shall see that at least the core part of an oscillon
of frequency $\omega$ can be extremely well described by a weakly localized
quasi-breather of the same frequency.

We emphasize, that the three chosen near-periodic states are typical,
i.e.\ the near-periodic states of all fine-tuned oscillons
(including all peaks belonging to $\phi_c=1$) are qualitatively and
quantitatively very similar to them.
Actually, the frequency of all calculated near-periodic states fell into the
interval $[1.379,1.413]$ spanned by the frequency of the two chosen first peaks
belonging to $\phi_c=-0.4$ and $\phi_c=-0.7$.

\begin{figure}[!ht]
\includegraphics[width=12cm]{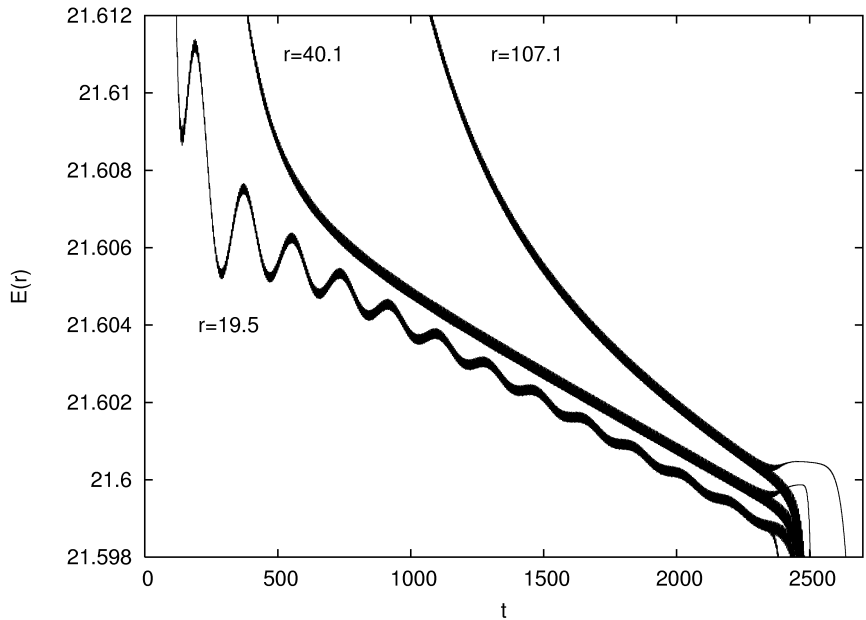}
\caption{\label{f:en}
The time dependence of the energy contained in spheres of radii $r=19.5$,  $r=40.1$
and $r=107.1$ for the $\phi_c=-0.4$ first peak.
The thickness of the curve indicates that the high frequency
oscillation is still not negligible at these high radii. The different endings of
the curves correspond to sub and supercritical states.
}
\end{figure}

On the third plot of Fig.\ \ref{f:fr} we can also see a slow but steady decrease
of the frequency $\omega$.
For the other two states, with frequencies closer to $\sqrt{2}$ no such behavior
is apparent in the time interval we could simulate.
This slow decrease is also manifested in the energy of the configuration.
On Fig.\ \ref{f:en} we plot the time dependence of the energy contained inside
spheres with three subsequent radii.
The decrease of the energy indicates that the near-periodic state slowly loses its
energy, consequently it cannot be exactly time-periodic.
Looking at the energy in the sphere at $r=40.1$, we can see that the decrease
of energy from $t=1000$ to $t=2000$ is $\Delta E=0.0041$.
Taking into account that the total energy is $E=21.60$, we can give a naive
estimate on the lifetime by calculating when the energy would decrease into its
half value at this rate, getting $\tau_e=2.6\cdot 10^6$.
We expect that all near-periodic states radiate, but this radiation is becoming
considerably weaker as the frequency get closer to $\sqrt{2}$.
This expectation is confirmed by the direct analysis of periodic solutions
in the next sections.

On Fig.\ \ref{f:en} the curve corresponding to the largest radius ($r=107.1$) indicates
how slowly the energy moves outwards because of the massive character of the field.
The different endings of the curves belonging to sub and supercritical states
illustrate the two kinds of decay mechanism of near-periodic states.
In the first, subcritical mechanism, the field quickly disperses, while energy
moves essentially outwards only.
In the second supercritical way, the energy of the field first collapses to a
smaller region near the origin and then disperses to infinity.
This resembles to the behavior of some unstable spherical shell, although no
shell structure is visible on the density plots, being the highest energy density
always at the center.
The instability of the near-periodic states with two distinct kinds of decay
mechanisms gives a qualitative explanation on why it is possible to reach
long lifetimes by fine-tuning the initial parameters.

\subsection{Fourier decomposition of the evolution results}
\label{s:fourier}

Since during the time evolution the field $\phi$ becomes approximately time-periodic for any longer
living oscillon state, it is natural to look at the Fourier decomposition of the results
provided by our evolution code.
Since the Fast Fourier Transform algorithm is very sensitive to the size of
the time step with which our evolution code writes out data, we use an alternative
direct method which turns out to be significantly more precise in determining
the basis frequency and the amplitude of the higher modes.
As a first step we determine the oscillation period by locating two subsequent maxima,
at instants $t_1$ and $t_2$, of the field at the origin $r=0$.
Since $t_1$ and $t_2$ fall in general between two consequent time
slices written out by our evolution code, we approximate their position by fitting
second order polynomials on the data.
It is apparent from our results that for near-periodic
states these maxima correspond to two consecutive
moments of time symmetry not only at the center but also in a large region around the
center to a very good approximation.
After determining the oscillation frequency $\omega=2\pi/(t_2-t_1)$, we obtain the
$n$-th Fourier mode at a radius $r$ by calculating the following integral using the
output $\phi(t,r)$ of the evolution code
\begin{equation}
\phi_n(r)=\int^{t_2}_{t_1}(\phi(t,r)+1)\exp(i n\omega t)dt \ . \label{fint}
\end{equation}
We note that care must be taken for the first and last incomplete time steps when
evaluating the integral (\ref{fint}).
In the case of exact periodicity and time symmetry the imaginary part of this integral would
be zero for all $n$.
In order to check the deviation from time symmetry for various radii at the moments
$t_1$ and $t_2$ we also evaluate the imaginary part of the integral for each $n$ and
verify whether it is small compared to the real part.

\begin{figure}[!ht]
\includegraphics[width=12cm]{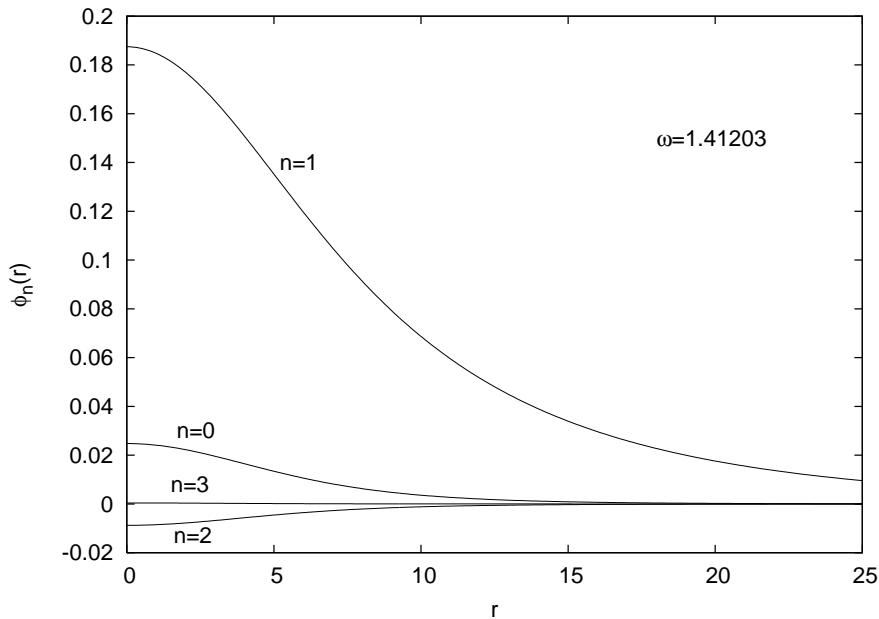}
\caption{\label{f:nolog2}
Radial dependence of Fourier modes obtained by decomposition of a
near-periodic state with frequency $\omega=1.41203$.
}
\end{figure}

As an example, on  Fig.\ \ref{f:nolog2} we give the Fourier decomposition
of the $\phi_c=-0.7$, $r_0=9.38283$ near-periodic state presented on
Fig.\ \ref{f:fr}.
The integral (\ref{fint}) was calculated between two subsequent maximums
of the field at the center at $t_1=5420.22$ and $t_2=5424.67$, yielding
an oscillation frequency of $\omega=1.41203$.
For the decomposition of a near-periodic state with a lower frequency
see Fig.\ 9 of \cite{Honda}, where the frequency is $\omega=1.38$
(page 107. of \cite{HondaPhD}).
From the two plots one can already see the general tendencies.
As the basis frequency increases towards the upper limit $\sqrt{2}$
the oscillon becomes wider, although with a decreasing amplitude.
The influence of the higher modes (the relative amplitude ratio)
is also getting smaller when the frequency grows.

%%%%%%%%%%%%%%%%%%%%%%%%%%%%%%%%%%%%%%%%%%%%%%%%%%%%%%%%%%%%%%%%%%%%%%%%%%%%%%%%%%%%%%%%%%%%%

\section{Fourier decomposition of the quasi-breathers}\label{s:mode}

%%%%%%%%%%%%%%%%%%%%%%%%%%%%%%%%%%%%%%%%%%%%%%%%%%%%%%%%%%%%%%%%%%%%%%%%%%%%%%%%%%%%%%%%%%%%%

 This section is concerned with a direct search for time-periodic solutions
of the NLWE Eq.\ (\ref{e:evol}) by means of a Fourier-mode decomposition
 of the scalar field $\phi$.
%with an investigation the properties of the oscillon-type solutions
A simple parameter counting already indicates that an infinite number of
parameters would be necessary in order to obtain an exponentially
localized breather, so even if some do exist, it seems to be difficult
to produce them directly.
Therefore we look for weakly localized solutions of Eq.\ (\ref{e:evol}),
and by analyzing their behavior in function of the free parameters we expect
to obtain a clear indication about the existence of truly localized breathers.

\subsection{Equations and their Asymptotic behaviors}\label{ss:asymptot}
Since we seek periodic solutions of Eqs.\ (\ref{e:evol}) we make a Fourier mode
decomposition of the form
\be\label{e:period}
\phi\l(t,r\r) =-1+ \tilde\phi=-1+\phi_0\l(r\r) + \sum_{n=1}^\infty \phi_n\l(r\r)
\cos \l( n \omega t \r).
\ee
We have assumed here that by a suitable choice of the time origin the solutions
are time-symmetric.
Inserting this form in Eq. (\ref{e:evol}) gives rise to a system of coupled
elliptic equations
\bea\label{e:system_bis}
\l[ \Delta -2 \r] \phi_0 &=& \phi_0^2 \l(\phi_0-3\r) +
\frac{3}{2} \l(\phi_0-1\r) \sum_{m=1}^\infty \phi_m^2
+\frac{1}{4}\sum_{m,p,q =1}^{\infty} \phi_m \phi_p \phi_q\,
\delta_{m , \pm p \pm q} \\
\nonumber
\l[\Delta - \lambda_{n}^2\r] \phi_n &=& 3\phi_0 \l(\phi_0-2\r)\phi_n
+\frac{3}{2}\l(\phi_0-1\r) \sum_{m,p = 1}^\infty \phi_m \phi_p\,
\delta_{n , \pm m \pm p} \\
\nonumber &&  +
\frac{1}{4}\sum_{m,p,q = 1}^{\infty} \phi_m \phi_p \phi_q\,
\delta_{n , \pm m \pm p \pm q}
\eea
where $\lambda_n^2=(2-n^2\omega^2)$. In fact we have already regrouped together
all the linear (first order) terms of the corresponding Fourier mode, $\phi_n$
to the left hand site of Eqs.\ (\ref{e:system_bis}).
We will make a systematic search for localized solutions of the
system (\ref{e:system_bis}).
We shall not assume
any a priori connection with the oscillons presented in the previous section
for the possible frequencies, $\omega$.
As we shall see below the eventual existence of a localized solution
can only be possible if system (\ref{e:system_bis}) exhibits some very special properties.

In order to look for regular solutions of Eqs.\ (\ref{e:system_bis})
we analyze the various asymptotic behaviors of the
homogeneous solutions of the corresponding operator on the left hand side
of Eqs.\ (\ref{e:system_bis}).

\begin{itemize}
\item If $\lambda_n^2>0$ then there is one regular homogeneous
solution, which falls off exponentially : $\exp\l[-\vert\lambda_n\vert r \r]/r$.
This function clearly decays sufficiently fast at infinity so that the energy
would  always be convergent.
\item $\lambda_n^2= 0$ is a degenerate case when the operator
is the usual Laplacian. The non-singular (decaying) homogeneous solution is simply
$1/r$ which, however, does not decay sufficiently fast for the energy to be bounded.
\item If $\lambda_n^2 < 0$, both homogeneous solutions tend
to zero at infinity, $\cos\l[\vert\lambda_n\vert r\r]/r$ and
$\sin\l[\vert\lambda_n\vert r\r]/r$.  As for the previous case, these
functions do not decrease fast enough to make the energy convergent. Apart from that,
both functions are regular at infinity and this
implies that there is no uniqueness of the solution which is only defined up to a given phase
(for more details, see Sec. \ref{ss:raccord_inf}).
\end{itemize}
Observe that no matter what the value of
$\omega$ is, there always exists an $n_\omega$, such that
{\sl all modes}, $\phi_n\; n>n_\omega$ will be of the third type (i.e.\ with $\lambda_n^2 < 0$).
It is now easy to understand why in the generic case, one cannot expect to find
exponentially localized ( with finite energy) solutions of Eqs.\ (\ref{e:system_bis}).
To ensure localization one has to suppress all slowly decaying oscillatory modes
which would normally require an infinite set of freely tunable parameters.
The problem of finding a localized solution can be seen as a matching problem
between the set of modes regular in a neighborhood of
the origin, $\{\phi_n^0\}_{n=0}^{\infty}$  and the set of modes with fast (exponential)
decay for $r\to\infty$, $\{\phi_n^\infty\}_{n=0}^{\infty}$.
In this case each mode $\phi_n^0$ regular at the origin, has a single
freely tunable parameter whereas none of the modes with fast fall off, $\phi_n^\infty$
for $n>n_\omega$ have any free parameters and therefore a sort of a miracle is needed
that the two sets could be matched.
This counting implies that while one can expect to find time-periodic
solutions Eqs.\ (\ref{e:system_bis}) (even a whole family), these solutions have generically
oscillatory tails for large values of $r$.
This clearly reflects the argument
{\sl ``anything that can radiate does radiate''} transposed to the stationary case.
On the other hand, it is not excluded that a localized
solution might exist for very particular values of $\omega$, for which the oscillatory
tails are absent.

\section{Numerics for the mode decomposition}\label{s:numerics}
\subsection{PDE solver}
We solve the system (\ref{e:system_bis}) using the LORENE library
\cite{lorene}. The basic features of LORENE are the use of spectral methods and multi-domain
decomposition. The problem we are facing here being purely spherical, the fields are
expanded on Chebyshev polynomials. The physical space is decomposed in various spherical
domains.

Using such techniques, solving differential equations can be reduced to, in each domain,
inverting
a matrix on the coefficients space. Then, a linear combination of the
particular solutions with the homogeneous ones is done, in order to impose regularity at the
origin, appropriate boundary conditions and continuity of the overall solution. We refer the
reader to \cite{GrandBGM01} for more details on the algorithm, in the case of a Poisson
equation.

For this work, we have extended the operators presented in \cite{GrandBGM01} and included the
Helmholtz operators
appearing in Eqs. (\ref{e:system_bis}), i.e. $\Delta - \lambda_n^2$ with both signs of
$\lambda_n^2$. This is rather
straightforward because, as for the Laplacian, they produce two homogeneous solutions. Only
the case of infinity has to be treated differently as we will see in Sec. \ref{ss:raccord_inf}.

\subsection{Description of the sources}\label{ss:sources}
As we have seen in Sec. \ref{ss:asymptot}, for all value of $\omega$, there exists a value of
$n$ after which all the
modes are dominated by homogeneous solutions of the type
$\sin\l(\l|\lambda_n\r|r+\varphi_n\r)/r$.
Such functions are not easily dealt with by our solver. Indeed, in order to treat
spatial infinity,
LORENE usually uses a simple compactification by means of the variable $u=1/r$ in the
exterior domain.  In the
past, this has enabled us to impose exact boundary conditions at infinity. When waves
are present, it is well known \cite{NovakB03,Somme49} that
such a compactification is not practical. Indeed,
no matter what scheme is used, there is always a point after which the distance between
computational points (grid or collocation) is greater than the characteristic
length of change of the wave, thus causing the scheme to fail. This issue is dealt with by
imposing boundary conditions at a finite radius $R_{\rm lim.}$ for the slowly decaying
oscillatory modes.
The reader should refer to
\cite{BonazGGN04} where such methods have been applied to gravitational waves.
For the exponentially decaying modes in Eqs.\ (\ref{e:system_bis})
we decide to keep in the right hand side (in the sources)
only those terms that are dominated by the exponentially decaying homogeneous solutions,
i.e., only those modes for which $\lambda_n^2 >0$.
As it will be seen later, the range of interesting pulsations
is $\omega < \sqrt{2}$.
Therefore the only two exponentially decaying modes are
$\phi_0$ and $\phi_1$.
In the Eqs.\ for $\phi_0$ and $\phi_1$ for $r>R_{\rm lim.}$,
we set all the higher modes ($n\geq2$) to zero
(including terms of the type $\phi_3\phi_1^2$).
So, we effectively solve for large values of the radius the following equations:
\bea
\label{e:sources_ext}
\l[ \Delta -2 \r] \phi_0&=& \frac{3}{2} \l(\phi_0-1\r) \phi_1^2 +  \phi_0^2 \l(\phi_0-3\r)\\
\label{e:sources_ext2}
\l[\Delta - \lambda_{1}^2\r] \phi_1&=& 3 \phi_0\l(\phi_0-2\r)\phi_1+\frac{3}{4}\phi_{1}^3\,.
\eea
This method yields solutions for $\phi_0$ and $\phi_1$ which are correct for ``intermediately
large'' values of $r>R_{\rm lim.}$ where the oscillatory and slowly decaying terms induced by
the nonlinearities do not dominate.
It is clear that for sufficiently large values of $r$  $\phi_0$ and $\phi_1$
do not decay exponentially, since their behavior will be dominated by the slowly decaying
oscillatory nonlinear source terms.
We have carefully checked that changing the value
of $R_{\rm lim.}$ does not influence the oscillatory modes, therefore we can conclude
that the back reaction of the $\phi_0$ and $\phi_1$  is negligible on $\phi_n$, $n\geq2$
for $r\gg R_{\rm lim.}$.

\subsection{The operators}\label{ss:raccord_inf}

When $\lambda_n^2>0$, the Helmholtz operator $\Delta - \lambda_n^2$ admits two homogeneous
solutions of which only one tends to zero at infinity. This situation is exactly the same as the one
for the standard Laplace operator and all the techniques presented in \cite{GrandBGM01} can
be used. Once again, as we will be working for $\omega<\sqrt{2}$, this happens only for $\phi_0$ and
$\phi_1$ and the associated sources have been given in the previous section \ref{ss:sources}.

The situation is quite different when dealing with the Helmholtz operator with
$\lambda_n^2<0$. Apart from the compactification problem previously mentioned,
we have to note that now there are two homogeneous solutions that are regular
at infinity. There is no reason to prefer one to the other. This means that there is no unique
solution to our problem. Indeed, one can get a solution by doing the matching with
any homogeneous solution of the type $\sin \l(\l|\lambda_n\r| r + \varphi_n\r)/r$, where
$\varphi_n$ can take any value in $\l[0, 2\pi\r[$. So to summarize,
for all modes such that $\lambda_n^2<0$ we have to match, at a finite radius
$R_{\rm lim.}$, the solution with a homogeneous one of the type :
\be
\phi_{n\geq 2} \l(r>R_{\rm lim}\r) = A_n \sin \l(\l|\lambda_n\r| r + \varphi_n\r)/r.
\ee
Clearly we need additional conditions to fix the values of the phases $\varphi_n$.
In order to do so, let us recall that we are mainly interested in finite energy solutions.
Such solutions should not contain any oscillatory behavior in $1/r$ at infinity,
i.e. {\sl all} the coefficients $A_n$
of such homogeneous solutions should be zero. One can hope to achieve that by searching, in the
parameter space
of the phases $\l(\varphi_2, ...., \varphi_n\r)$, the values that minimizes the absolute value
of coefficient of the first oscillatory homogeneous solution that appears : $\l|A_2\r|$.
This solution being ``quite close'' to a localized breather, for that reason we refer to it
as a quasi-breather.

Given the non-linearity of the problem, the location of the minimum can not be found analytically and
one has to rely on a numerical search.
We make use of the
multi-dimensional minimizer provided by the {\em GSL} numerical library \cite{gsl}.
The algorithm is
based on the simplex algorithm of Nelder and Mead \cite{NelderMead}.
We start by setting all the phases
to $\pi/2$ and iterate the procedure until the value of the minimum converges with a given
threshold. Let us mention that the phases are searched in $\l[0, \pi\r[$, the rest of the interval, being
described by changing the sign of the amplitudes. The simplex solver converges rapidly, the function
$\l|A_2\r|$ being very smooth, with no local extrema.

So the final situation is the following :
\begin{itemize}
\item For $\phi_0$ and $\phi_1$ the space is compactified and the sources in the exterior
region are given by Eqs.
(\ref{e:sources_ext}) and (\ref{e:sources_ext2}).
\item For $\phi_n$, $n\geq 2$, we only solve for $r<R_{\rm lim}$ and match the solution with a
homogeneous solution
$A_n \sin\l(\l|\lambda_n\r|r + \varphi_n\r) / r$, the $\varphi_n$ being determined by the
simplex solver by minimizing
$\l|A_2\r|$.
\end{itemize}

\subsection{Avoiding the trivial solution}\label{ss:trivial}

The system (\ref{e:system_bis}) is solved by iteration but a problem arises from the fact that
the trivial solution $\phi_n = 0$ is a solution of the equations (then $\phi_0$ is a constant
whose value is either $0$, $1$ or $2$). No matter what the initial guess for the different
modes is, the code always converges to a trivial static solution of this type.

So, one needs to find a way to prevent this to happen. Given that the system
(\ref{e:system_bis}) is coupled, it is sufficient to impose $\phi_1\l(r=0\r) \not= 0$,
in order to avoid the trivial solution. To do so, after each step of the iteration,
 we rescale
$\phi_1$ everywhere, by a factor $\alpha$ in order to impose that $\phi_1\l(r=0\r)$ has a
certain value. In the
general case, after convergence of the code, this scaling parameter is different from $1$
meaning that the value of $\phi_1$ we found is no longer solution of the system
(\ref{e:system_bis}). However for some values of $\omega$
(being in an interval, see Sec. \ref{s:results}), it is possible to find, exactly one value of
$\phi_1\l(r=0\r)$ such than the scaling factor is one. Thus, it is possible, at least
for some values of $\omega$ to find the appropriate value of $\phi_1\l(r=0\r)$ such that the
obtained modes are non-trivial solutions of Eqs. (\ref{e:system_bis}).

In practice, after convergence to a threshold level of typically $10^{-3}$ we switch on the
convergence
toward the true value of $\phi_1\l(r=0\r)$. With a technique already use, for example to get neutron stars
of appropriate mass (see Sec. IVD3 of \cite{GourgGTBM}), at each step, we change the value of $\phi_1$ at the
center, in order to make $\alpha$ closer to one. The value of $\phi_1\l(r=0\r)$ {\em to which one wants
to converge} is modified according to :
\be
\phi_1\l(r=0\r) \rightarrow \phi_1\l(r=0\r) \l[\frac{2-\log\alpha}{2-2\log\alpha}\r]^{0.1}.
\ee
Doing so, the value of $\phi_1$ at the center converges to the only value such that the scaling parameter
$\alpha$ is one, providing us with the real, non-trivial, solution of the system (\ref{e:system_bis}).

\subsection{Influence of the phases}\label{s:phase}
\begin{figure}
\includegraphics[height=6.5cm]{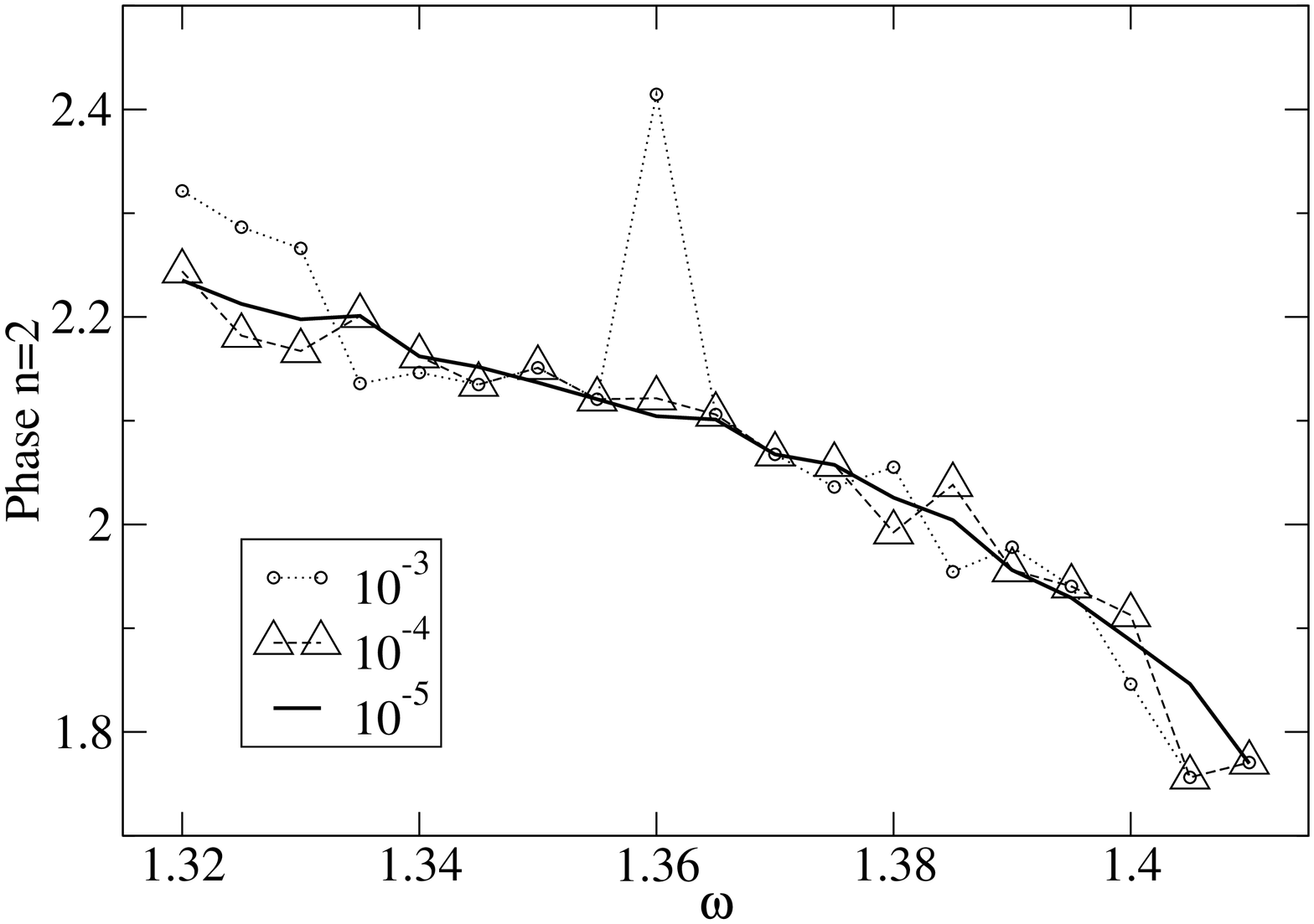}
\includegraphics[height=6.5cm]{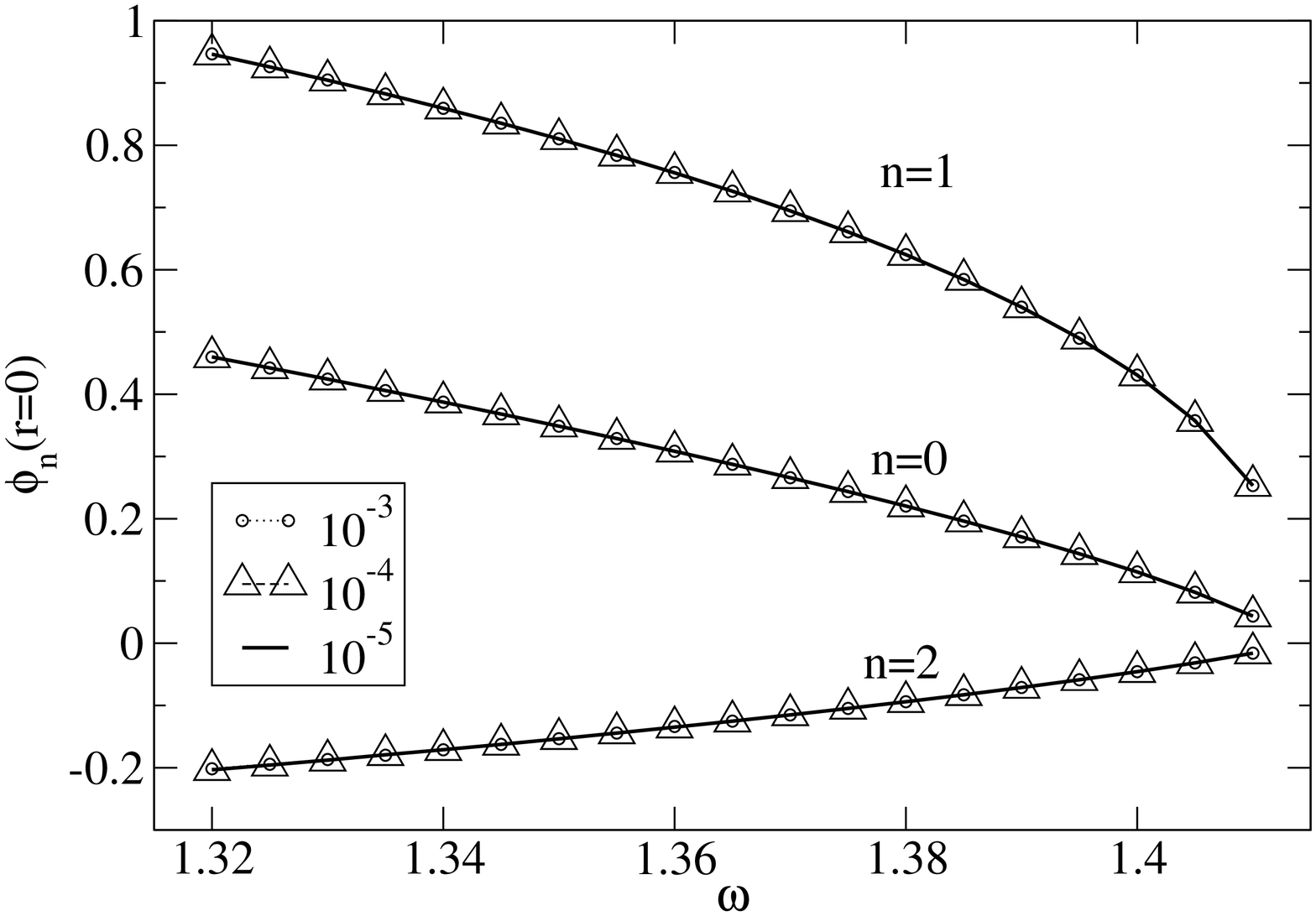}
\includegraphics[height=6.5cm]{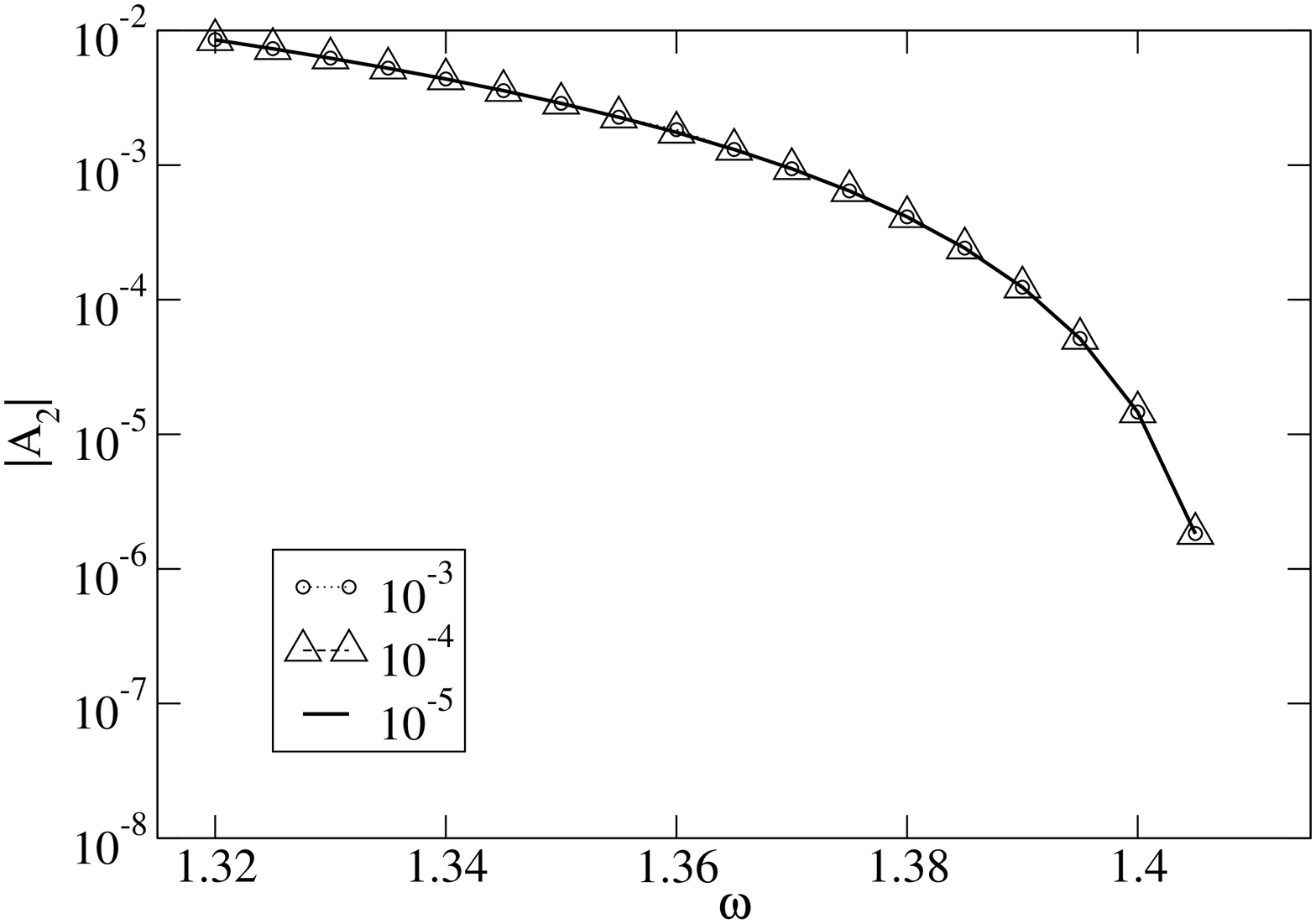}
\caption{\label{f:conv_prec}
Convergence of the various results  with respect to the precision required when finding the
minimum of $\l|A_2\r|$.
}
\end{figure}

In order to verify that the simplex solver converges to the proper minimum of $\l|A_2\r|$,
 we show, on Fig.
\ref{f:conv_prec} how various quantities vary when one changes the precision at which the
extremum is found.
Fig. \ref{f:conv_prec} presents the values of :
\begin{itemize}
\item the phase of the second mode $\varphi_2$ at which the minimum is found,
\item the values of the first modes at the origin $\phi_n\l(r=0\r)$,
\item and the value of the minimum of $\l|A_2\r|$.
\end{itemize}
Those three quantities are shown as a function of $\omega$, for three different values of
the precision required for the simplex solver, $10^{-3}$, $10^{-4}$ and $10^{-5}$.

From the values of the phase $\varphi_2$, it appears that a precision of $10^{-3}$ or $10^{-4}$
is not good enough, the curve being quite noisy. The curve obtained with a precision of
 $10^{-5}$ is smoother and we will assume that this level of precision is sufficient for
 the purpose of this paper and this value will be chosen for all the rest of this work.
The situation is even better for both the values of the modes at the origin and the actual
value of the minimum. Indeed, as can be seen on Fig. \ref{f:conv_prec}, those quantities
show almost not dependence on the level of precision.
This is simply related to the fact that the values of the fields depend very weakly
on the values of the phases of the homogeneous solutions in the external region.
 This is true for the phase of the second mode $\varphi_2$ but even more for the higher order phases
$\varphi_{n>2}$.

\begin{figure}
\includegraphics[height=6.5cm]{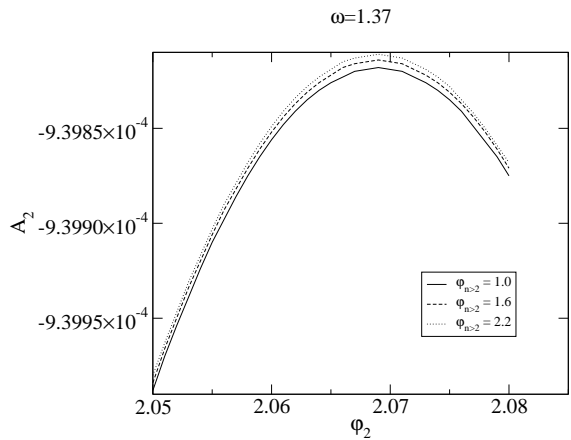}
\caption{\label{f:low_dep}
Value of $A_2$ as a function of the phase $\varphi_2$, for $\omega=1.37$.
 The other phases are maintained equal and set to
1.0 (solid line), 1.6 (dashed line) and 2.2 (dotted line).
}
\end{figure}

The very moderate dependence of $A_2$ on the phases is illustrated by Fig. \ref{f:low_dep}
where the value of the amplitude is shown, as a function of the phase
$\varphi_2$, for $\omega = 1.37$. All the other phases $\varphi_{n>2}$ are fixed
to the same value, each of them corresponding to one curve on Fig. \ref{f:low_dep}.
It is clear that the influence of the phases of the modes $n>2$ is very weak.
The extrema on the three curves of Fig. \ref{f:low_dep} are very close to
the real extremum found by the simplex solver. Therefore we are quite confident that we
can find the minimum of $\l|A_2\r|$ with a good accuracy, even if the associated values
of the phases are slightly less accurate.

\subsection{Convergence tests}\label{ss:convergence}
In order to further check the validity and accuracy of our code with respect to the
computational
parameters, we present in this section various additional convergence tests.
The idea is to take a reference
set of computational parameters and to change one of those at a time, in order to check that the
obtained results do not change much. The first three radial domains are chosen with boundaries
$\l[0, 1\r]$, $\l[1,2\r]$ and $\l[2, 4\r]$. After $r=4$, we keep the size of the domain constant
to $4$ (i.e. the boundaries of the domain $i$ are $\l[4\l(i-3\r), 4\l(i-2\r)\r]$).
We found that this is necessary to ensure that in every domain, we have enough collocation
points to resolve the oscillatory homogeneous solutions.

The various computational parameters are the following : $n_z$ is the number of radial domains
which
relates directly to the value of $R_{\rm lim.}$ given the setting of the domain mentioned above.
$N_r$ is the number of coefficients in each domain.
Finally we will also checked the convergence of the results with respect to the finite number of modes
we consider in the system (\ref{e:system_bis}).

Our standard setting consists of $n_z=14 \Rightarrow R_{\rm lim.} = 44$, $N_r = 33$,
and the use of 6 modes. We then change one parameter at a time to verify that
the obtained results are indeed meaningful.
The convergence is presented by showing the same three
quantities as in Sec. \ref{s:phase} : the phase $\varphi_2$, the values of the first modes at the
origin and the amplitude $\l|A_2\r|$.

\begin{figure}
\includegraphics[height=6.5cm]{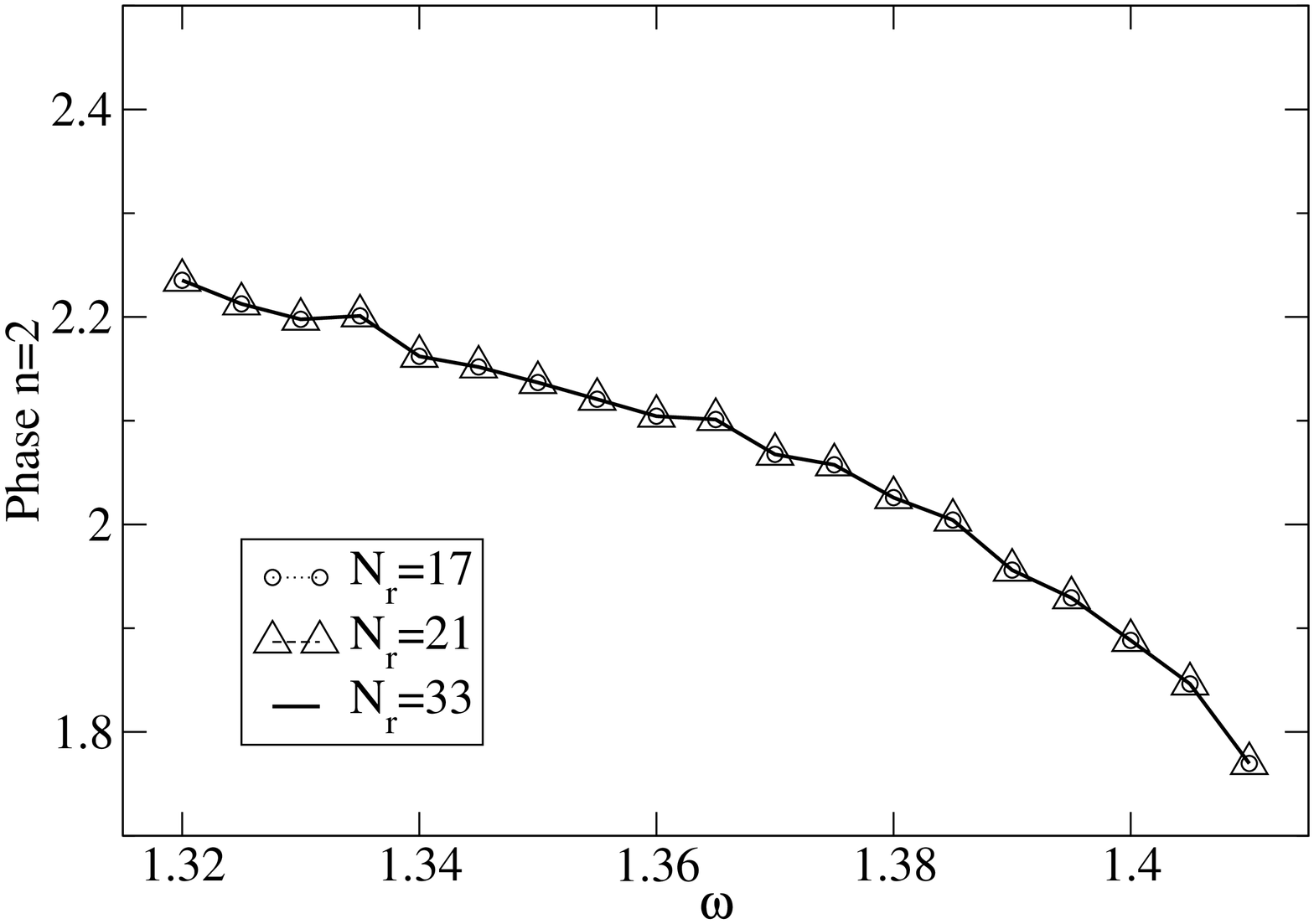}
\includegraphics[height=6.5cm]{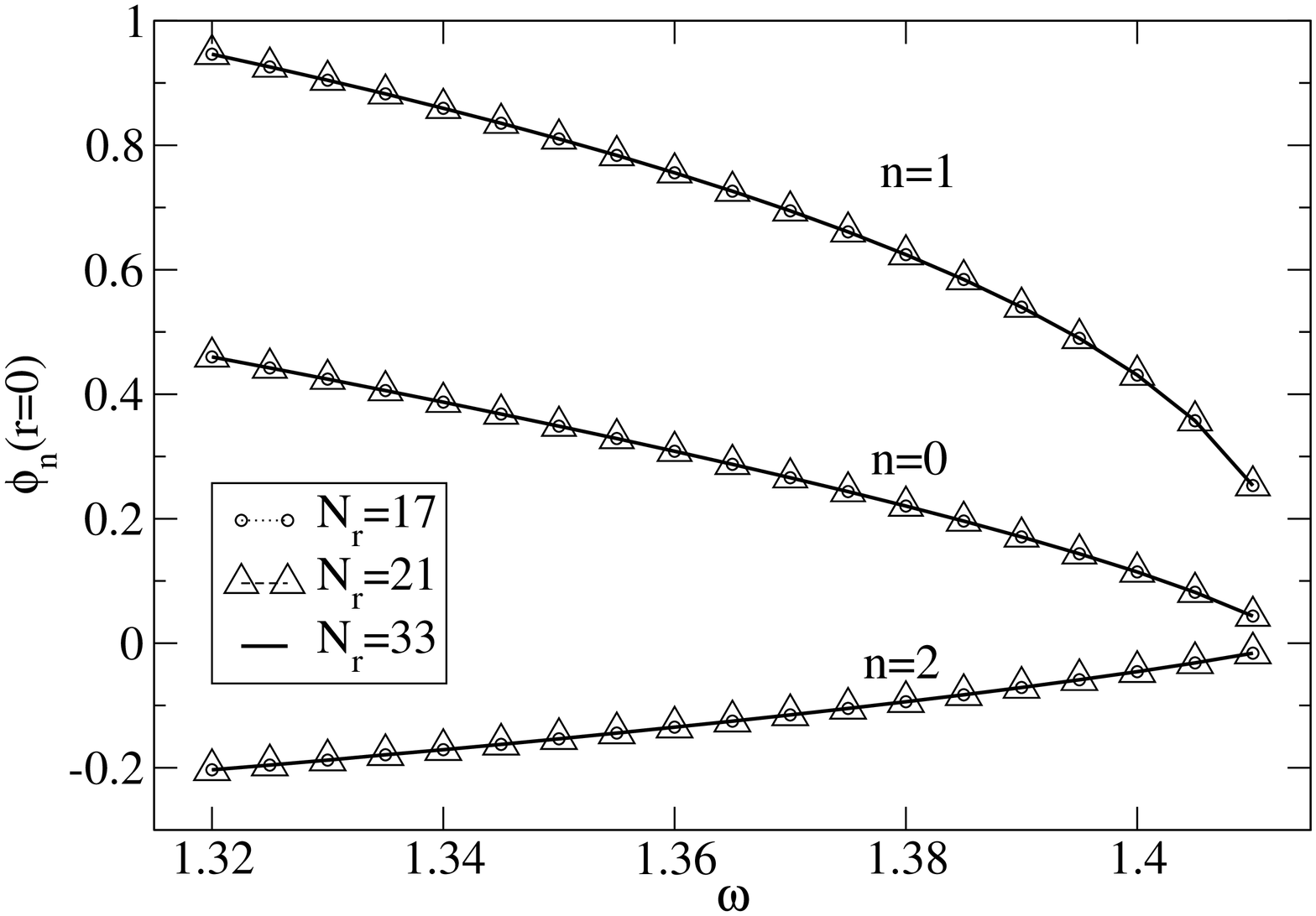}
\includegraphics[height=6.5cm]{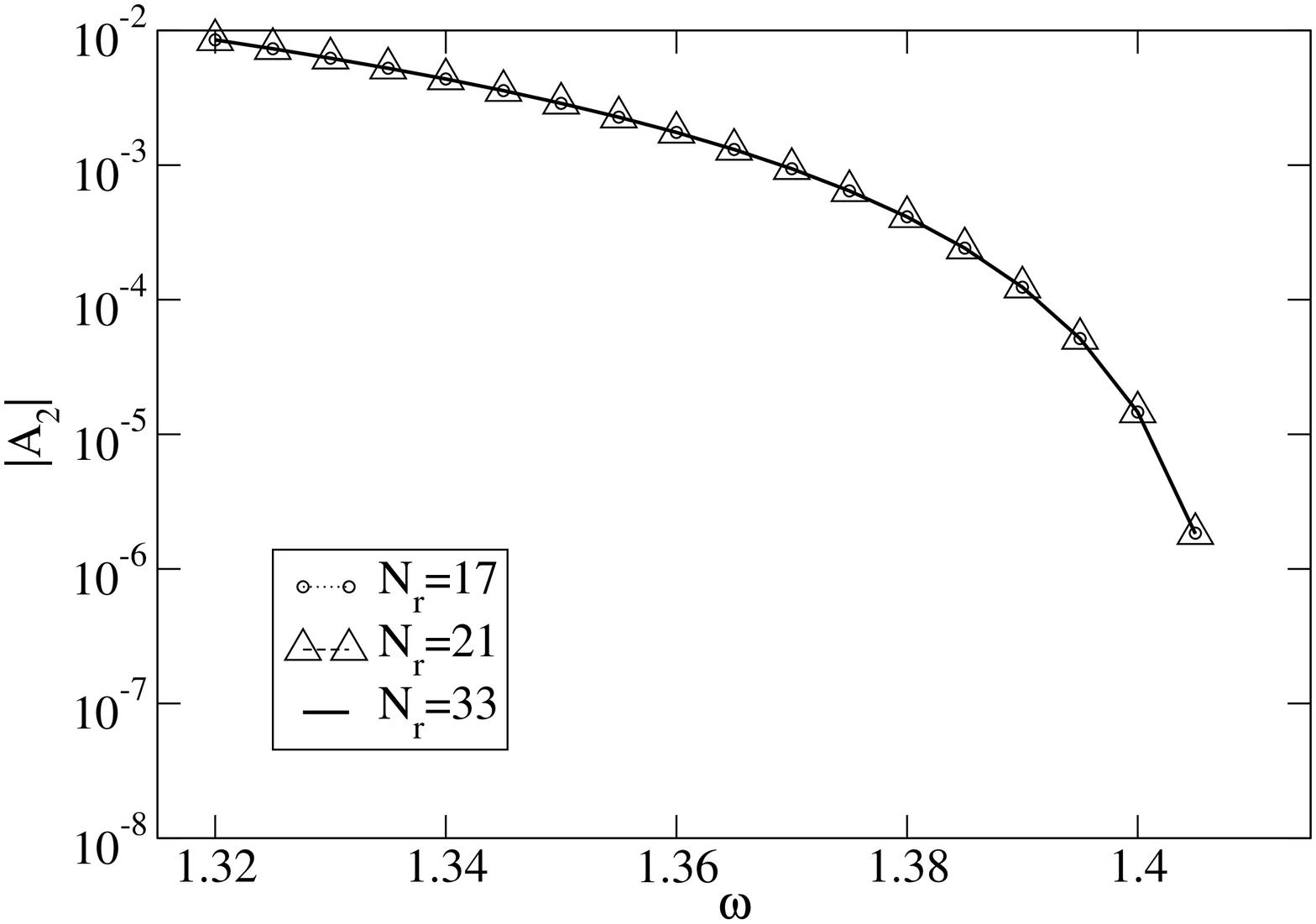}
\caption{\label{f:conv_nr}
Same as Fig. \ref{f:conv_prec} but when varying the $N_r$ in each domain.
}
\end{figure}

The three plots of Fig.\ \ref{f:conv_nr} show the dependence of the results when varying the
number of
coefficients in every domain. Since for the values of $N_r$ the curves of Fig.\ \ref{f:conv_nr}
change only slightly, this is a strong
sign that $N_r=33$ is greatly sufficient to get accurate results. It is also supported by the
fact that,
after the end of the iterative scheme we find that the relative difference between the right and left
hand sides of Eqs. (\ref{e:system_bis}) is smaller than $10^{-9}$.

\begin{figure}
\includegraphics[height=6.5cm]{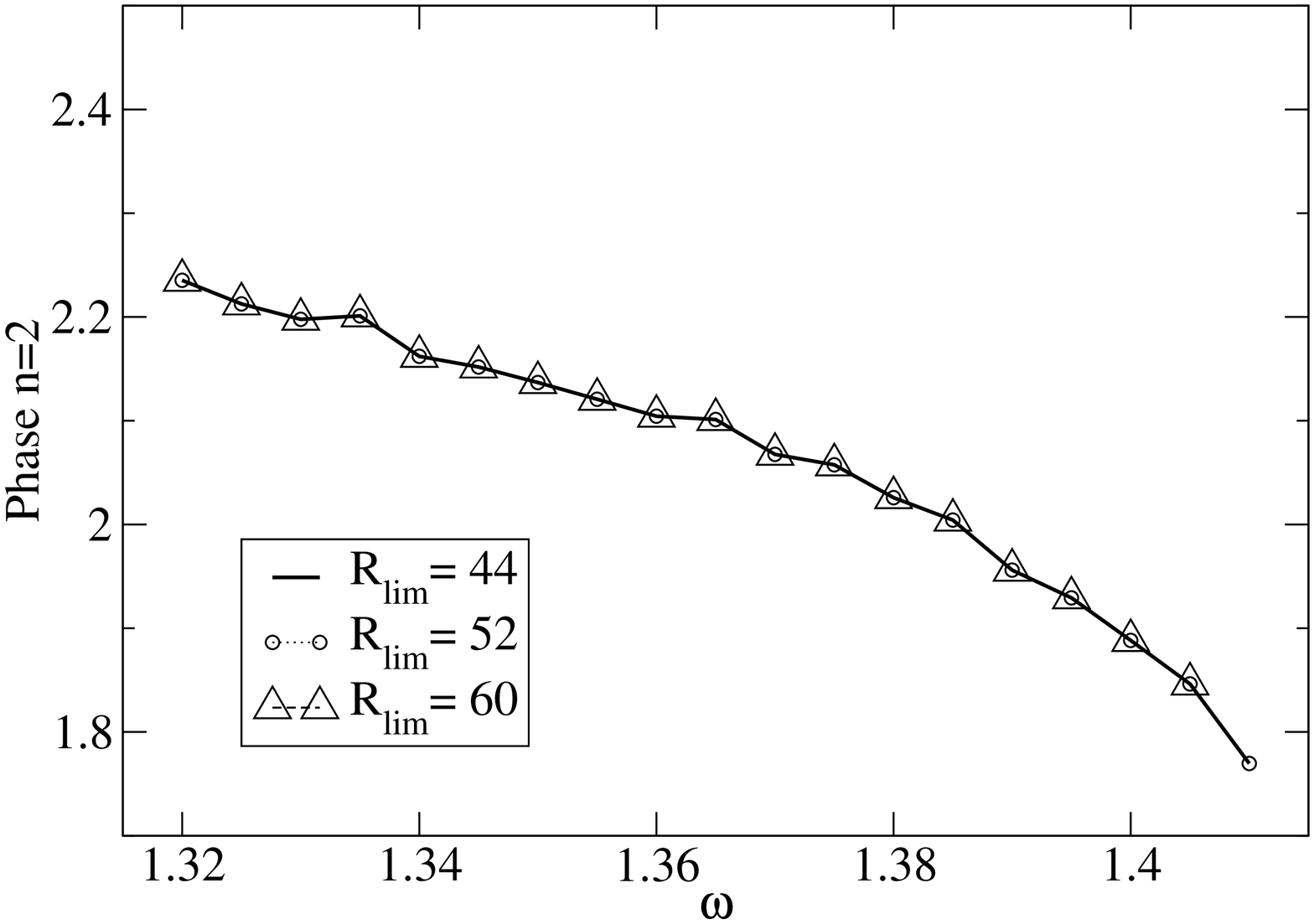}
\includegraphics[height=6.5cm]{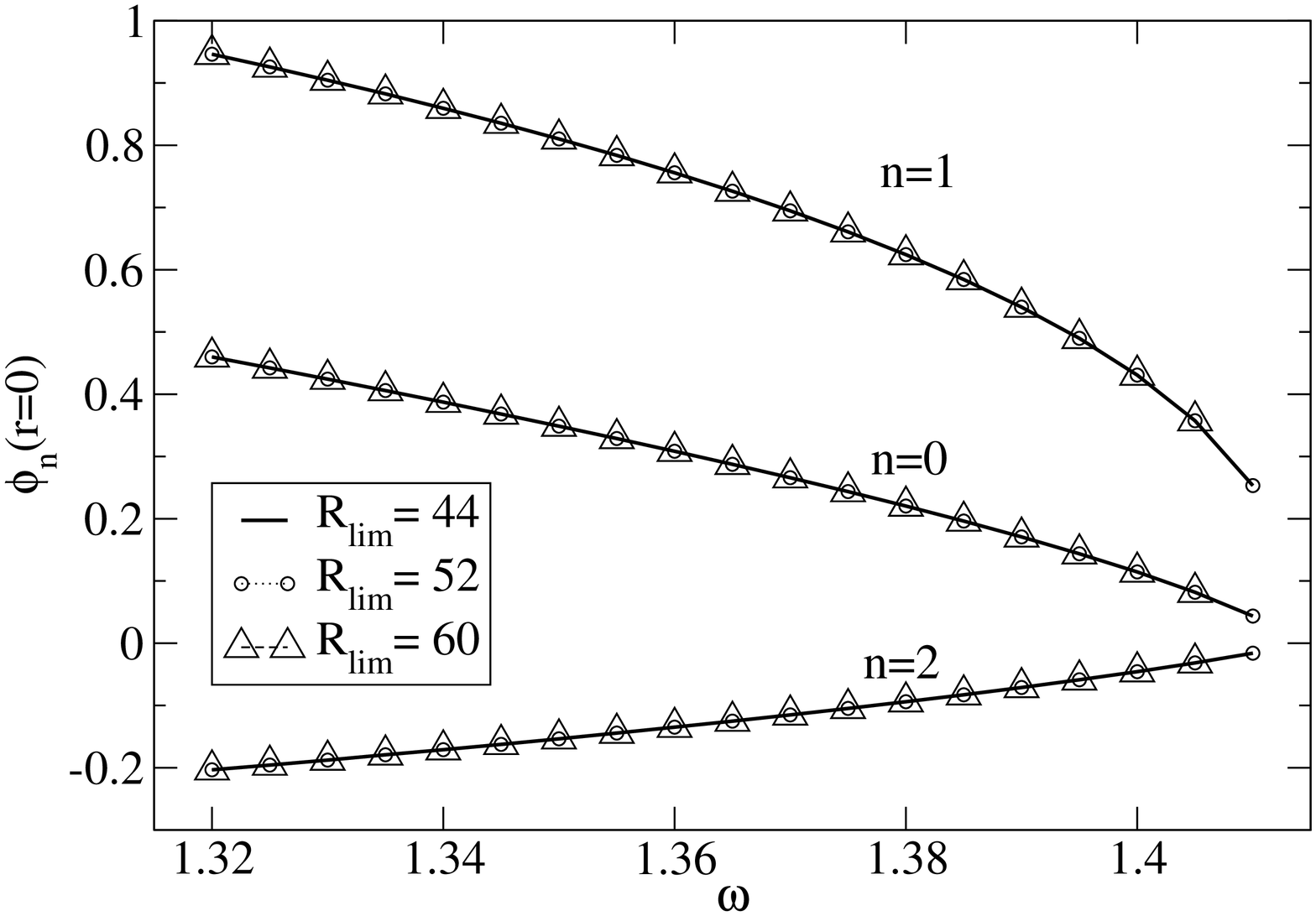}
\includegraphics[height=6.5cm]{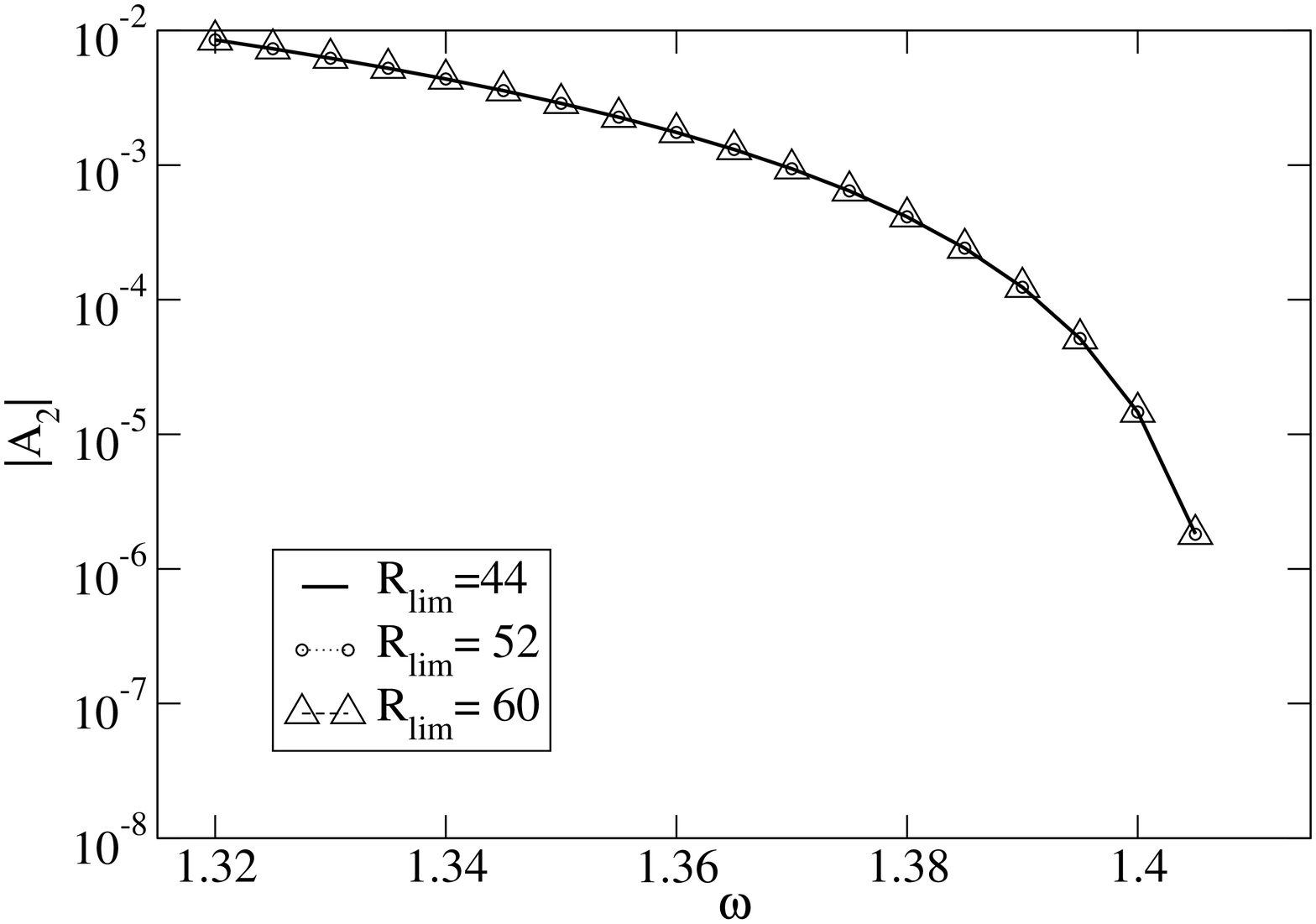}
\caption{\label{f:conv_rlim}
Same as Fig. \ref{f:conv_prec} but changing the value of $R_{\rm lim.}$.
}
\end{figure}

\begin{figure}
\includegraphics[height=6.5cm]{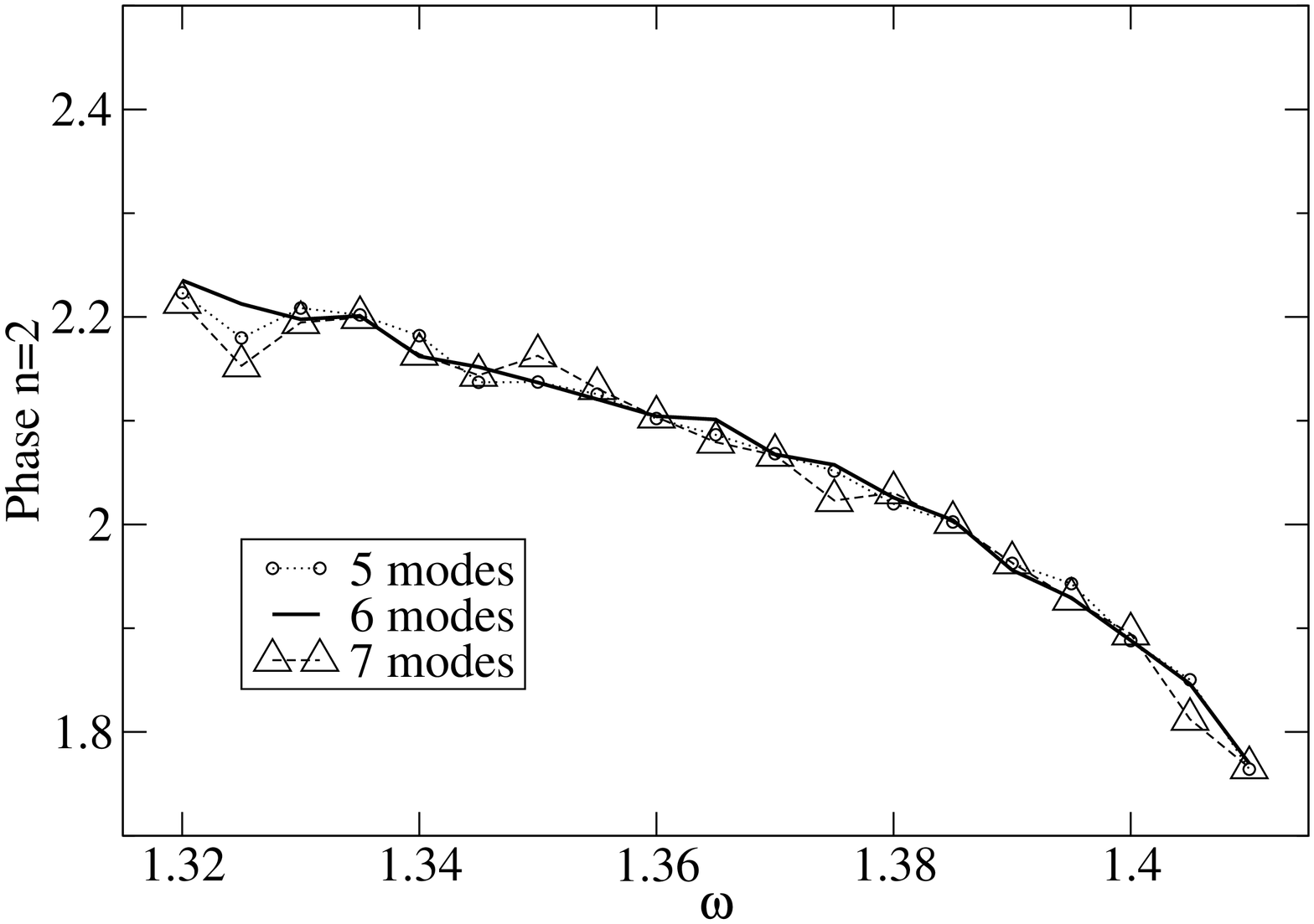}
\includegraphics[height=6.5cm]{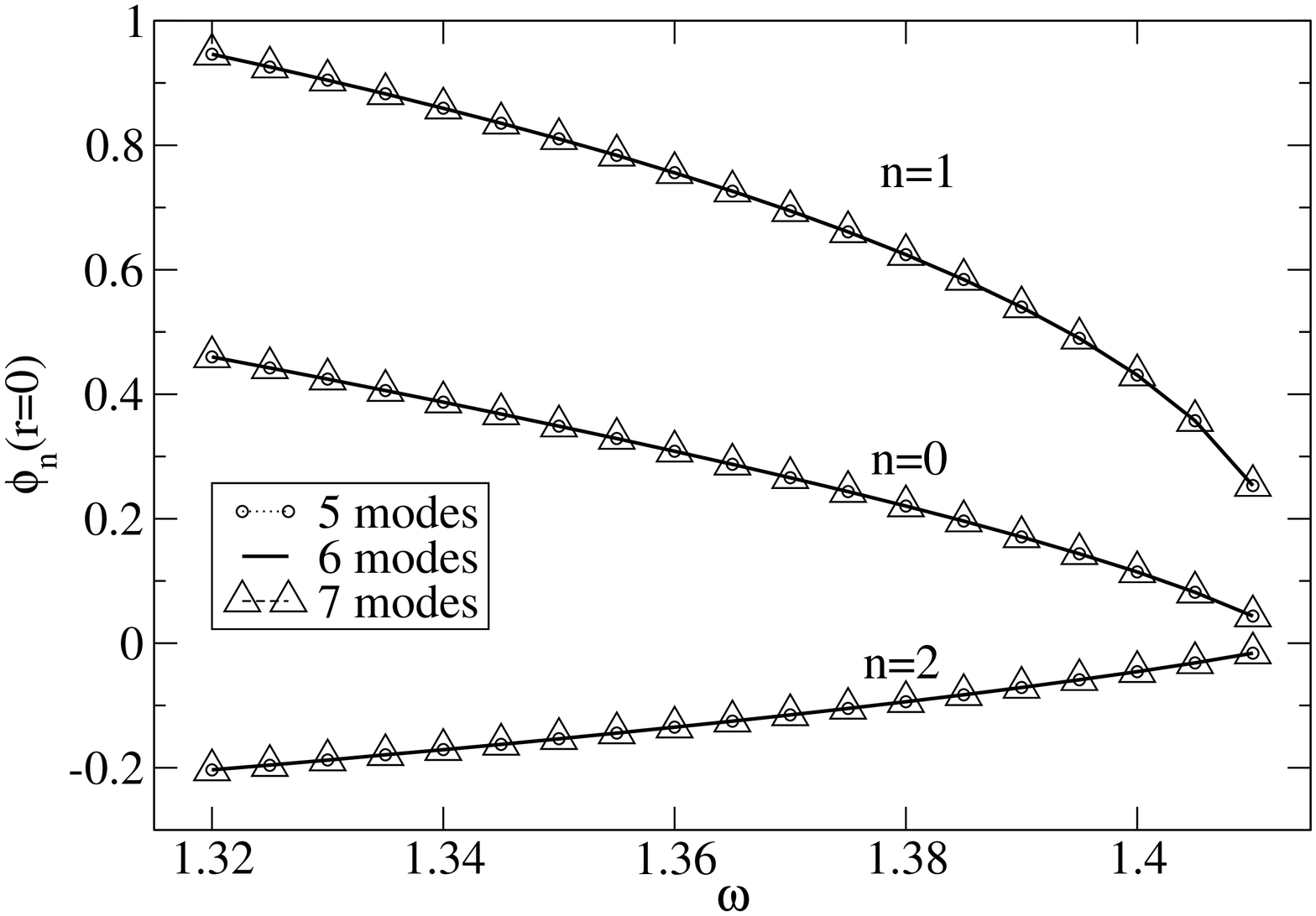}
\includegraphics[height=6.5cm]{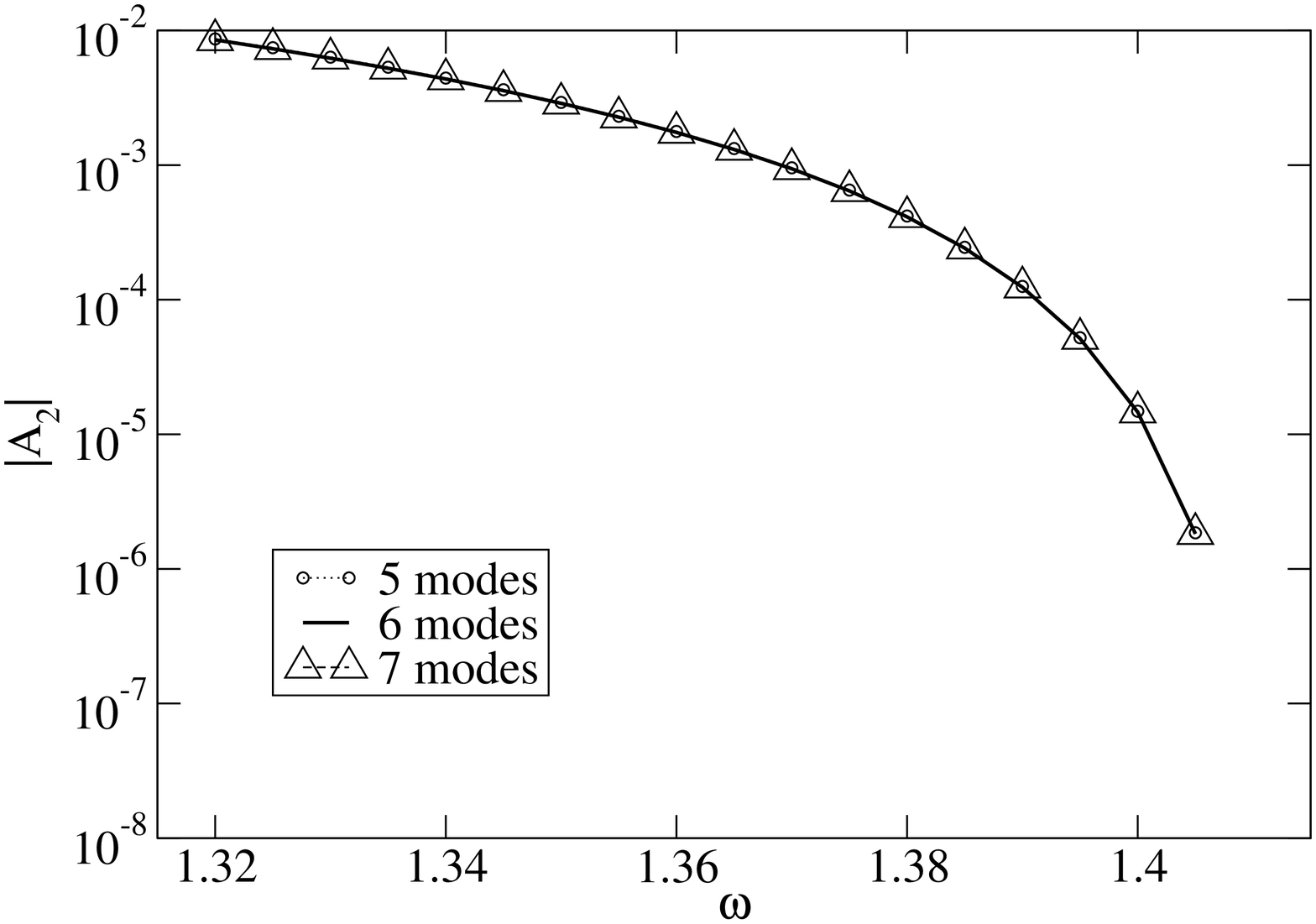}
\caption{\label{f:conv_modes}
Same as Fig. \ref{f:conv_prec} but varying the number of modes.}

\end{figure}

A strong test is provided by the plots of Fig. \ref{f:conv_rlim}. They show that the results
are
independent of the value of the outer computational radius $R_{\rm lim.}$ thus validating the
matching
procedure with the oscillatory homogeneous solutions.
Finally Fig.\ \ref{f:conv_modes} illustrates the fact that the results do not
change substantially when increasing the number of modes.

\begin{figure}
\includegraphics[height=8cm]{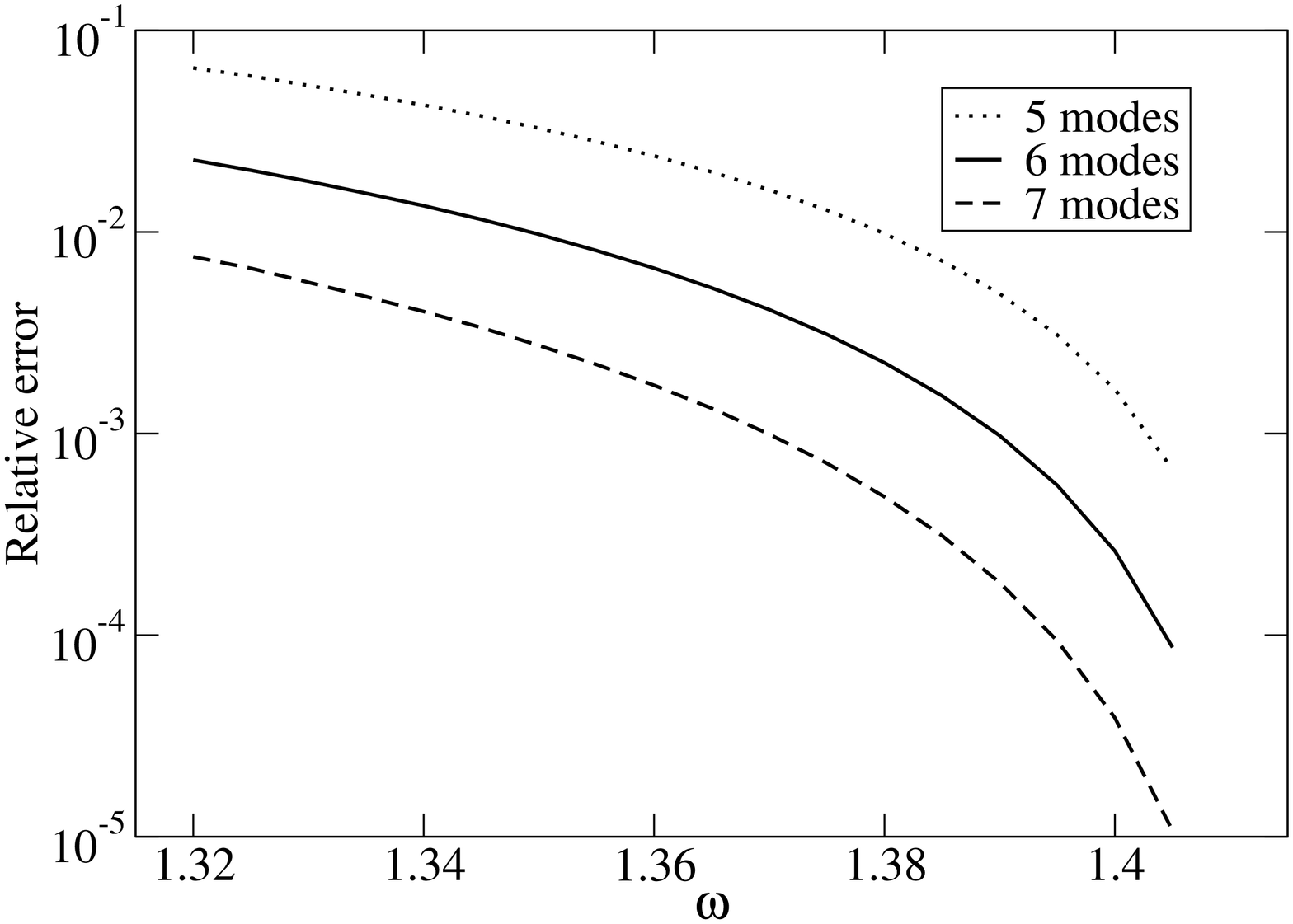}
\caption{\label{f:error}
Relative difference between the l.h.s. and r.h.s. members of Eq. (\ref{e:evol}). The result is
averaged over one
period and we present the error as a function of $\omega$, for 5, 6 and 7 modes.
}
\end{figure}

The last test of our code has been a comparison of the r.h.s. and l.h.s. of Eq. (\ref{e:evol}),
using the Fourier
expansion (\ref{e:period}). We have computed the relative difference between the r.h.s. and l.h.s.
after averaging
over one period. The results for various number of modes are depicted on Fig. \ref{f:error}
as a function of $\omega$. As expected, the error decreases as the number of modes increases.
This is not
surprising, given that Eq. (\ref{e:system_bis}) could be satisfied only for an infinite number
of modes.
The error also increases when $\omega$ decreases. This can be understood by recalling that the
modes are more important when $\omega$ is small (i.e. see the behavior of $\phi_n\l(r=0\r)$ as a function
of $\omega$ on Figs. \ref{f:conv_nr}, \ref{f:conv_rlim} and \ref{f:conv_modes}). Thus,
the effect of the missing modes is more important for smaller $\omega$.

Finally we would like to emphasize that the computational parameters exhibited in this section are
sufficient to compute solutions in the regime of ''moderate'' frequencies (i.e. frequencies
around $1.37-1.38$). If one wishes to go to much lower frequencies, one would need to include
more modes, for higher order modes will be more important, as indicated by
Fig. \ref{f:error}. On the other hand, as will be seen in the next section,
the matching point $R_{\rm lim.}$ must be increased when one approaches the critical value $\omega_c = \sqrt{2}$.

\section{Results}\label{s:results}

As illustrated on
Figures \ref{f:conv_nr}, \ref{f:conv_rlim} and \ref{f:conv_modes}
we do find time-periodic quasi-breather solutions of Eqs.\ (\ref{e:system_bis}) for any
value of the pulsation frequency $1.32\leq\omega\leq1.41$, and there is little doubt that
such solutions exist for all frequencies in the range $]0, \sqrt{2}[$.
It is also clear from the very same Figures that for
$\omega\to\omega_c = \sqrt{2}$ the family of solutions we consider converges pointwise to
the trivial solution.
We do not expect such quasi-breather solutions
of the system (\ref{e:system_bis}) to exist for $\omega>\omega_c$.
The critical value $\omega_c = \sqrt{2}$ is expected to be related to
the change of nature of the Helmholtz operator for $\phi_1$,
$\sqrt{2}$ being the value at which $\phi_1$ ceases to decay like an exponential.
In the $1+1$ dimensional case Coron \cite{Coron} has proved that the allowed frequencies
are indeed constrained by $\omega<\omega_c$.
We have not attempted to prove the analogous statement for our case, as it would
certainly require some sophisticated mathematical tools.

\begin{figure}
\includegraphics[height=8cm]{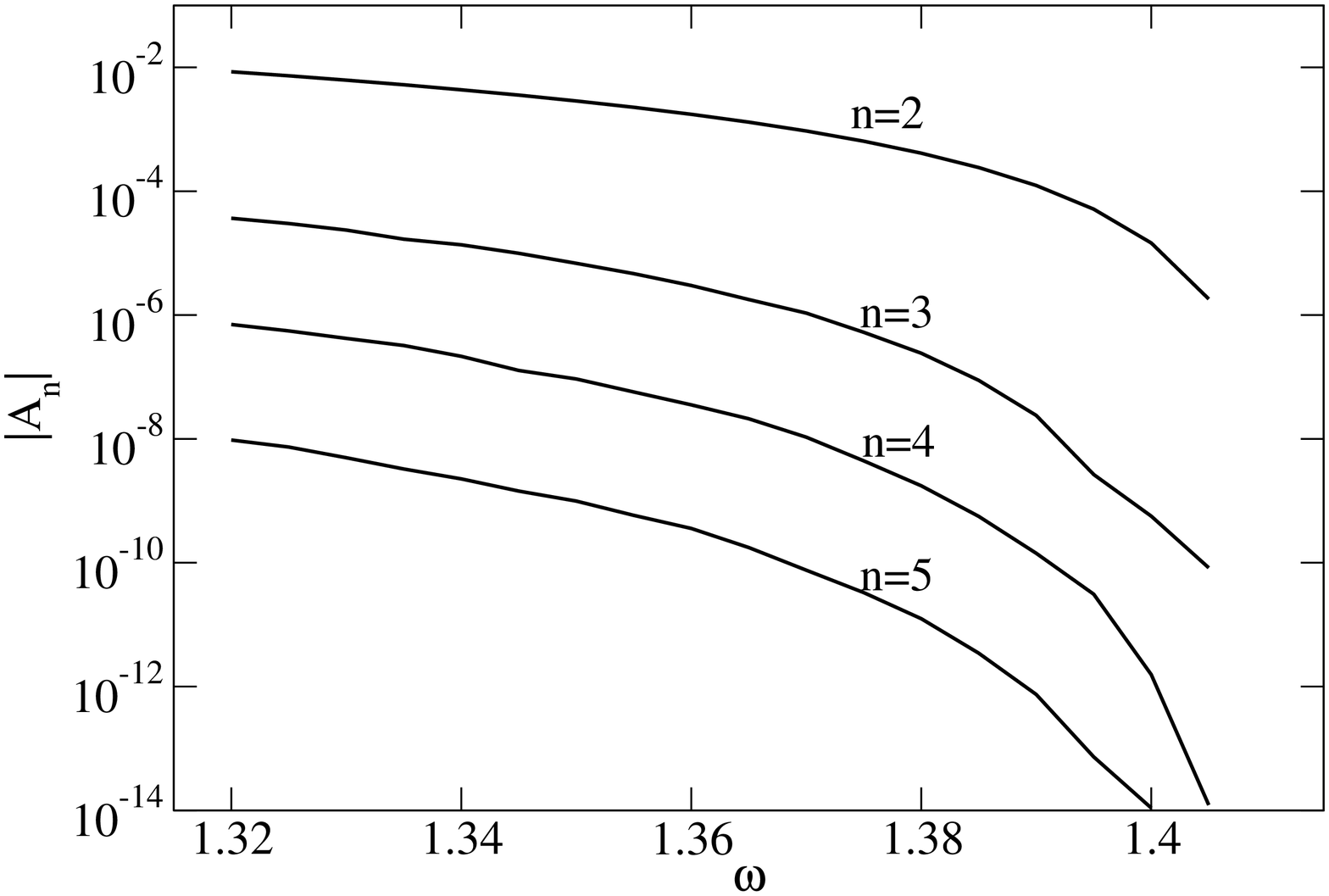}
\caption{\label{f:coefs}
Coefficients of the oscillatory homogeneous solutions $\varphi_n$ ($n=2,3,4,5$)
as a function of
$\omega$.
}
\end{figure}

The quasi-breather solutions obtained this way are not well localized
in space because of their slowly decaying $\propto1/r$ oscillatory tail.
Consequently none of these solutions has finite energy.
In fact we have selected a special class of time periodic solutions
by minimizing the amplitude of their oscillatory tail. This is, in some
sense, the closest one can get to a breather and we have called those configurations
``quasi-breathers''.
There is no special value of $\omega$ for which the amplitude of the oscillatory tail
would show any tendency to become very small or going to zero.
This is illustrated on Fig. \ref{f:coefs} where the logarithm of the coefficients
of ``tails'' are shown for the modes $\phi_2$, $\phi_3$, $\phi_4$ and $\phi_5$,
as a function of $\omega$.
The only value of $\omega$ for which the curves go to
zero is the critical one for which the solution tends to the trivial one.
The many convergence tests of Sec. \ref{ss:convergence} show that we can accurately
compute the value of those coefficient and that they are
not numerical artifacts.

\begin{figure}
\includegraphics[height=6.5cm]{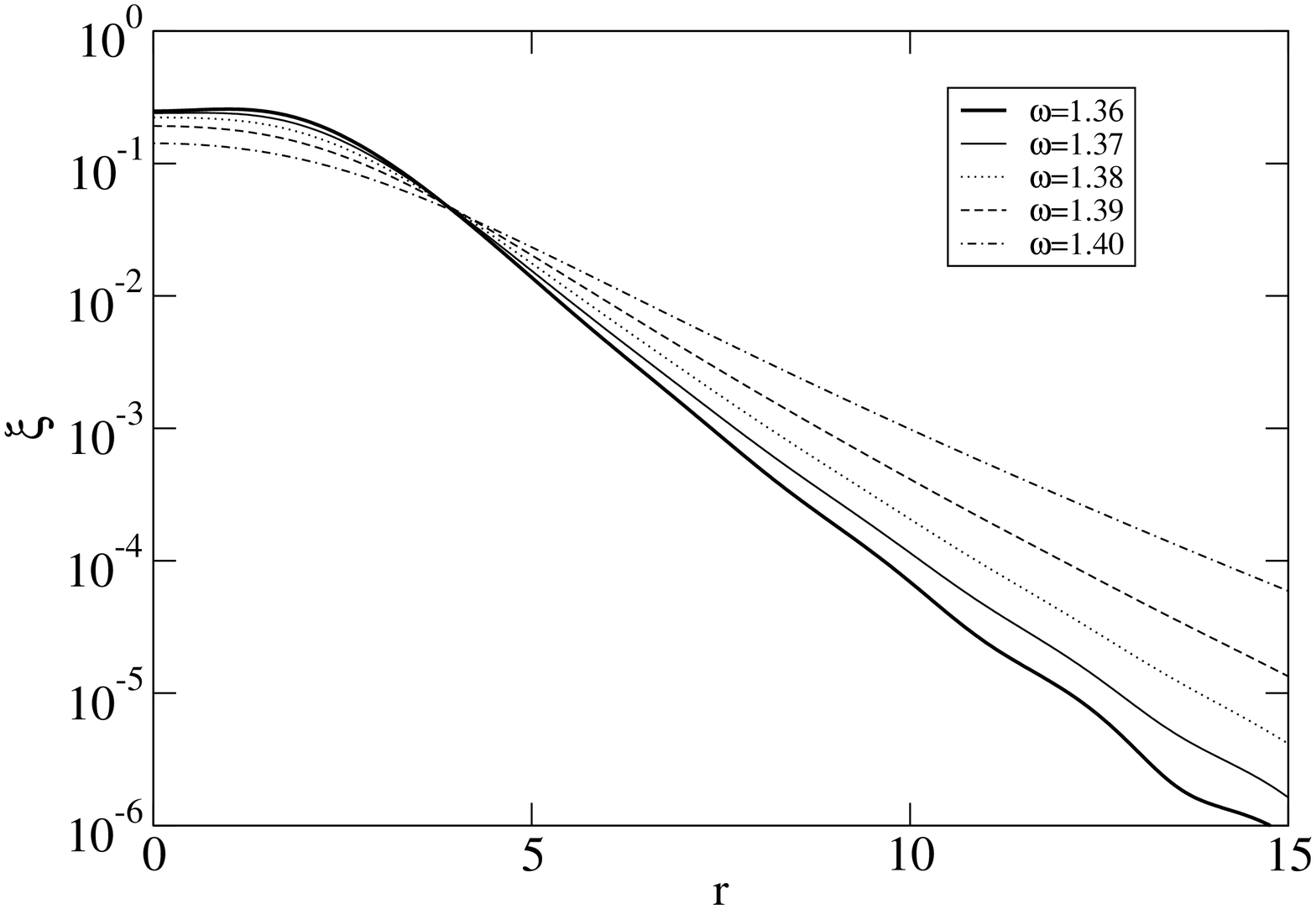}
\includegraphics[height=6.5cm]{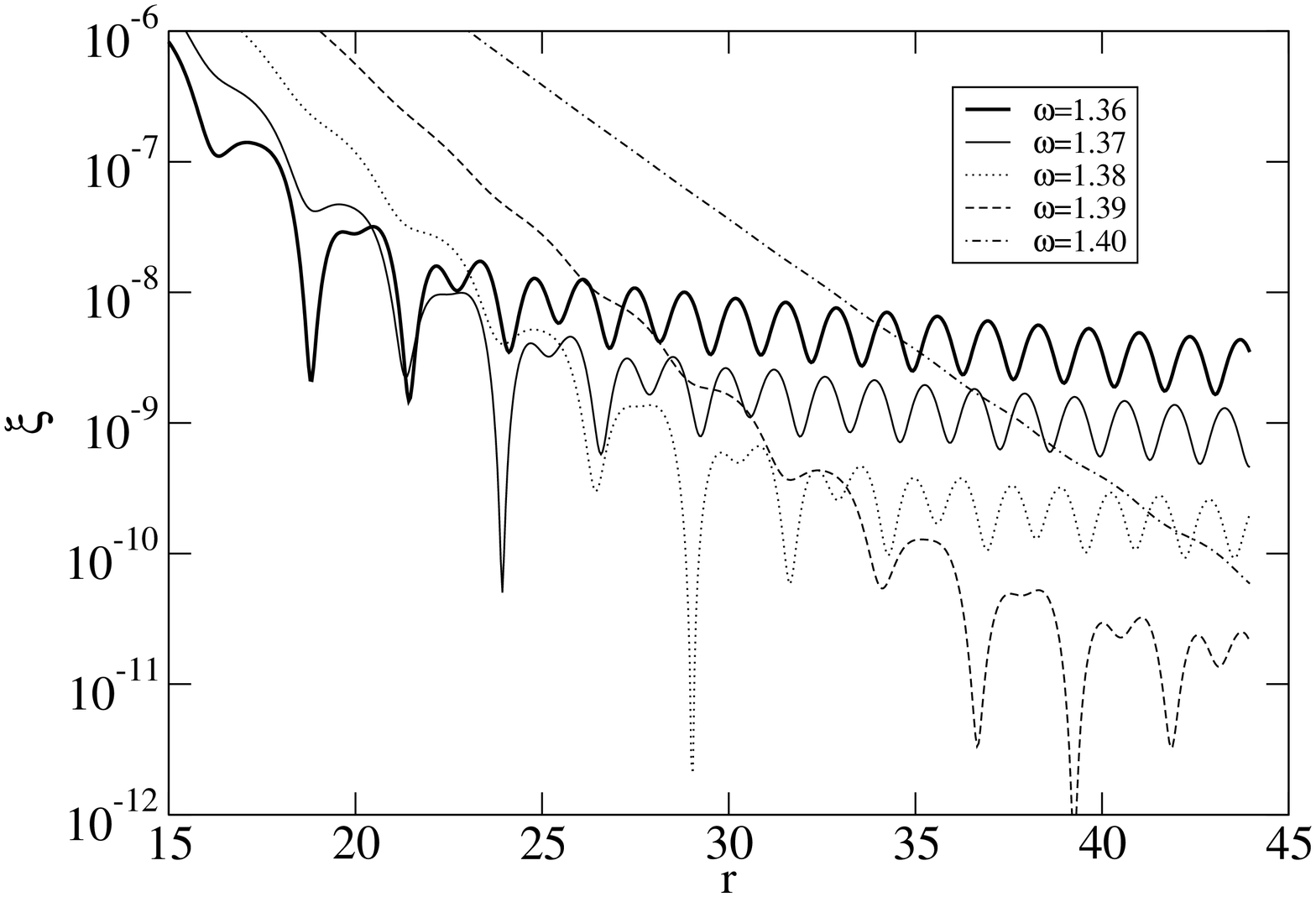}
\caption{\label{f:dens}
Energy density for various values of $\omega$, near the origin (left panel) and in the region where the oscillatory tails
dominate (right panel). $R_{\rm lim.}$ is equal to 44.
}
\end{figure}

\begin{figure}
\includegraphics[height=6.5cm]{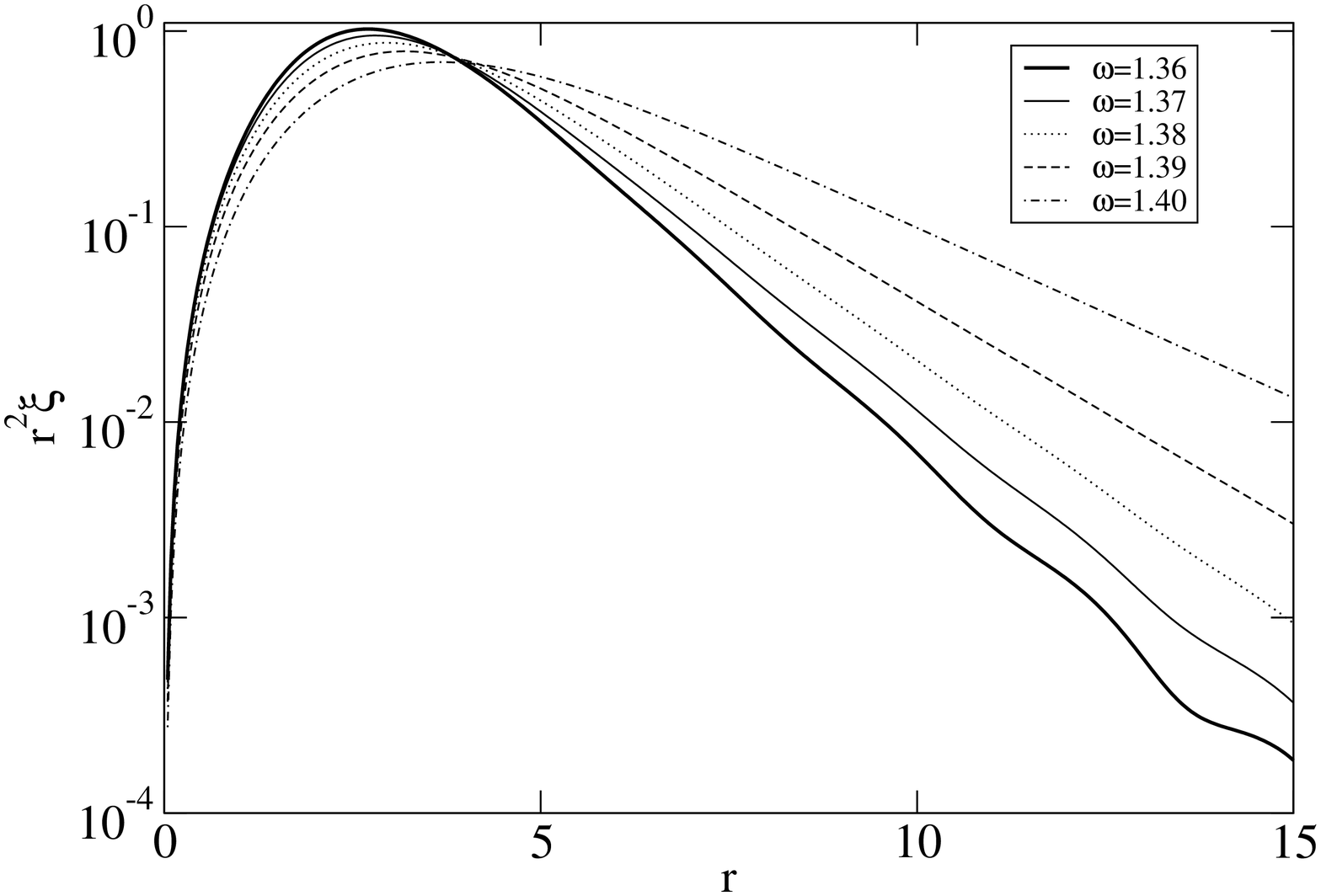}
\includegraphics[height=6.5cm]{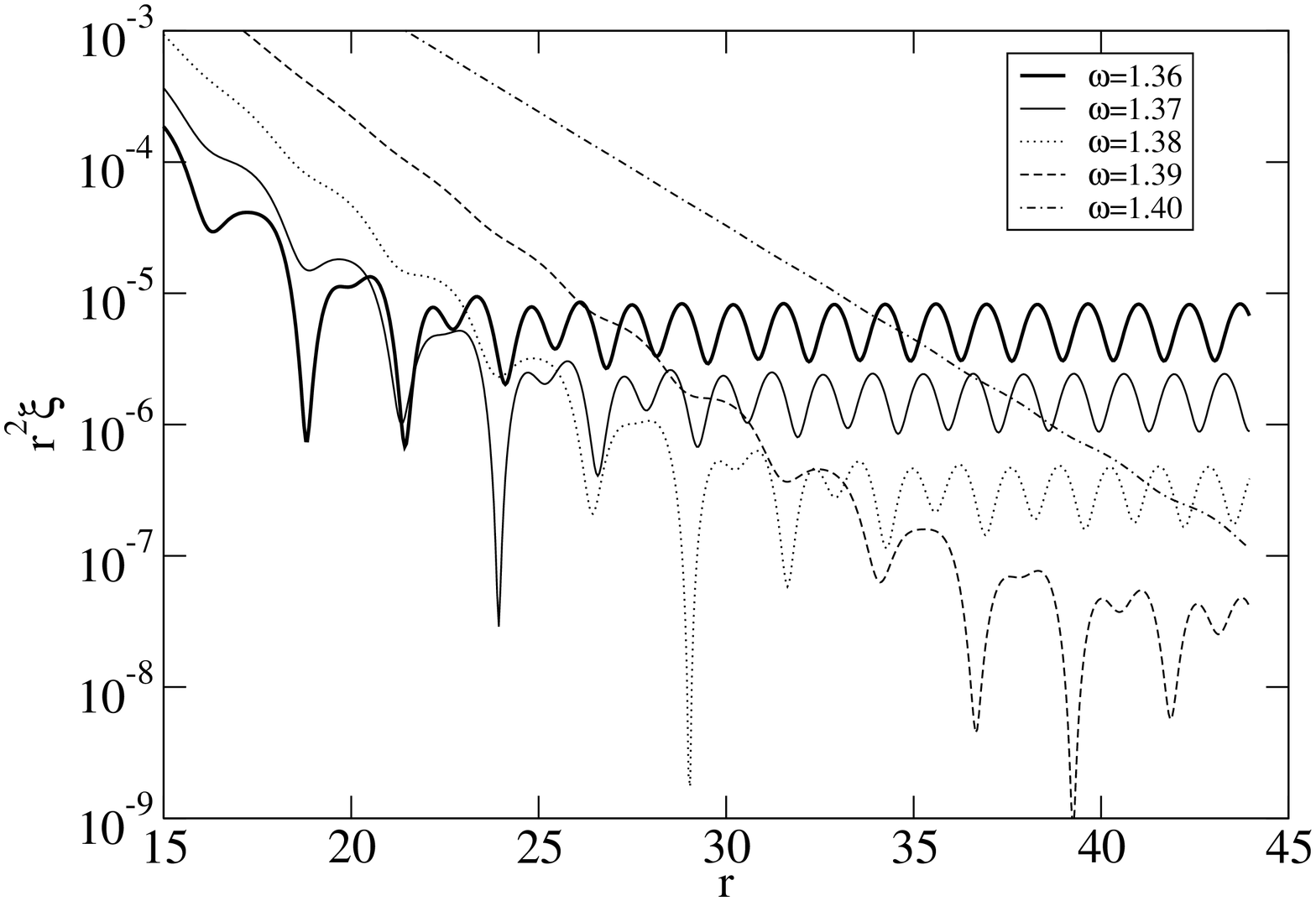}
\caption{\label{f:lin}
Same as Fig. \ref{f:dens} but with the volume element included : $r^2 \xi$.
}
\end{figure}

%As already mentioned, the presence of such homogeneous solutions causes the energy to diverge
%at infinity.
To illustrate the behavior of the quasi-breathers
we show on Fig. \ref{f:dens} and \ref{f:lin} the energy density $\xi$ and the one including
the volume element,
i.e.\  $r^2 \xi$ as a function of the radius, for values of $\omega$ going from
$1.36$ to $1.40$.
We can clearly see two qualitatively different behaviors:
i) the solutions have a well defined ``core'', where the
behavior is dominated by the exponential decay of the fields and so the density goes to zero.
This core is getting larger when $\omega\to\sqrt{2}$.
ii) However, inevitably at some point, the oscillatory tails start to dominate and
the density reaches a plateau, ultimately causing the total (integrated) energy to be infinite.
Let us mention that if for high values of $\omega$, the plateau is not seen, it comes solely
from the fact
that the value of $R_{\rm lim}$ used on Fig. \ref{f:dens} and \ref{f:lin} is not sufficient.
In spite of the numerical smallness of the amplitude of this plateau,
we are quite confident that its value is reasonably accurate, and also
that it really corresponds to a physical effect. Indeed, its value is very stable
when varying the various computational parameters.

As $\omega$ increases, the value of the
plateau diminishes. This is related to the fact
that the coefficients of the homogeneous solutions are decreasing functions of $\omega$ and to
the quadratic nature of the energy density in $\phi$, i.e.\ the value of the plateau
is roughly the square of the greatest coefficient (i.e. the one for $\phi_2$).

\begin{figure}
\includegraphics[height=6.5cm]{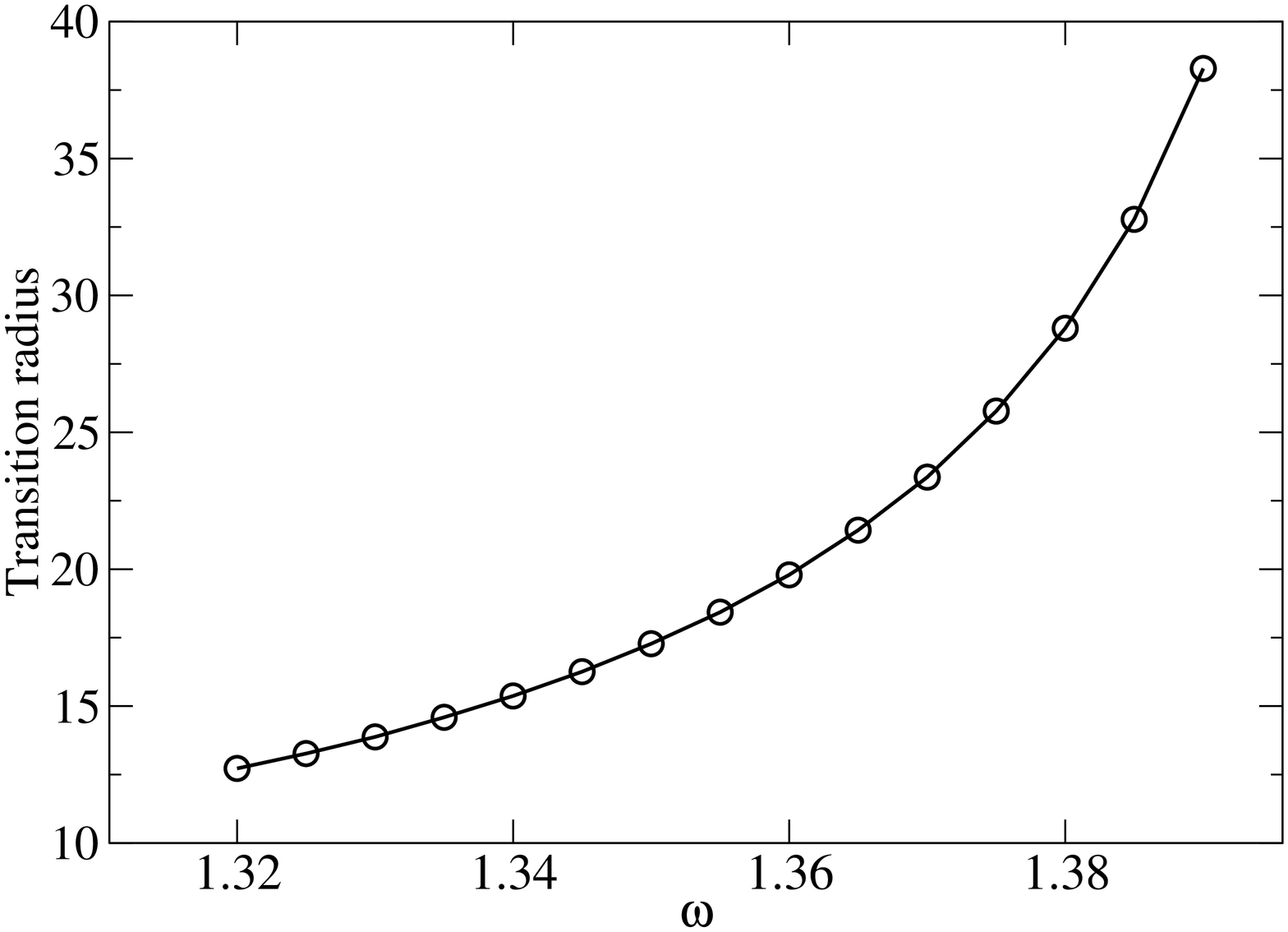}
\includegraphics[height=6.5cm]{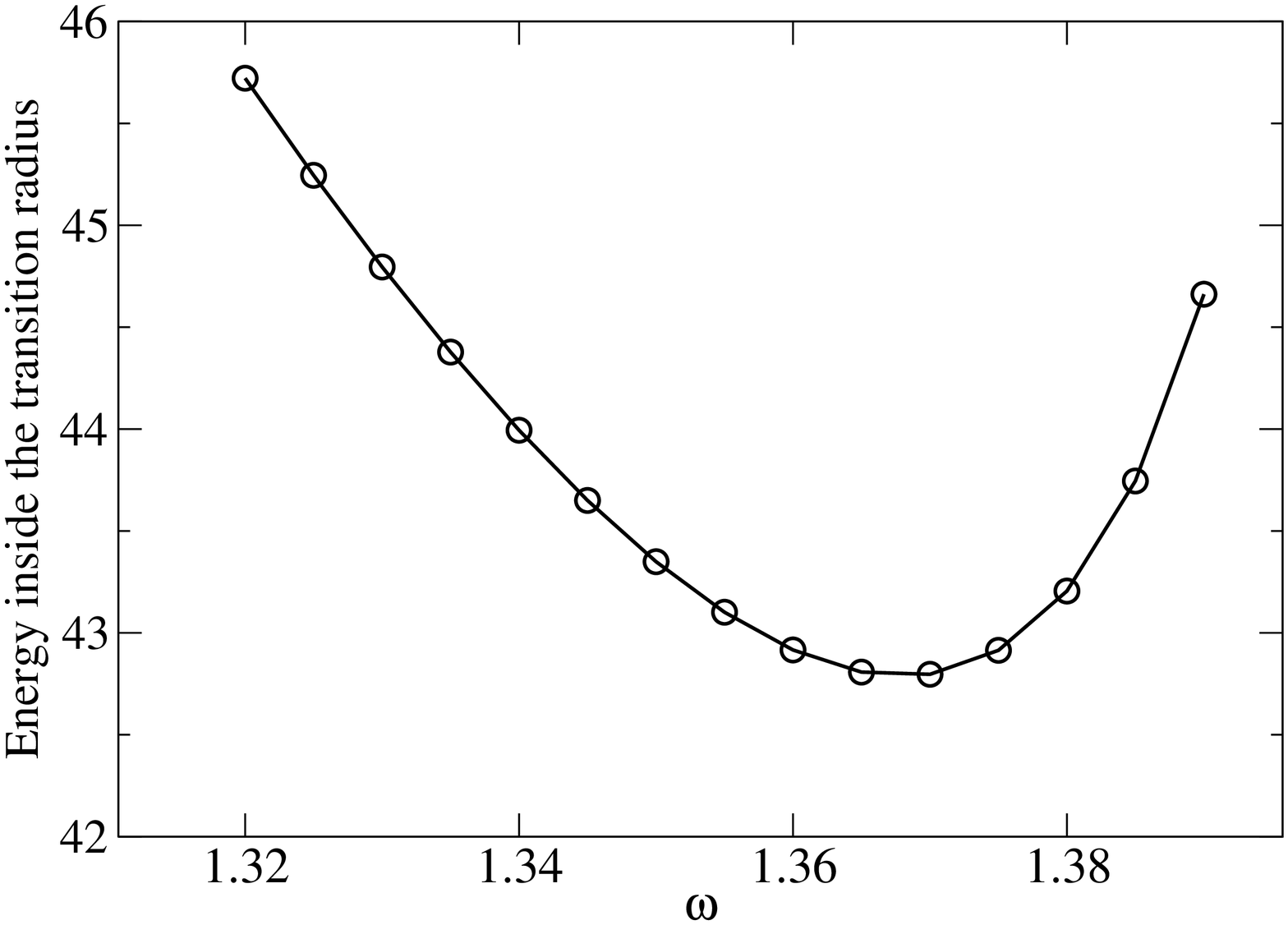}
\caption{\label{f:trans}
The transition radius between the exponential decay and the oscillatory tail (left panel)
and the energy inside
this radius (right panel), as a function of $\omega$.
}
\end{figure}

As already stated, we can also confirm that the transition radius between the core and the
plateau increases as one approaches $\omega = \sqrt{2}$. To be more quantitative, we define
the transition radius
$R_{\rm trans}$ as the first value of $r$ for which $\phi_1$ reaches the amplitude of the
oscillations :
\be
\phi_1\l(r=R_{\rm trans}\r) = \frac{\l|A_2\r|}{r}.
\ee
It is the radius at which the oscillatory behavior
starts to dominate. $R_{\rm trans}$ as a function of $\omega$ is shown on the left panel of
Fig. \ref{f:trans}.
The precise location of the transition radius can not be computed for $\omega>1.39$, given our
nominal choice for $R_{\lim} = 44$. Once again, this illustrates
the fact that when going to higher values of $\omega$, one needs to increase the boundary
radius $R_{\lim}$.
On the right panel of Fig. \ref{f:trans} we show the total energy inside the transition radius,
i.e. the total energy of the core of the configuration. Contrary to the transition radius,
this is not a monotonic function and there is a configuration of minimum energy.
This non trivial
behavior is due to two competing effects when increasing $\omega$, first the increase of the
transition radius, and second the decrease of the magnitude of the modes.
We also note that this minimum is very close to
the value of the frequency which Honda and Choptuik \cite{Honda} claim to be the one of
their conjectured breather solution (i.e. they
quoted $\omega \approx 1.366$) but this may very well be just a coincidence.
It seems likely that the
 energy contained in the core diverges when $\omega \rightarrow \sqrt{2}$.

\begin{figure}
\includegraphics[height=6.5cm]{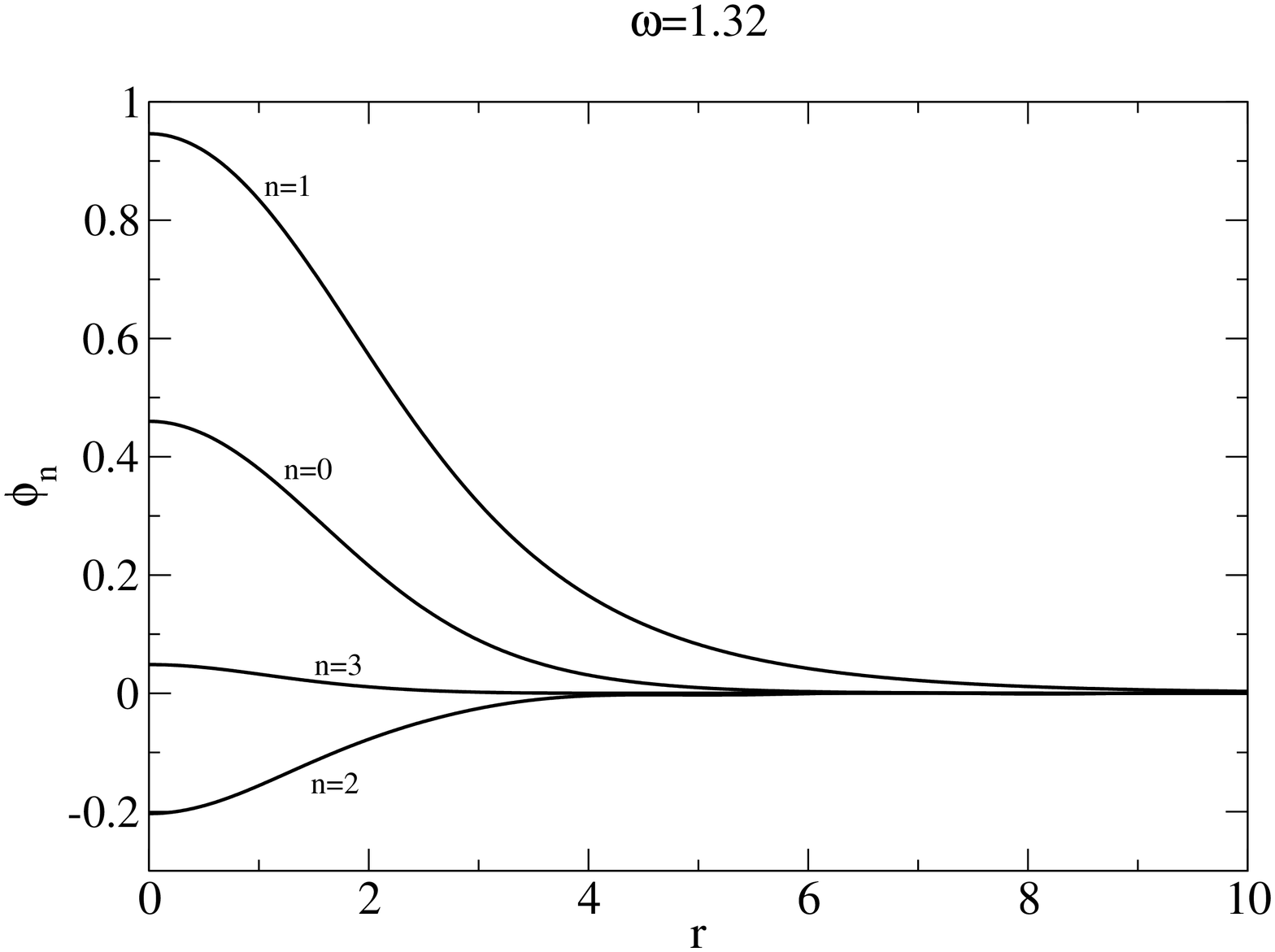}
\includegraphics[height=6.5cm]{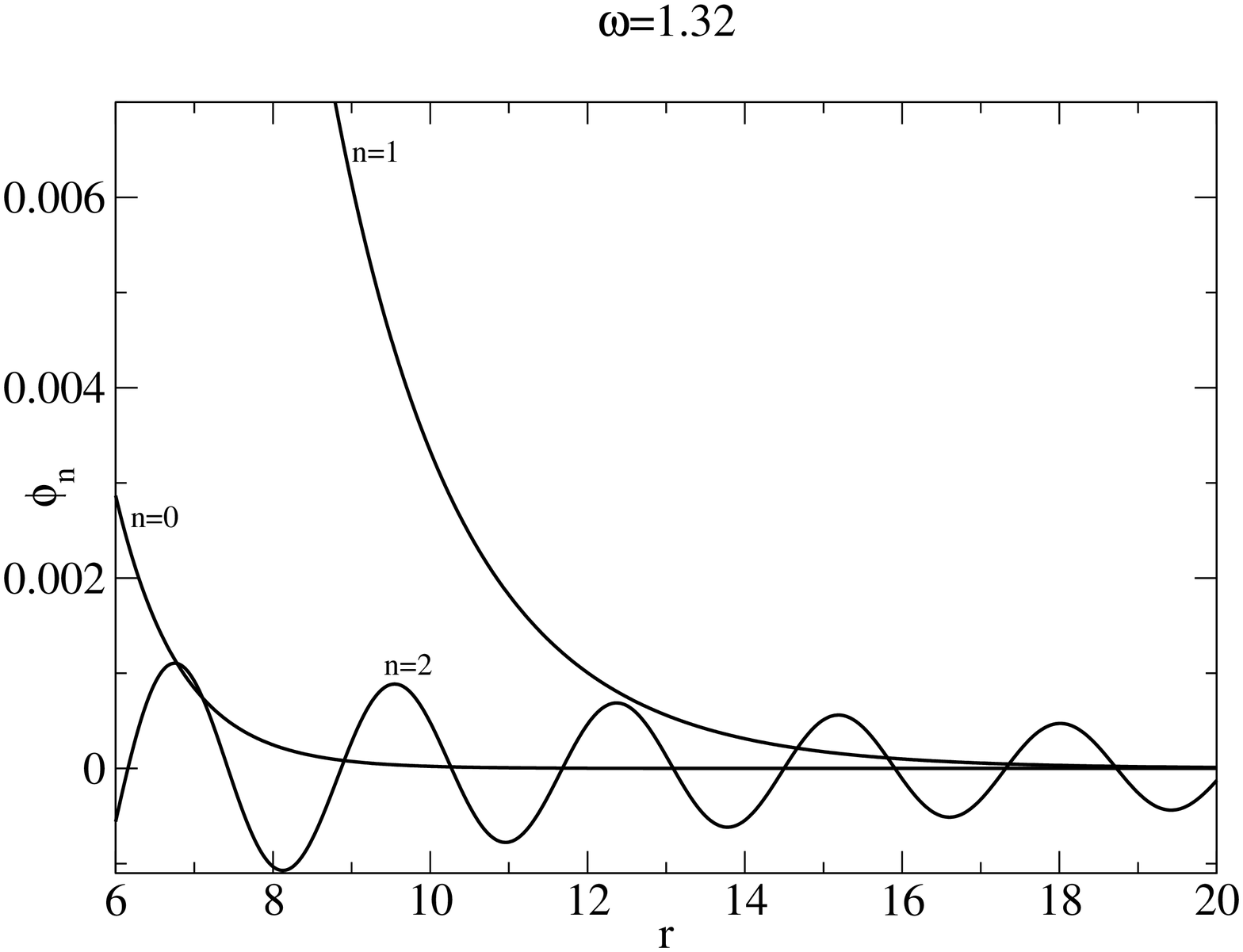}\\
\includegraphics[height=6.5cm]{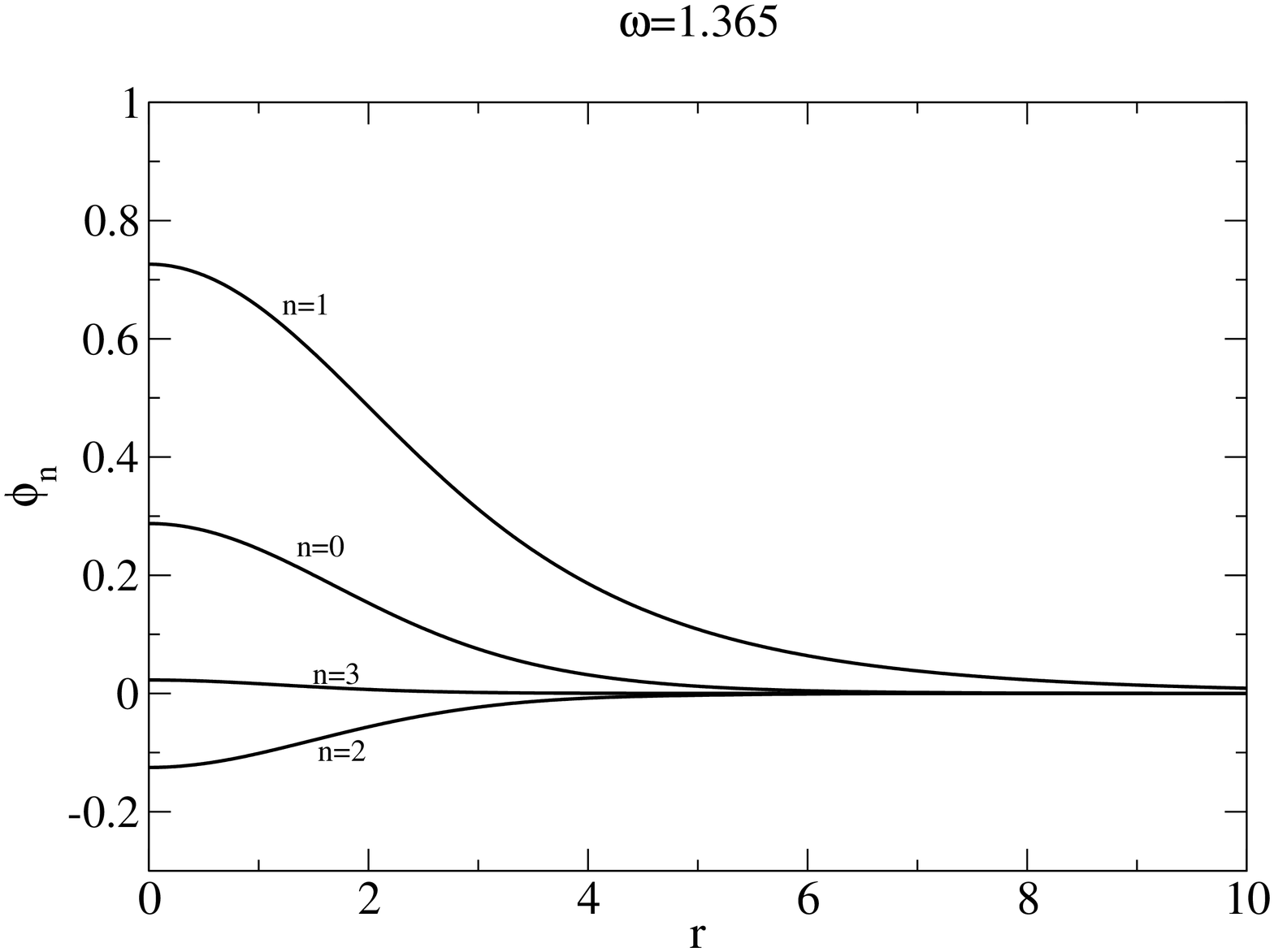}
\includegraphics[height=6.5cm]{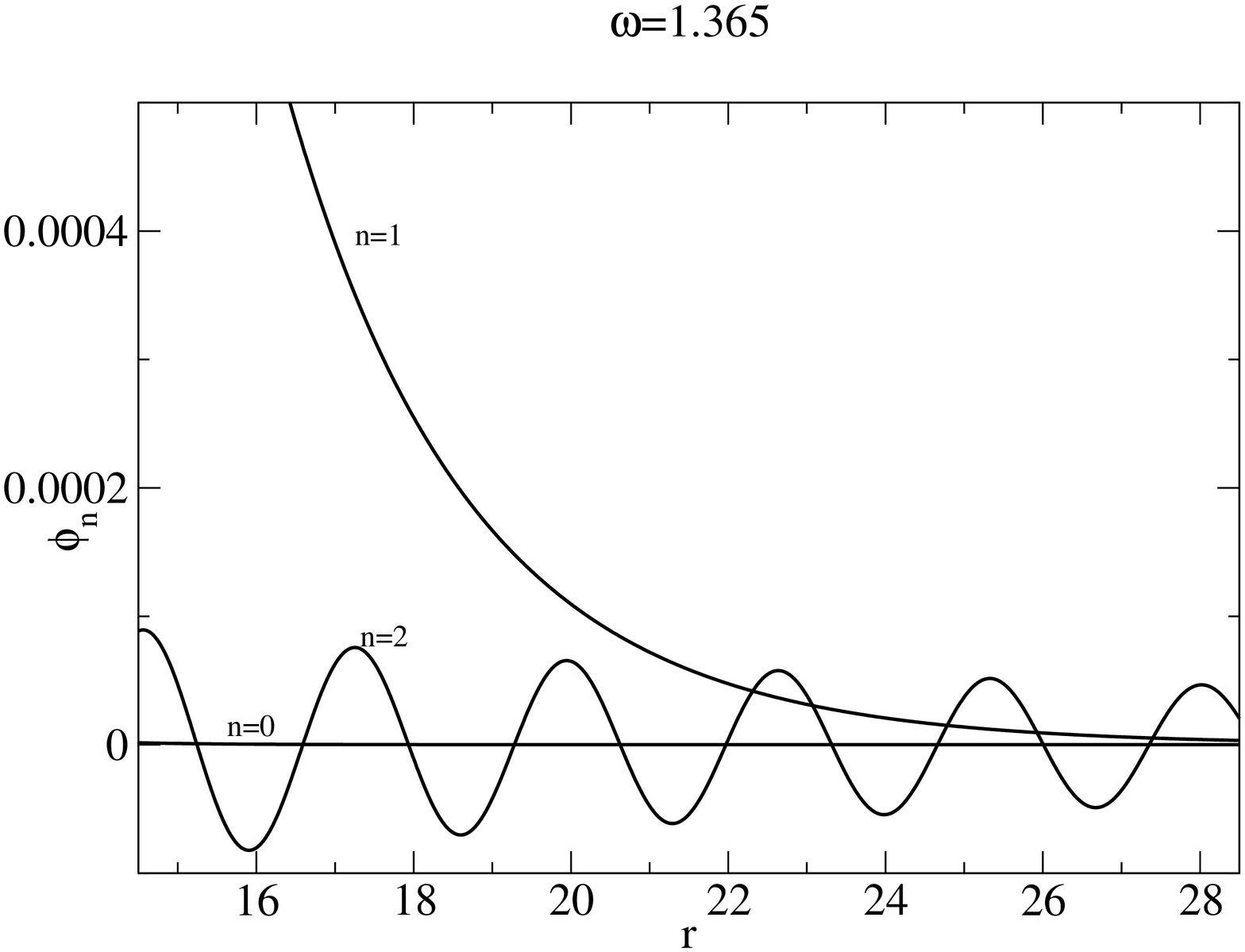}\\
\includegraphics[height=6.5cm]{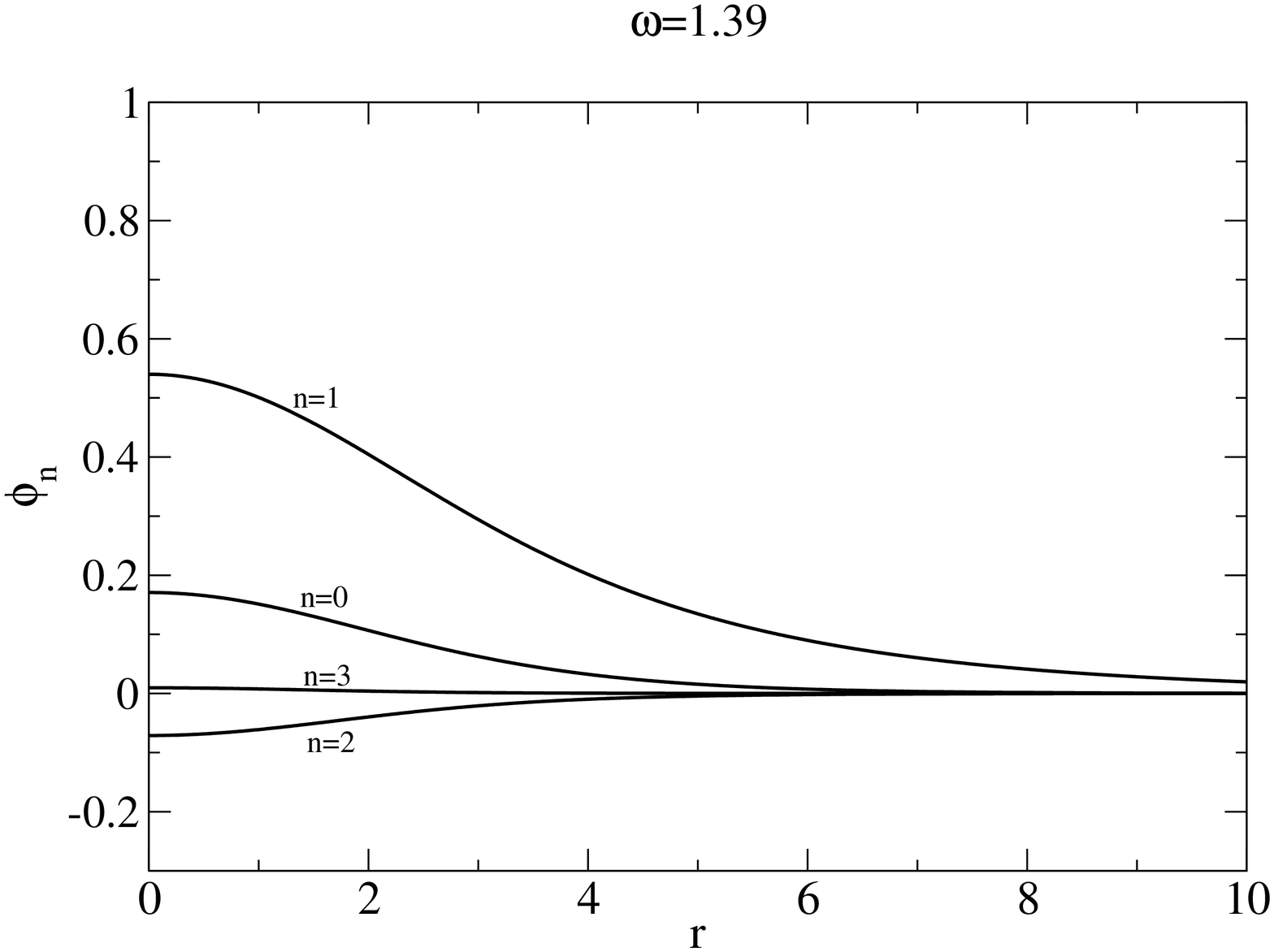}
\includegraphics[height=6.5cm]{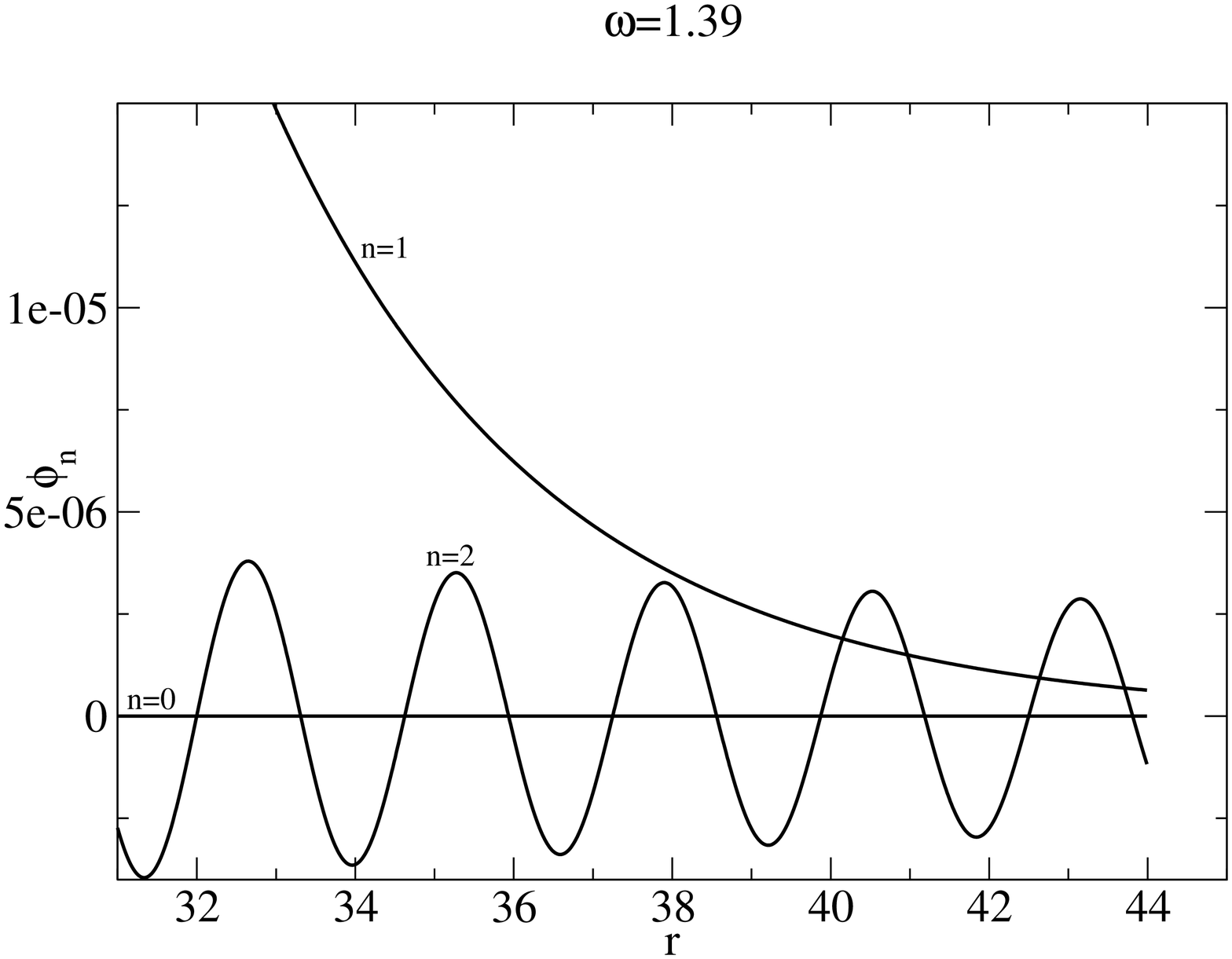}
\caption{\label{f:oscillons}
The first modes near the origin and in the transition region for
$\omega=1.32$, $1.365$ and $1.39$.}
\end{figure}

Finally, on Fig. \ref{f:oscillons} we show the first modes near the origin (left panels)
and in the region of the transition radius (right panels). Three different values of
$\omega$ are shown, corresponding to a low value ($1.32$), the value corresponding to the
minimum of energy ($1.365$, cf Fig. \ref{f:trans}) and one rather high frequency ($1.39$).

\clearpage

%%%%%%%%%%%%%%%%%%%%%%%%%%%%%%%%%%%%%%%%%%%%%%%%%%%%%%%%%%%%%%%%%%%%%%%%%%%%%%%%%%%%%%%%%%%%%

\section{The Quasi-breather content of the oscillons}\label{s:compar}

%%%%%%%%%%%%%%%%%%%%%%%%%%%%%%%%%%%%%%%%%%%%%%%%%%%%%%%%%%%%%%%%%%%%%%%%%%%%%%%%%%%%%%%%%%%%%

\subsection{Fourier decomposition of evolution results}

In this Section we present our oscillon scenario, based on the existence of the
quasi-breathers.
In particular we present some convincing evidence, that an oscillon
state of main frequency $\omega$ is {\sl quantitatively described} by a
quasi-breather of the same frequency, whose tail is cut off at some value of $r$.
In order to do this we perform Fourier decomposition of various
a long-lived oscillon solutions found by doing the evolution
(see Subsection \ref{s:fourier} for the actual method).
This gives us the fundamental pulsation frequency $\omega$ and the various modes.
Then, those modes are compared to the ones found by directly solving the system
(\ref{e:system_bis}) for this particular value of $\omega$.
The same kind of comparison is done in Ref. \cite{Honda} as to support the
existence of a periodic solution of frequency $\omega\approx 1.366$.
An important difference is that in Ref.\ \cite{Honda} the comparison
is performed only for higher frequency near-periodic states in an interval of
$[0\leq r\leq10]$ which is comparable to the core part of the quasi-breather
and therefore its oscillatory tail is not yet apparent.

It is apparent from the results provided by our evolution code that the
solutions corresponding to the time evolution data are very close to being time
symmetric not only at the center but
also in the intermediately far region, justifying the choice of only cosine terms in the
mode decomposition (\ref{e:period}) when constructing QBs.

\subsection{Near-periodic states}

\begin{figure}[!ht]
\includegraphics[width=12cm]{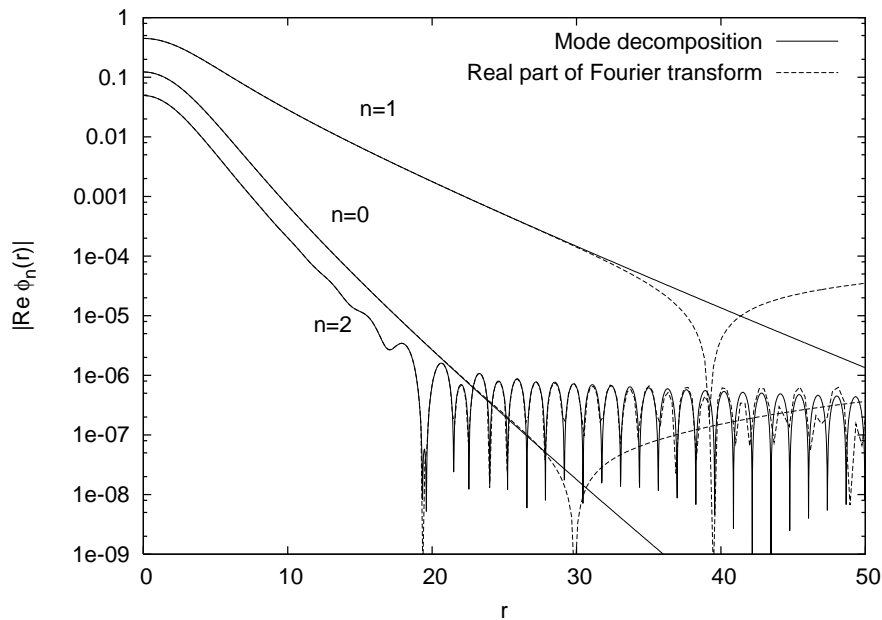}
\caption{\label{f:comp_139a}
Comparison between the modes coming from solving the system (\ref{e:system_bis}) directly
and the real part of the modes obtained by performing a Fourier transform of the
evolution for $\omega = 1.398665$.
In order to be able to represent on the same graph the large values in the central
oscillon region along with the small tails at high radii the absolute values are
plotted logarithmically.
We note that the central value $\phi_n(0)$ is positive for $n=0$ and for
odd $n$ while it is negative for other $n$.
The peaks pointing downwards on the figure correspond to places where the functions
change signature.
}
\end{figure}

\begin{figure}[!ht]
\includegraphics[width=12cm]{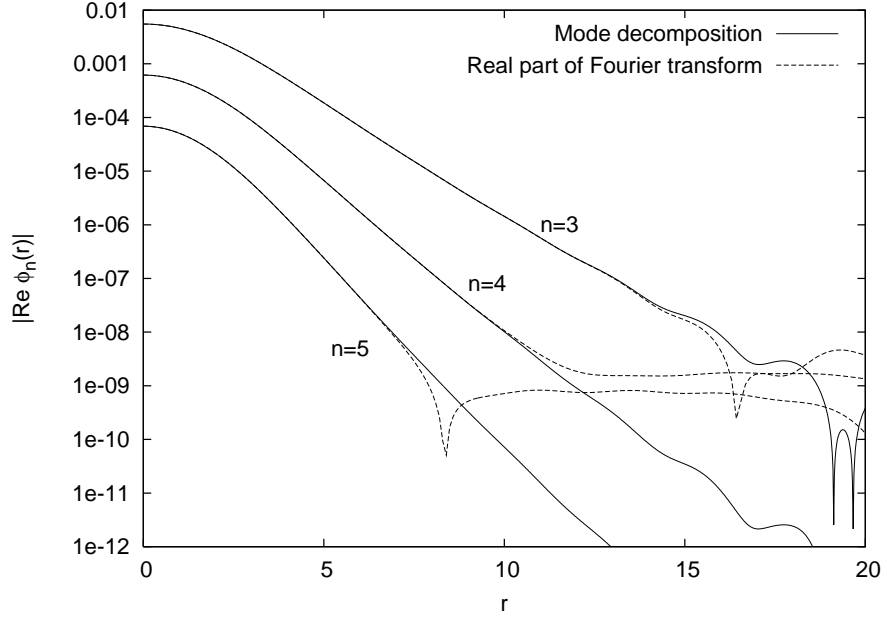}
\caption{\label{f:comp_139b}
Comparison of the higher modes of the two systems presented on the previous
figure.
}
\end{figure}

\begin{figure}[!ht]
\includegraphics[width=12cm]{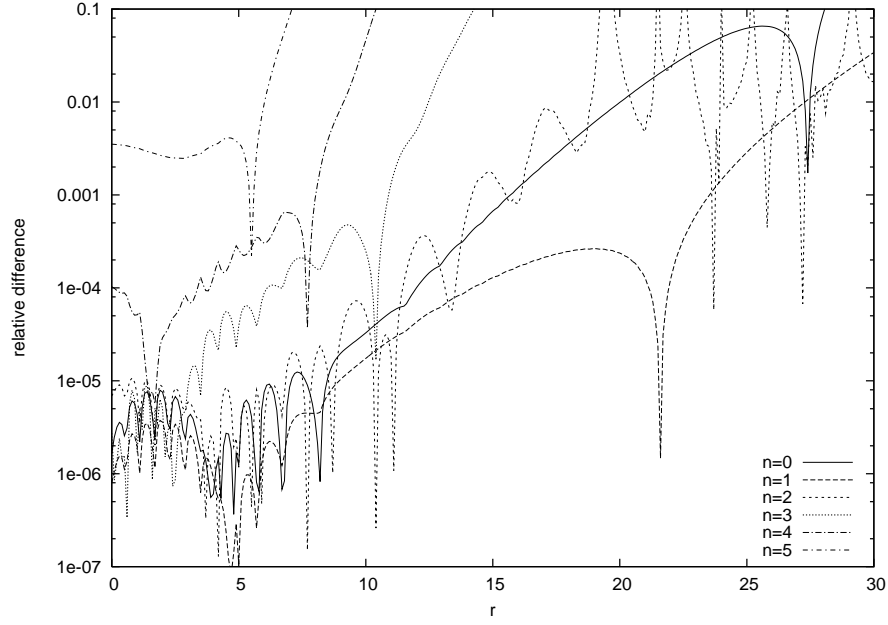}
\caption{\label{f:compd}
Dependence on the radius $r$ of the relative difference of the modes coming from the
mode decomposition and the Fourier transformation for $\omega=1.398665$.
The relative difference is defined as the absolute value of the ratio of the difference
and the mode decomposition value.
}
\end{figure}

First we perform the Fourier decomposition of various near-periodic states
obtained by fine-tuning the initial parameter $r_0$, since these states appear
to be periodic and time symmetric to a very high degree.
For a specific case
Figures \ref{f:comp_139a} and \ref{f:comp_139b} show the real part of the modes coming
from the Fourier transform of the evolution and the modes obtained by the direct
solution of (\ref{e:system_bis}).
The comparison is done with a near-periodic state corresponding to the first peak on
the lifetime curve with initial data $\phi_c=1$.
The Fourier decomposition is performed between the two maxima at
$t_1=1592.29$ and $t_2=1596.78$.
The frequency calculated from the position of the maxima is $\omega=1.398665$,
which is used to calculate the corresponding periodic solution by solving
(\ref{e:system_bis}).
The agreement for the first oscillating mode $n=2$ is remarkably good even in the
oscillating tail region.
The curves obtained by the two different decompositions are
indistinguishable in the central region.
In order to show the excellence of the agreement the radial dependence
of the relative difference of the corresponding modes obtained by the two
methods is shown on Fig.\ \ref{f:compd}.
We have obtained similarly good agreement not only for $\omega=1.398665$ but for
all frequencies for which near-periodic states exist
(i.e. for $1.365<\omega<\sqrt{2}$).

This comparison between the results coming from the
evolution code of Sec.\ \ref{s:evol} and from the Fourier-mode decomposition of the
quasi-breathers is a very strong consistency check.
Let us recall that the two codes have been developed completely independently.

\begin{figure}[!ht]
\includegraphics[width=12cm]{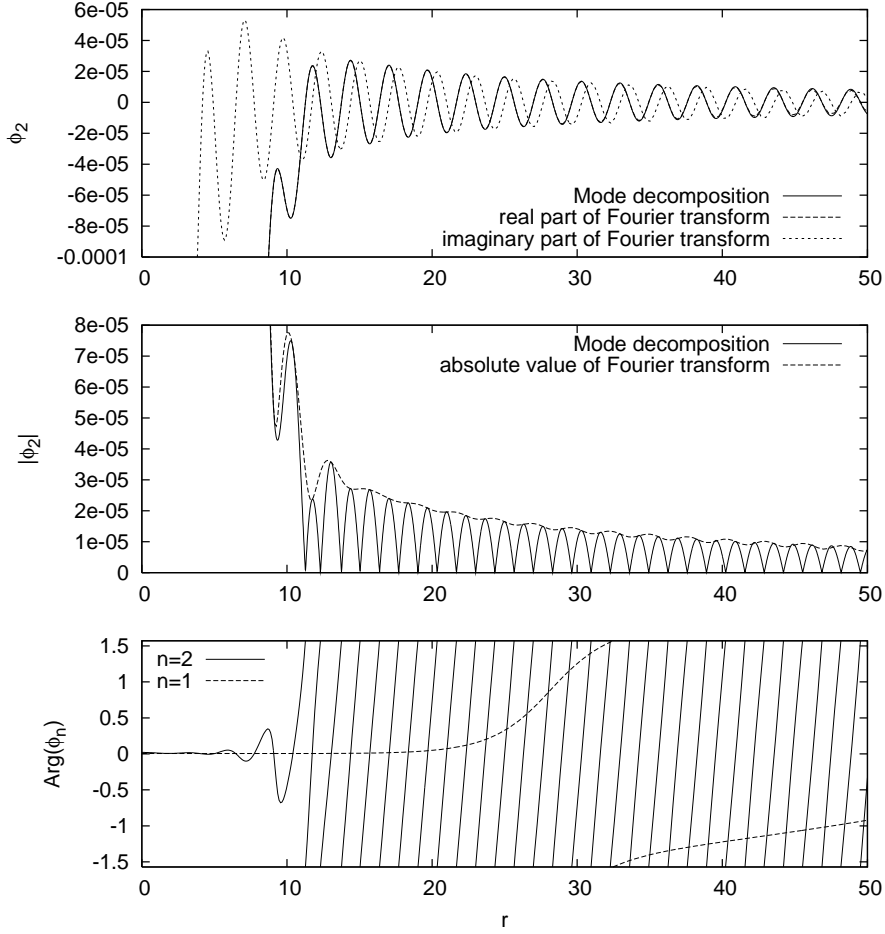}
\caption{\label{f:zoom}
The first two graphs compare the oscillatory region of the mode $\phi_2$ coming from
solving the system (\ref{e:system_bis}) directly and the complex $\phi_2$ obtained by
performing Fourier transform on the result of the evolution code, for $\omega=1.38$.
The first graph shows the real and imaginary parts separately, while the second
compare the absolute values.
The third graph plots the radial dependence of the complex argument of the modes
$\phi_1$ and $\phi_2$ obtained by the Fourier transform.
}
\end{figure}

Figure \ref{f:zoom} shows, for large radii, the first mode for which the oscillatory term
appears, i.e. $\phi_2$, in the case $\omega=1.38$.
It can be seen from the first of the three plots that the real part of the Fourier
transform agrees very well with the decomposition of the corresponding QB,
the two curves can hardly be distinguished.
However, just at those radii where the oscillatory behavior appears the imaginary
part starts being comparable to the real part, breaking the time symmetry in this
outer region.
The second plot shows that the absolute value of the complex $\phi_2$ obtained by the
Fourier decomposition essentially behaves like a smooth envelope curve covering the
standing wave like peaks of $|\phi_2|$ obtained by the mode decomposition.
The presence of the complex argument of the Fourier transform shown on
the third plot indicates that the oscillating tail obtained by the evolution
code is composed of outgoing waves carrying out energy from the core region.

\subsection{Oscillons from Gaussian initial data}

\begin{figure}[!ht]
\includegraphics[width=12cm]{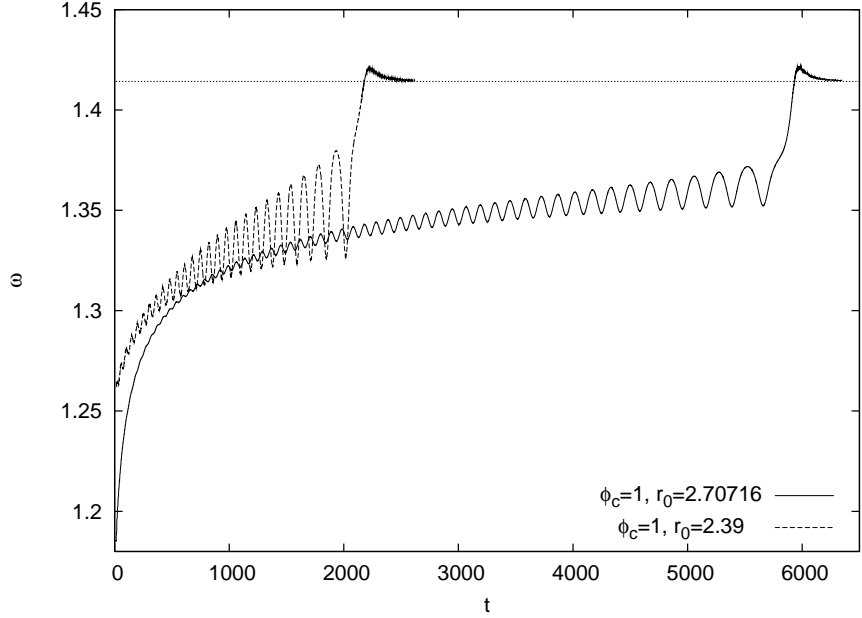}
\caption{
\label{f:nontune2}
Time evolution of the underlying high frequency oscillation of the oscillon state
of Fig.\ \ref{f:nontune1} is presented.
}
\end{figure}

\begin{figure}[!ht]
\includegraphics[width=12cm]{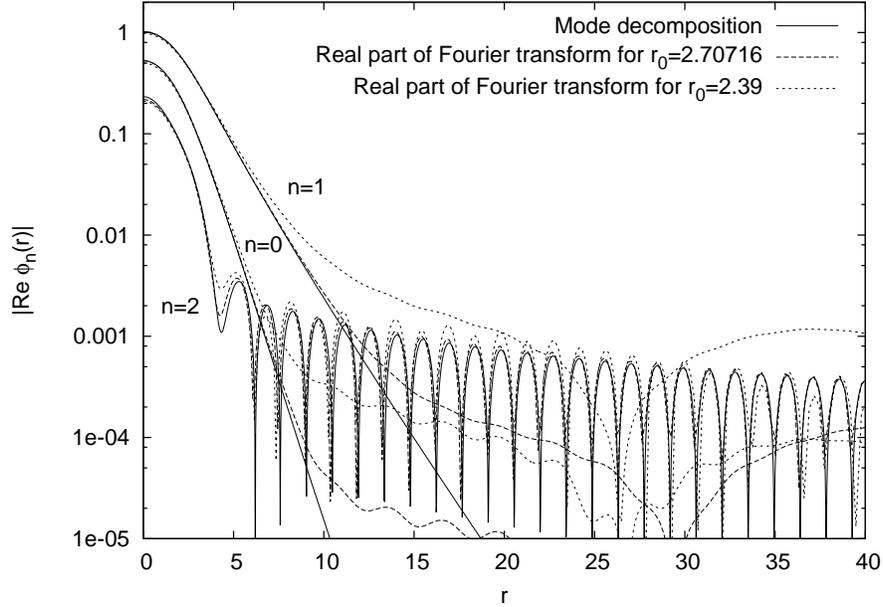}
\caption{\label{f:ntmode30a}
The modes of the periodic solution with frequency $1.3$ is compared with the
Fourier decomposition of two different oscillon states.
The first oscillon state with initial data $\phi_c=1$, $r_0=2.70716$ is decomposed
between the two subsequent maxima after $t=471.47$, where the calculated
frequency is $1.2995$.
The second state with $\phi_c=1$, $r_0=2.39$ is decomposed just after $t=298.72$,
where the frequency is $1.3011$.
}
\end{figure}

\begin{figure}[!ht]
\includegraphics[width=12cm]{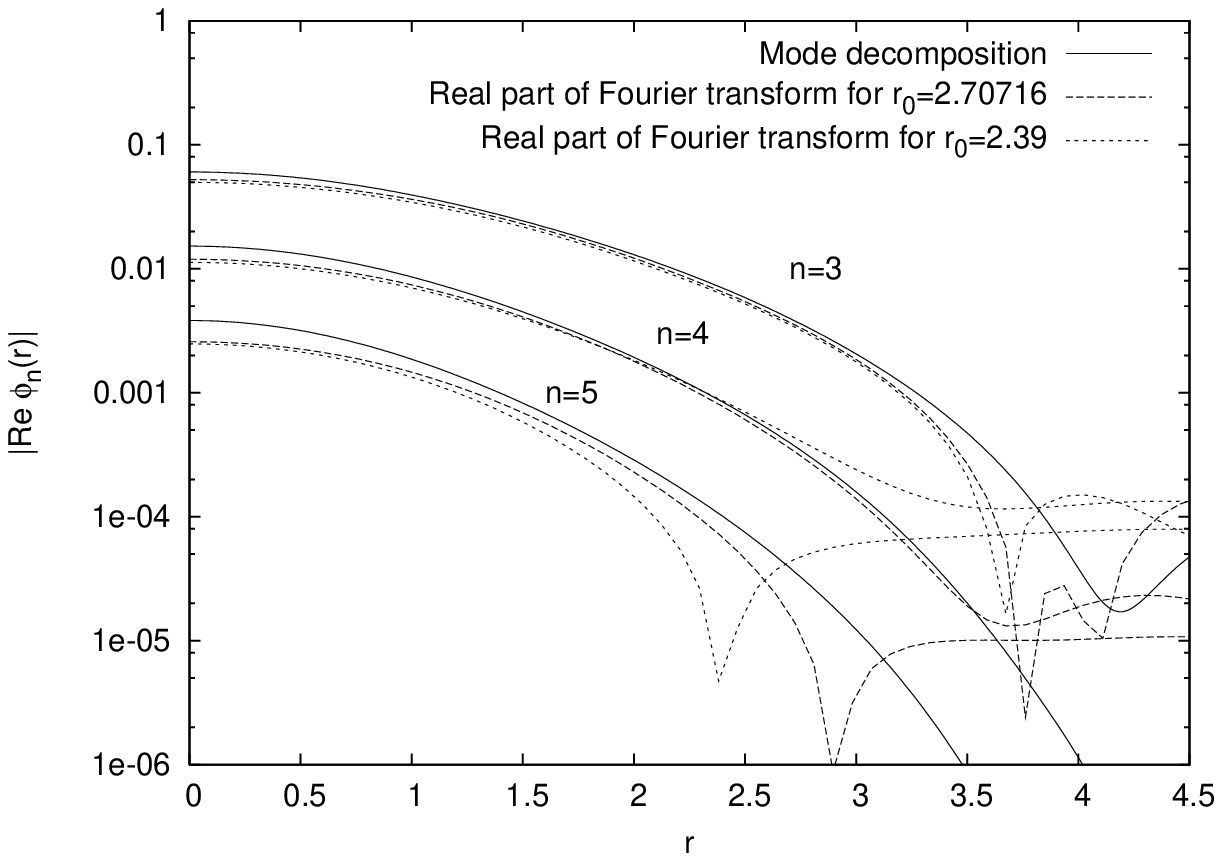}
\caption{\label{f:ntmode30b}
Comparison of the higher modes of the two systems with frequency
$\omega=1.3$ presented on the previous figure.
}
\end{figure}

\begin{figure}[!ht]
\includegraphics[width=12cm]{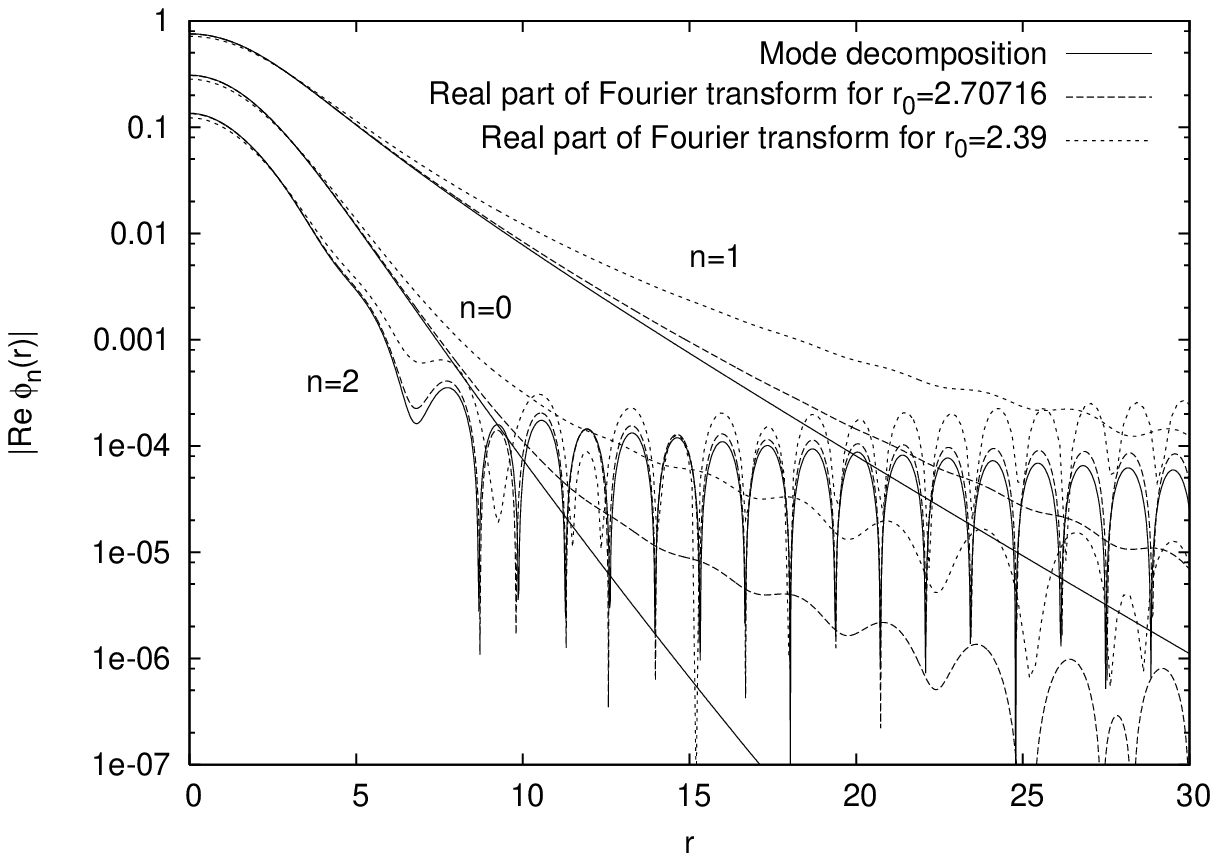}
\caption{\label{f:ntmode36a}
The modes of the periodic solution with frequency $1.36$ is compared with the
Fourier decomposition of two different oscillon states.
The first oscillon state with initial data $\phi_c=1$, $r_0=2.70716$ is decomposed
between the two subsequent maxima after $t=4158.76$, where the calculated
frequency is $1.3596$.
The second state with $\phi_c=1$, $r_0=2.39$ is decomposed just after $t=1526.58$,
where the frequency is $1.3613$.
%The order of the modes is the same as on the previous figure.
}
\end{figure}

\begin{figure}[!ht]
\includegraphics[width=12cm]{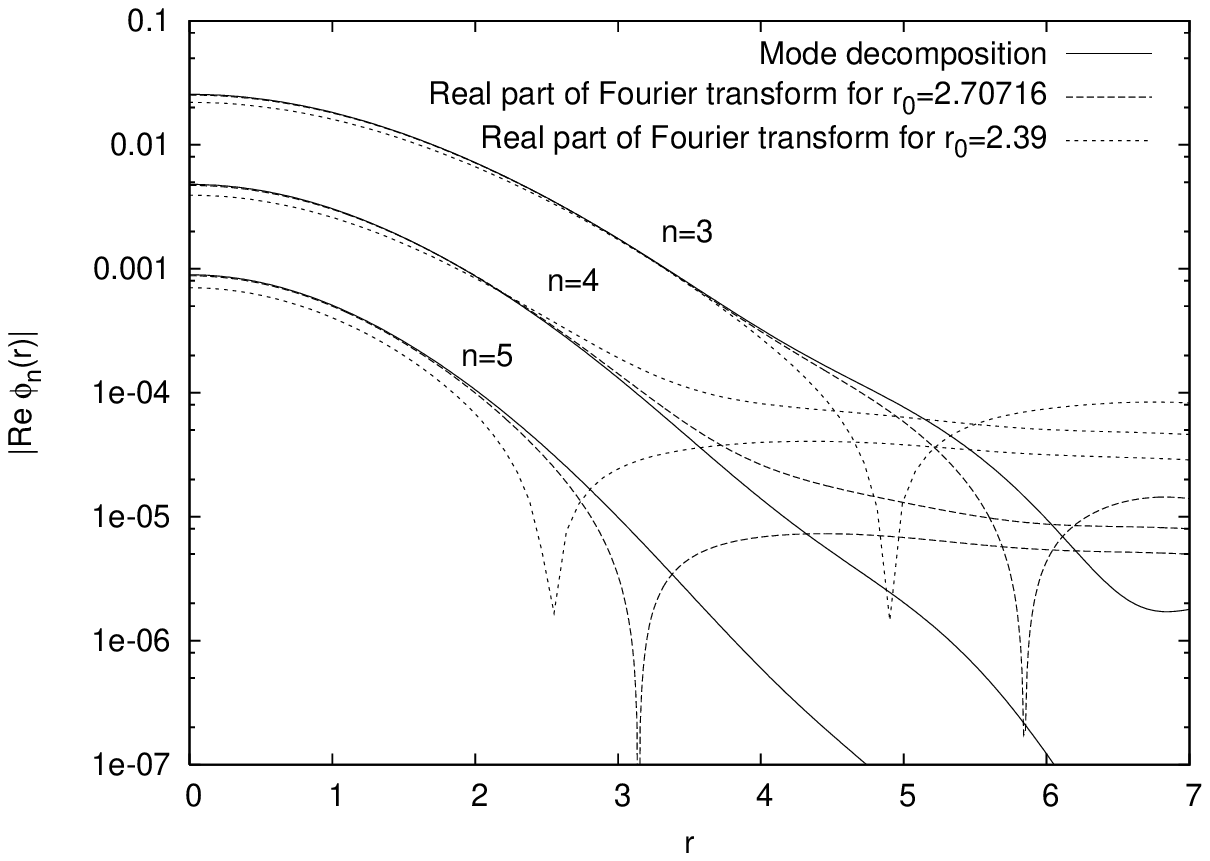}
\caption{\label{f:ntmode36b}
Comparison of the higher modes of the two systems with frequency
$\omega=1.36$ presented on the previous figure.
}
\end{figure}

Oscillons obtained by the evolution of generic Gaussian initial data are
relatively long living for a large set of the possible values of the initial
parameter $r_0$.
On Fig.\ \ref{f:nontune2} we exhibit the time evolution of the frequency of
the two typical non fine-tuned states presented on  Fig.\ \ref{f:nontune1}.
Although these states are clearly non-periodic, their lifetime is still very large
compared to the period of the basic high frequency oscillation mode.
We have performed the Fourier decomposition of these states at
various moments of time and compared the results to the mode decomposition of the
corresponding QB.
On Figs.\ \ref{f:ntmode30a},  \ref{f:ntmode30b}, \ref{f:ntmode36a} and
\ref{f:ntmode36b} the Fourier decomposition of the
two oscillon states shown on Figs.\ \ref{f:nontune1} and \ref{f:nontune2} is given
for moments of time
where the frequency is approximately $1.30$ and $1.36$.
For comparison the corresponding modes calculated by solving (\ref{e:system_bis})
using the given frequencies is also shown on the figures.
Although the agreement of the modes is not as excellent as for the
near-periodic states, the curves calculated by the two methods are
still remarkably close.
Especially important is the similarity at the tail section of the first oscillating
mode, i.e. mode $2$.
This indicates that the periodic quasi-breather solutions can describe, at least
qualitatively, even the generic long living oscillons.
In particular, the existence of the oscillating tail region is responsible for
the slow but steady energy loss of these configurations.

\subsection{Initial data obtained by mode decomposition}

\begin{figure}[!ht]
\includegraphics[width=12cm]{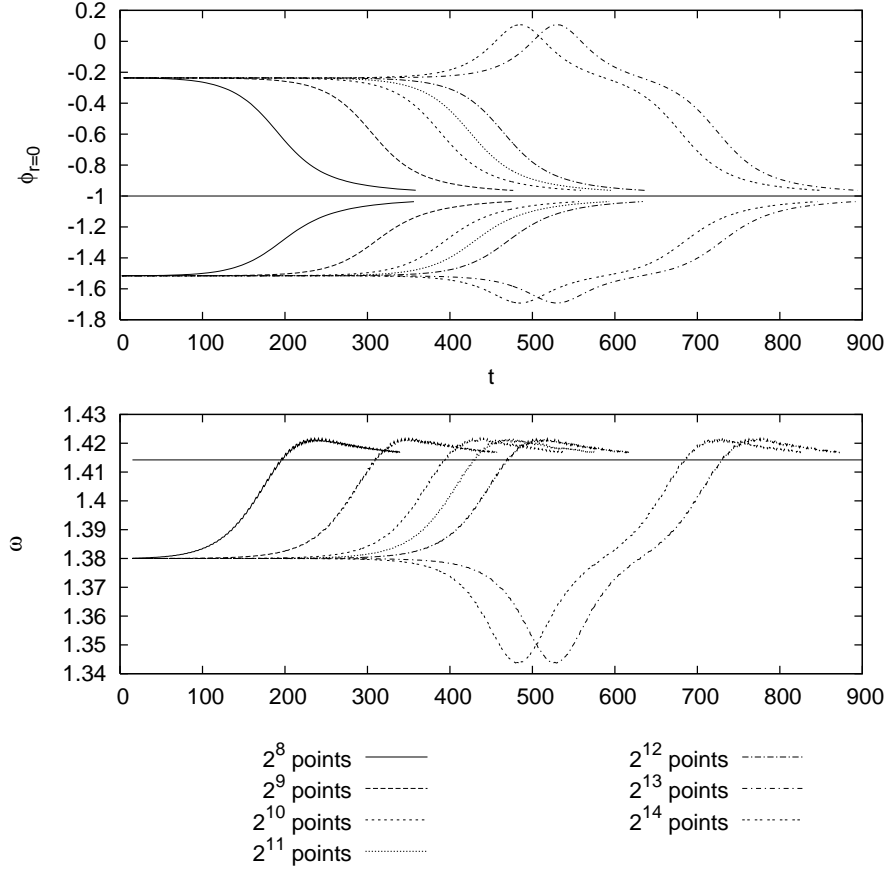}
\caption{\label{f:evolve}
On the first graph the upper and lower envelopes of the oscillations of the field at
the origin are shown as functions of time for numerical simulations with
various number of spatial grid points.
The initial data is generated by solving the system (\ref{e:system_bis}) for the
modes at the pulsation frequency $\omega = 1.38$.
On the second graph the time dependence of the oscillation frequency is plotted
for each simulation.
}
\end{figure}

We have also tested our scenario in the reverse way.
First, for a given $\omega$ (1.38 in the first example), we have solved
the system (\ref{e:system_bis}) for the modes of the quasi-breather.
In particular, we can compute $\phi\l(t=0\r)$ at the moment of time symmetry, and use
this as initial data for the evolution code.
The evolution of such initial data is shown on Fig. \ref{f:evolve} for numerical
simulation with various spatial resolutions.
The field (here its value at the origin) oscillates at the appropriate
frequency for a relatively long period (approx.\ t=300, with about 66 oscillations)
before it collapses in a sub or supercritical way very similar to the collapse of
the $\phi_c=-0.4$ states shown on Fig.\ \ref{f:max}.
The initial state is so close to the ideal configuration that it even depends
on the chosen numerical resolution whether in the final stage the configuration
collapses in a subcritical or supercritical way.
Strangely, the longest living state is not the one with the highest resolution,
which is probably connected to the fact that the initial data still contain
small numerical errors.
Although the lifetimes attainable by this method are very long compared to the
dynamical time scale of the linearized problem, they are still shorter than the
$\tau\approx 2000$ achievable by fine-tuning the first peak of the
$\phi_c=-0.4$ initial data.
Actually, it should not be surprising that the numerically determined initial data
corresponding to a QB cannot be as long-living as a configuration
which is fine-tuned to 32 decimal digits.

\begin{figure}[!ht]
\includegraphics[width=12cm]{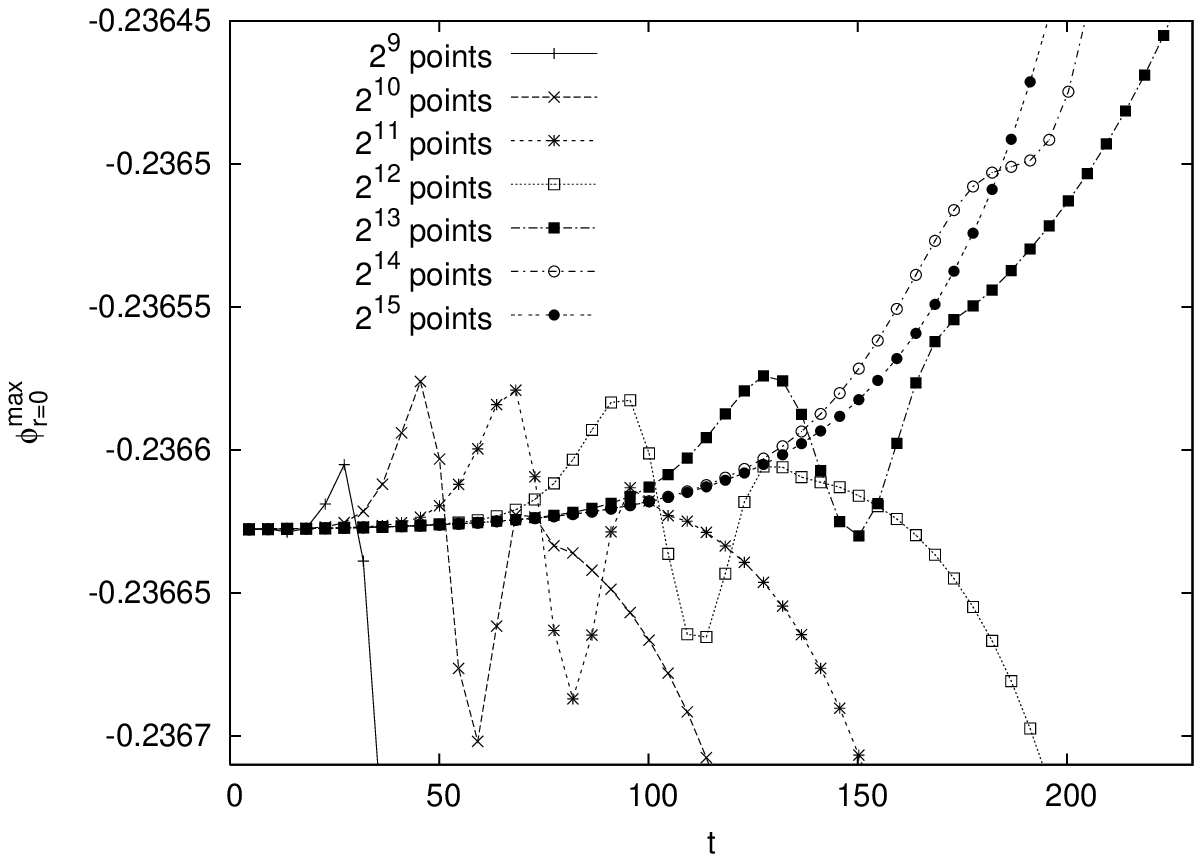}
\caption{\label{f:evolve2}
Initial behavior of the maximum curve of the evolution of the $\omega = 1.38$
initial data with numerical resolutions corresponding to various number of spatial
grid points.
The points on the curves represent the actual maximum points of the oscillating
field $\phi$ at the origin $r=0$.
%The time dependence of the frequency obtained from the highest resolution
%calculation is also given for reference.
}
\end{figure}

Since the modes are matched to oscillating tails at large radii, the initial
data provided by the mode decomposition of a QB is valid up to arbitrarily large
$r$.
However, because the slow falloff of the oscillating tails, the energy contained
in balls of radius $r$ diverges as $r$ goes to infinity.
Using such configuration as initial data for our evolution code does not cause
serious problems, since the physical distance between grid points in our numerical
representation increases with the distance from the center, and consequently
the high frequency tails cannot be represented numerically above a certain
radius.
This provides a cutoff in the initial configuration and an effective outer
boundary during the evolution.
An advantage of our conformal compactification method is that this outer boundary
moves to higher and higher radii when increasing the numerical resolution.
On Fig.\ \ref{f:evolve2} we show the initial part of the upper envelope curve
for numerical runs with various spatial resolutions.
We emphasize that this strong resolution dependence is entirely due to the
inadequate representation of the infinite energy initial data.
When we used initial data with compact support or sufficiently fast falloff,
such as the Gaussian initial data in (\ref{e:ff}), then the code remained
convergent up to much larger time periods (at the order of $t=10000$), and
the curves with various resolutions agreed to very high precision at the initial
stage (at around $t=100$).

\begin{figure}[!ht]
\includegraphics[width=12cm]{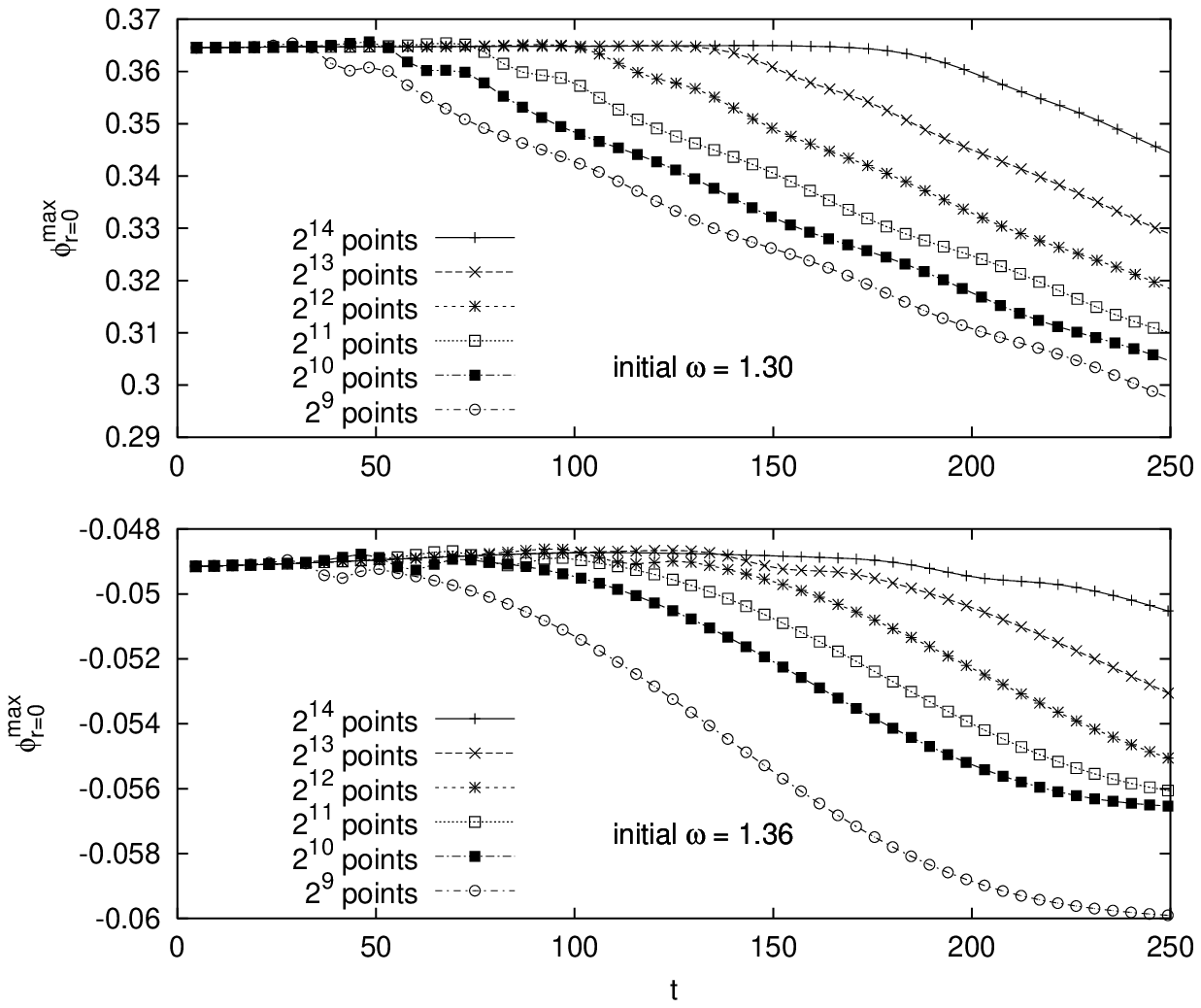}
\caption{\label{f:evolve3}
Initial behavior of the upper envelope of the oscillations at $r=0$ of the
evolution of the  $\omega=1.30$ and $\omega=1.36$ initial data for numerical
simulations with various number of spatial grid points.
}
\end{figure}

\begin{figure}[!ht]
\includegraphics[height=8cm]{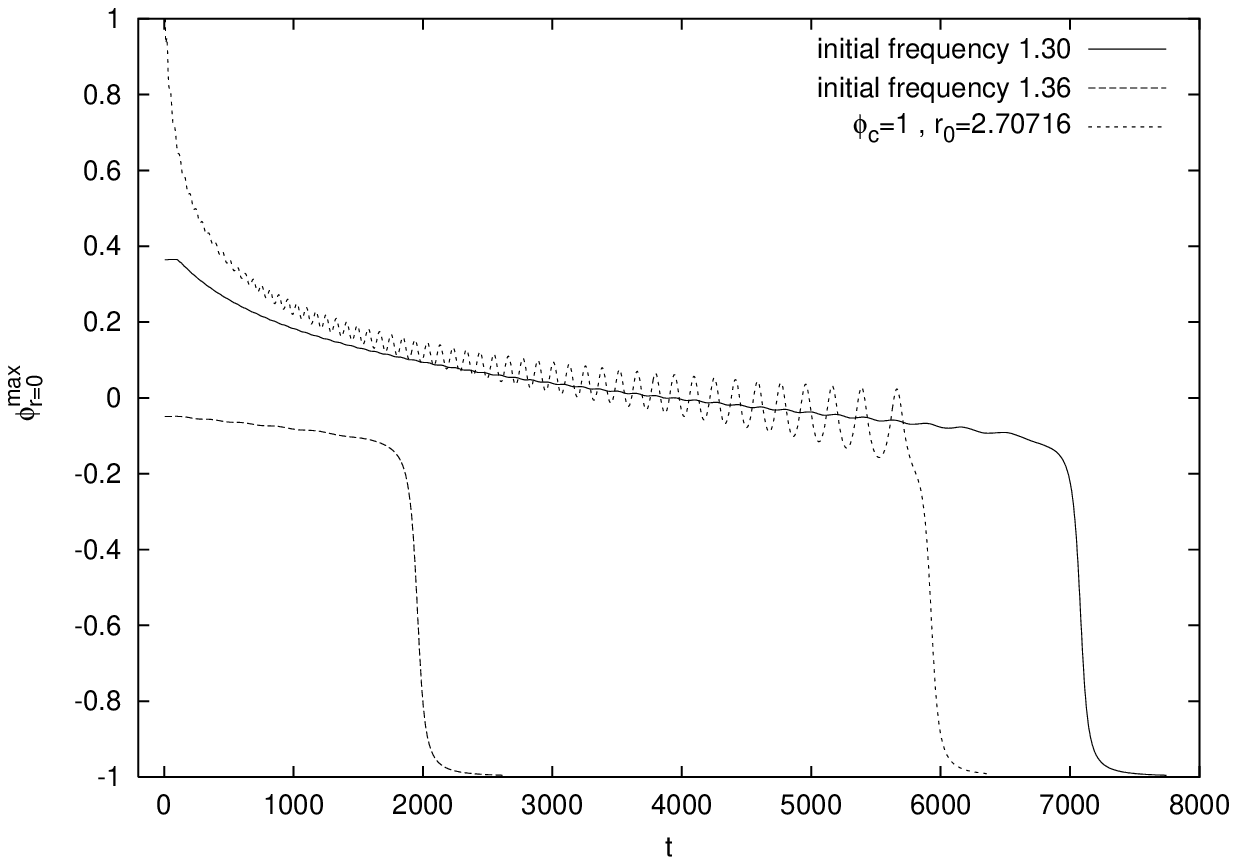}
\caption{\label{f:maxev}
Time evolution of the upper envelope curve of the oscillation of $\phi$ at $r=0$
generated by three different initial data.
The first two initial data is provided with the mode decomposition method,
solving the system (\ref{e:system_bis}) by choosing the
frequencies $\omega=1.30$ and $\omega=1.36$.
The third evolution corresponds to Gaussian-type initial data (\ref{e:ff})
with $\phi_c=1$ and $r_0=2.70716$.
}
\end{figure}

\begin{figure}[!ht]
\includegraphics[width=12cm]{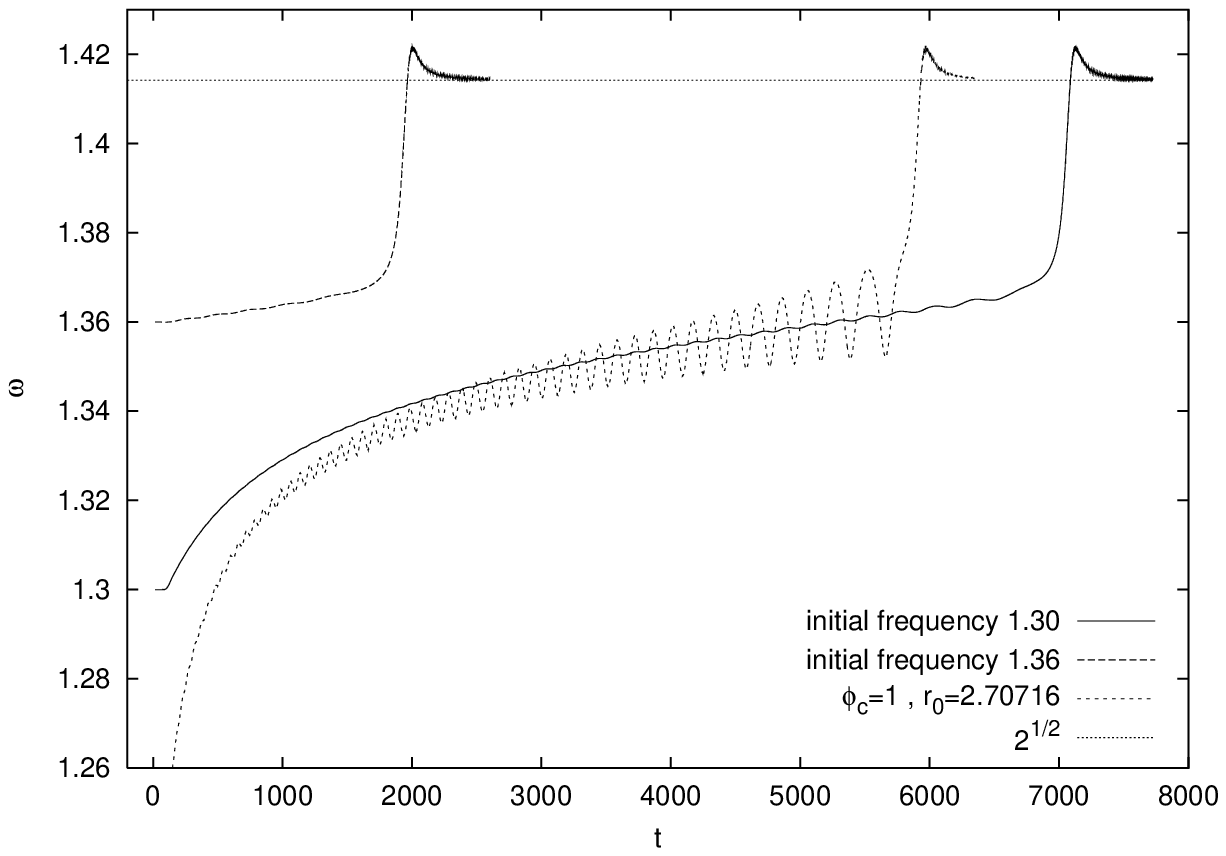}
\caption{\label{f:frev}
Time evolution of the frequency of the oscillation of $\phi$ at $r=0$ for the
three kind of evolution simulations shown on the previous figure.
}
\end{figure}

An important point is that, with this method, we can generate almost periodic
oscillons at frequencies which are difficult or impossible to attain by fine-tuning
initial data of the form (\ref{e:ff}).
On Fig.\ \ref{f:evolve3} we show the upper envelope of the oscillations at
$r=0$ in the initial stage of the evolutions when using initial data provided
by the mode decomposition method with frequencies $\omega=1.30$ and
$\omega=1.36$.
Again, as the numerical resolution is increased, the size of the
oscillating tail that can be taken into account in the numerical representation
of the initial data gets larger, and the evolution will remain nearly
periodic for longer times.
The field can oscillate truly periodically only if the energy lost by radiation
through the dynamically oscillating tail is balanced out by an incoming radiation
already present at the tail section of the initial data.
If the tail is cut off above some radius $r$ then the non-periodic region
moves inwards from that radius essentially at the speed of light.
Getting into the non-periodic domain, for the frequencies $\omega\leq 1.36$ the
tail amplitude appears to be large enough to cause a small but significant energy
loss, a decrease in amplitude, and consequently a steady frequency increase.
On Figs.\ \ref{f:maxev} and \ref{f:frev} we show the long-time behavior of the
amplitude and frequency of the oscillations generated by initial data with
frequencies $\omega=1.30$ and $\omega=1.36$.
Apart from a relatively short stable initial stage, the main characteristic of
the evolution is a slow but steady decrease of the amplitude accompanied by a
simultaneous increase of the frequency up to a point
(approx.\ $\omega=1.365$) where the configuration quickly decays.
This evolution is very similar to the behavior of generic
(i.e.\ not fine-tuned)
oscillons started from Gaussian initial data of type (\ref{e:ff}).
A typical example of such evolution, with initial data corresponding to an
$r_0$ close to the top of the lifetime curve (but between two peaks) with
$\phi_c=1,$ is also shown on Figs.\ \ref{f:maxev} and \ref{f:frev} in order to
facilitate comparison.
It is also rather remarkable that the onset of the rapid decay seems to coincide
with the configuration which minimizes the energy inside the core (see Fig. \ref{f:trans}).

\begin{figure}[!ht]
\includegraphics[width=12cm]{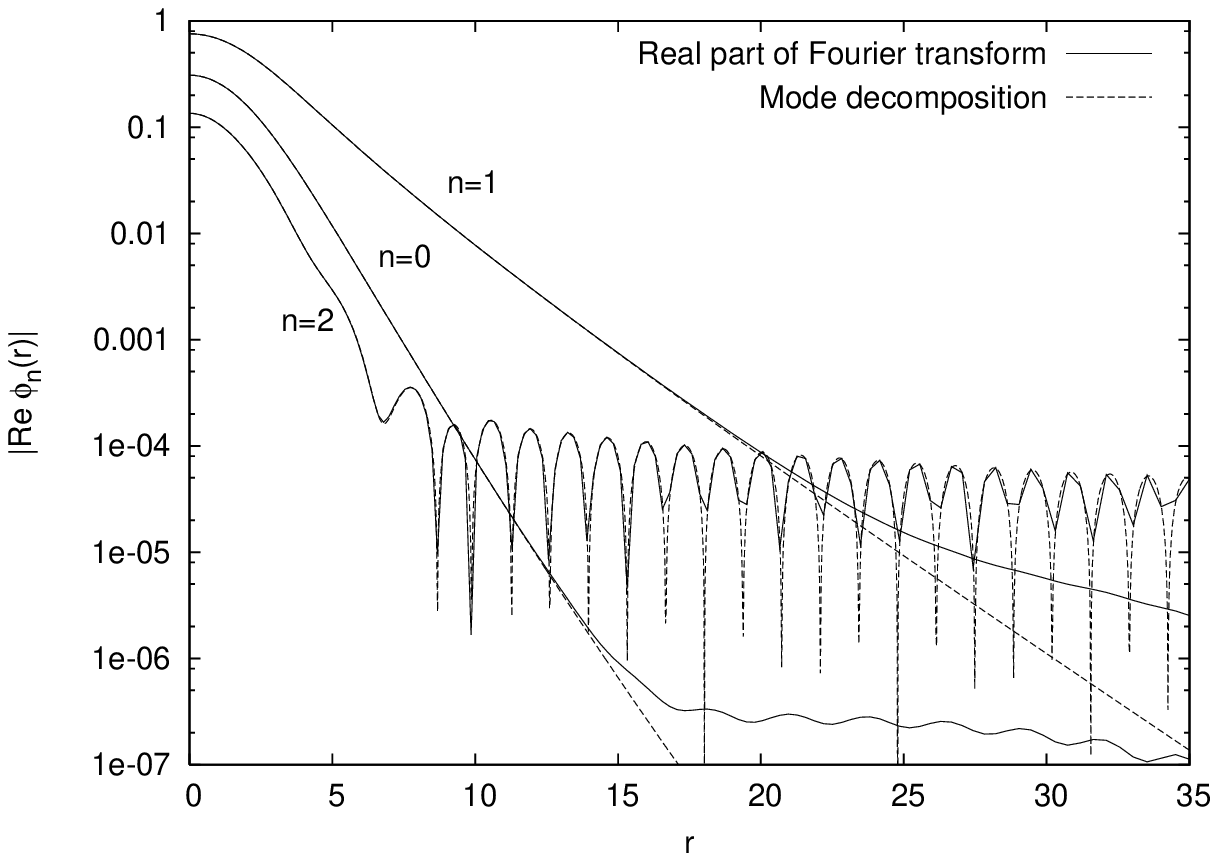}
\caption{\label{f:stable1}
Fourier decomposition of the evolution of the $\omega=1.3$ initial data close
to the moment of time when its frequency increases to $1.36$.
The results are compared to the modes of the periodic solution obtained
by the mode decomposition method for the $\omega=1.36$ frequency.
}
\end{figure}

\begin{figure}[!ht]
\includegraphics[width=12cm]{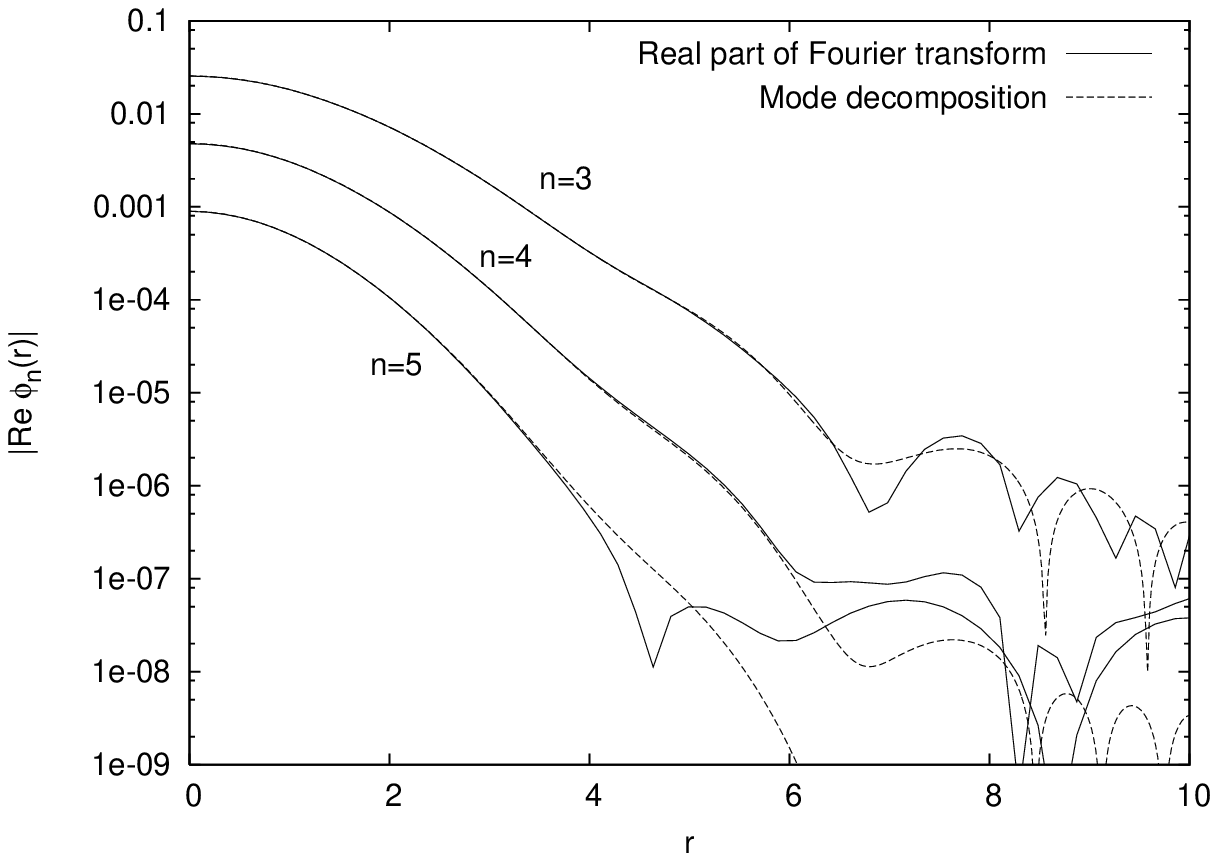}
\caption{\label{f:stable2}
Same thing as Fig. \ref{f:stable1} but for the higher order modes.
}
\end{figure}

The longtime evolution of a low frequency initial data provided by the mode
decomposition method is very similar to the evolution of generic oscillons
evolving from Gaussian initial data.
An important difference though, is the smaller low frequency modulation of the
envelope curve and of the time dependence of the frequency.
From this one can expect that the Fourier decomposition of these states is
even closer to that of the quasi-breather state with the corresponding frequency.
On Figures \ref{f:stable1}  and \ref{f:stable2} we give the Fourier
decomposition of the evolution
of the $1.3$ initial data between the two maxima after $t=5257.45$, where
the calculated frequency is $1.35997$.
The agreement is markedly better than on Figs. \ref{f:ntmode36a} and
\ref{f:ntmode36b}, which is most
likely due to the much lower amplitude of the low frequency modulations on the
envelopes of field value at the center.
These low frequency modes are excited to a higher amplitude by Gaussian type of
initial data, while they appear to be present to a much lower extent in the
lower frequency periodic initial data (with $\omega=1.3$ in this case).
We remind the reader that the existence of the peaks on the lifetime curve
(see Fig.\ 4 of \cite{Honda}) is closely related to these low frequency
oscillations.
Namely, the peaks separate domains of the initial data in $r_0$ with different
number of low frequency oscillations on the envelope curve.

\section{Conclusion}\label{s:conclu}
In this paper, we have adapted and applied spectral methods implemented in the
LORENE library to find time-periodic solutions of the spherically symmetric
wave equation of $\phi^4$ theory.
Our code passed numerous tests and we are quite
confident that it is sufficiently precise.
With our code we find that for frequencies $0<\omega<\sqrt{2}$ there is a whole family of
standing wave-type solutions with a regular origin having a well
localized core and an oscillatory tail.
Because of the slow decrease of the oscillatory tail the total energy of these solutions
in a ball of sufficiently large radius, $R$, is proportional to $R$ cf.\ Figs.\ (15,16).
We have constructed a special class of solutions, called quasi-breathers,
defined by minimizing the amplitude of the oscillatory tail of the solutions.
The size of the core of the QBs gets larger and larger as $\omega\to\sqrt{2}$ and the
amplitude of the oscillatory tail is getting smaller and smaller. Interestingly the energy
contained in the core of the QBs exhibits a minimum for $\omega\approx 1.365$.

Using our high precision time evolution code, we have also investigated the time evolution of Gaussian initial data.
We observe that a generic oscillon state can be characterized by a slowly varying frequency,
$\omega(t)$, increasing in time up to a critical one $\omega_{\rm c}$, when it decays
rapidly. We have also investigated in detail the near-periodic states described first
in Ref.\ \cite{Honda} which can be characterized by an almost constant frequency.
In contradistinction to Ref.\ \cite{Honda} we find such  near-periodic states
for any frequency $\sqrt{2}>\omega>\omega_{\rm c}$.
Moreover we find that the near-periodic states decay in time (i.e.\ they cannot be truly time periodic states)
cf. Figs.\ (\ref{f:fr},\ref{f:en}). In particular while loosing some of their energy
their frequency decreases very slowly with time.

By a careful comparison of the Fourier modes of an oscillon state of frequency $\omega(t)$
with that of the corresponding QB, we have obtained convincing evidence
that the localized part of the oscillon is nothing but the core
of the QB of the same frequency.
Our results demonstrate that the time evolution of an oscillon
state can be described to a good approximation as an adiabatic evolution through a sequence of QBs
with a slowly changing frequency  $\omega(t)$.
What is more, the oscillatory tail of the QB describes
very well the standing wave part of the oscillon.
Therefore any oscillon contains the core and a significant part
of the oscillatory tail of the corresponding QB.
The existence of near-periodic states is closely related to the fact that the energy of the oscillon
core exhibits a minimum for a critical frequency $\omega_{\rm c}\approx 1.365$ and for
$\sqrt{2}>\omega>\omega_{\rm c}$ a single unstable mode appears which can be suppressed
by fine-tuning the initial data.

\section{Acknowledgments}

This research has been supported by OTKA Grants No. T034337, TS044665,
K61636 and by the Pole Numerique Meudon-Tours.
G. F and I. R would like to thank the Bolyai Foundation for financial
support.

\end{document}